%% file: discord.tex
\documentclass[a4paper,11pt]{article}
\pdfoutput=1 % to allow compilation in JCAP
\usepackage{jcappub}
\usepackage{amsmath}
\usepackage{amssymb}
\usepackage{bm}
\usepackage{graphicx}
\usepackage{enumitem}
\usepackage{geometry}
\usepackage{xspace}
\usepackage{xcolor}
\usepackage{pdflscape}
\usepackage{enumerate}
\usepackage{soul} % Added to be able to strike through text instead of removing it

\allowdisplaybreaks

\makeatletter
\gdef\@fpheader{}
\g@addto@macro\bfseries{\boldmath}
\makeatother

\input{newcommands}

\def\setI{\mathbb{I}}

\newcommand{\kmk}{\pm\bm{k}}
\newcommand{\udS}{{}}

\subheader{}

\title{Discord and Decoherence}

\author[a]{J\'er\^ome Martin,}
\author[a,b]{Amaury Micheli,}
\author[c,a]{Vincent Vennin}

\affiliation[a]{Institut d'Astrophysique de Paris, UMR 7095-CNRS,
Universit\'e Pierre et Marie Curie, 98 bis boulevard Arago, 75014
Paris, France}

\affiliation[b]{  IJCLab, CNRS/IN2P3, Université Paris-Saclay, 91405 Orsay, France
}
  
\affiliation[c]{Laboratoire Astroparticule et Cosmologie, CNRS Universit\'e de Paris, 10 rue Alice Domon et L\'eonie Duquet, 75013 Paris, France}

\emailAdd{jmartin@iap.fr}
\emailAdd{amaury.micheli@ijclab.in2p3.fr}
\emailAdd{vincent.vennin@cnrs.fr}

\date{today}

\begin{document}

\sloppy

\abstract{In quantum information theory, quantum discord has been proposed as a tool to characterise the presence of ``quantum correlations'' between the subparts of a given system. Whether a system behaves quantum-mechanically or classically is believed to be impacted by the phenomenon of decoherence, which originates from the unavoidable interaction between this system and an environment. Generically, decoherence is associated with a decrease of the state purity, i.e. a transition from a pure to a mixed state. In this paper, we investigate how quantum discord is modified by this quantum-to-classical transition. This study is carried out on systems described by quadratic Hamiltonians and Gaussian states, with generalised squeezing parameters. A generic parametrisation is also introduced to describe the way the system is partitioned into two subsystems. We find that the evolution of quantum discord in presence of an environment is a competition between the growth of the squeezing amplitude and the decrease of the state purity. In phase space, this corresponds to whether the semi-minor axis of the Wigner ellipse increases or decreases, which has a clear geometrical interpretation. Finally, these considerations are applied to primordial cosmological perturbations, where we find that quantum discord can remain large even in the presence of strong decoherence.}

%thus allowing us to investigate how large-scale structures in our universe, which are believed to arise from quantum fluctuations, can exhibit classical properties.}

\maketitle

\section{Introduction}
\label{sec:intro}

A intriguing fact in modern science is that, sometimes, it is not
straightforward to decide whether a system behaves classically or
quantum-mechanically. This is for instance the case in Cosmology where
it is believed that the structures observed in our universe are
nothing but quantum fluctuations amplified to astrophysical
scales~\cite{Starobinsky:1979ty,Mukhanov:1981xt,Grishchuk:1990bj}. Even
if this hypothesis allows us to explain the properties of these
structures, acquiring evidence that would establish their origin
beyond any doubt turns out to be highly non-trivial. Indeed, assuming
that the primordial fluctuations are stochastic rather than quantum
leads to almost the same consequences up to corrections that, in
practice, are very difficult to reveal
experimentally~\cite{Polarski:1995jg,Martin:2015qta}.

  Recently, new methods have been developed to address the question of
  whether a system is classical or quantum-mechanical. A typical
  approach consists in dividing the system into two sub-systems and to
  study and characterise the nature of the correlations between these
  two sub-systems. As a matter of fact, there exist efficient tools to
  decide whether correlations are classical or quantum-mechanical in
  nature. Sometimes, indeed, correlations are impossible to understand
  in a classical framework (see, for instance, the Bell
  experiments~\cite{Bell:1964kc,Brune:1996zz}) which establishes
  unambiguously their quantum origin. This strategy leads to the
  concept of quantum
  discord~\cite{Henderson:2001,Zurek:2001}. However, the ability of
  quantum discord to precisely identify the quantum nature of some
  correlations has been challenged in the case of mixed states while,
  in the case of pure states, there is a one-to-one correspondence
  between quantum discord and entropy of
  entanglement~\cite{2018RPPh...81b4001B,2010arXiv1003.5256D}. On the
  other hand, the quantum-to-classical transition of a system is
  generically believed to be connected to the phenomenon of
  decoherence~\cite{Zurek:1981xq,Zurek:1982ii}. This mechanism, which
  has been observed in the laboratory~\cite{Brune:1996zz}, takes into
  account that any system is in fact always an open system, namely a
  system in interaction with other degrees of freedom that
  collectively constitute an environment. This interaction, when one
  is only interested in the properties of the system, is responsible
  for the appearance of classical properties.

  It is therefore interesting to study how the quantum discord
  ``responds'' to the presence of decoherence in a system and to
  investigate how quantum discord can track the ``classicalization''
  of a system. This is the main goal of the present paper. This study
  will be carried out in the generic case of a quadratic
  Hamiltonian. Physically, this is very relevant since many systems
  are described by this type of Hamiltonians. This is for instance the
  case for the Schwinger effect, the dynamical Casimir effect, the
  Hawking effect, inflationary fluctuations, etc. Technically, this is advantageous since the
  quantisation of these systems always leads to Gaussian states for
  which there exists an efficient formalism permitting the calculation
  of quantum discord. When it comes to concrete applications, we will
  consider the example of cosmological
  perturbations~\cite{Mukhanov:1990me}. In addition to the advantages
  mentioned above, this will also allow us to shed new light on the
  question of whether their quantum origin can be observationally
  revealed, a long-standing question in Cosmology that has recently
  been the subject of many new studies~\cite{Polarski:1995jg,Burgess:2006jn,CampoParentani,Choudhury:2016cso,Hollowood:2017bil,Martin:2018zbe,Martin:2018lin,Martin:2021qkg,Lin:2010pfa,Hsiang:2021qqo}.

  This article is organised as follows. In~\Sec{sec:DiscordQField}, we
  present a description of the quadratic systems considered in this
  paper and provide the formulas permitting the calculation of their
  quantum discord. In~\Sec{sec:timeevolutionwo}, as a warm-up, we
  explain how the time evolution of these systems and their quantum
  discord can be calculated in absence of an
  environment. In~\Sec{sec:caldleggeteom}, we introduce a simple
  model, based on the Caldeira-Leggett model, which allows us to study
  and calculate quantum discord in presence of
  decoherence. In~\Sec{sec:CosmoPerturbations}, we apply this
  formalism to the theory of cosmological perturbations of
  quantum-mechanical origin. At the end of the article,
  in~\Sec{sec:conclusions}, we present our conclusions. Finally, the
  technical details of our calculations are given in a series of
  appendices. In~\Sec{ap:partition}, we come back to the notion of
  partitions of a system and explain it in more
  details. In~\Sec{app:cov}, we calculate the covariance matrix of a
  Gaussian system for an arbitrary partition. In~\Sec{ap:discord}, we
  explain how the formula giving the quantum discord used in the main
  text is arrived at. In~\Sec{ap:sec:covmatcaldlegget}, we calculate
  the covariance matrix of the system in presence of an environment
  and derive efficient approximations for its components.

\section{Quantum discord of a Gaussian field}
\label{sec:DiscordQField}

\subsection{Quantum phase space}
\label{subsec:description}

In this work we consider the case of a real quantum scalar field with
a local quadratic Hamiltonian
\begin{align}
\label{eq:Hamiltonian:generic}
\hat{H} = \frac{1}{2} \int_{\setR^3} \dd^3 \bm{x}\,  
\hat{z}^\mathrm{T}(\bm{x})
\Lambda(\tau) \hat{z}(\bm{x})\, ,
\end{align}
where $\hat{z}(\bm{x})= \left( \hat{\phi}(\bm{x}) \, ,
\hat{\pi}_\phi(\bm{x}) \right)^{\mathrm{T}}$ contains the field $\hat{\phi}$
and its conjugate momentum $\hat{\pi}_\phi$, which satisfy the
canonical commutation relations
\begin{align}
\label{eq:Com:Rel:Fields}
\left[\hat{\phi}(\bm{x}), \hat{\pi}_\phi(\bm{y})\right]
= i \delta(\bm{x}-\bm{y}).
\end{align} 
We assume that the $2$$\times$$2$ symmetric matrix $\Lambda(\tau)$
does not depend on $\bm{x}$ but only on time $\tau$, which can result
from the invariance under spatial translations of the physical setup
on which the field is introduced. For instance, the field $\phi$ may
describe cosmological perturbations evolving on top of a homogeneous
and isotropic background, as further discussed in
\Sec{sec:CosmoPerturbations}, but for now the formalism we develop
remains generic and applies to any system described by a (possibly
infinite) collection of parametric oscillators.
%This is for instance the case for the Schwinger effect, the dynamical
%Casimir effect, the Hawking effect, \etc.
Note that $\Lambda_{\phi\phi}(\tau)$ can nonetheless contain gradient
operators $\partial/\partial\bm{x}$ to any positive (in agreement with
the locality assumption) and even (in order to preserve homogeneity
and isotropy) power. If the theory does not feature higher-than two
derivatives, which we assume here, then the other entries of
${\Lambda}$ cannot contain spatial gradients. In what follows we
introduce several successive canonical transformation, \ie changes of
variables that preserve the structure of the
commutators~\eqref{eq:Com:Rel:Fields}, which make the expression of
the Hamiltonian~\eqref{eq:Hamiltonian:generic} simpler.

Let us perform a first canonical transformation and introduce the
variables $\hat{v}(\bm{x})$ and $\hat{p}(\bm{x})$ defined as
\begin{align}
\begin{split}
\label{eq:canonicalTransform:vp}
\hat{\phi}(\bm{x}) &= \sqrt{\Lambda_{\pi\pi}}\hat{v}(\bm{x})\, \\
\hat{\pi}_\phi(\bm{x}) &=
\left(\frac{1}{2}\frac{{\Lambda}'_{\pi\pi}}{\Lambda_{\pi\pi}}
-\Lambda_{\phi\pi}\right)\frac{\hat{v}(\bm{x})}{\sqrt{\Lambda_{\pi\pi}}}
+\frac{\hat{p}(\bm{x})}{ \sqrt{\Lambda_{\pi\pi}}}\, ,
\end{split}
\end{align}
where a prime denotes derivation with respect to time, and where one
can easily check that $\hat{v}$ and $\hat{p}$ obey the same
commutation relations as the original fields, see
\Eq{eq:Com:Rel:Fields}. In terms of these new variables, the
Hamiltonian takes the simple form~\cite{Grain:2019vnq}
\begin{align}
\label{eq:Hamiltonian:realSpace}
\hat{H} = \frac{1}{2}\int_{\setR^3}\dd^3\bm{x} \left[\hat{p}^2(\bm{x})
  +\omega^2 \hat{v}^2(\bm{x})\right]\, ,
\end{align}
where
$\omega^2=\Lambda_{\phi\phi}\Lambda_{\pi\pi}+1/2({\Lambda}''_{\pi\pi}/
\Lambda_{\pi\pi})-3/4
({\Lambda}'_{\pi\pi}/\Lambda_{\pi\pi})^2
-{\Lambda}'_{\phi\pi}-\Lambda_{\phi\pi}^2
+\Lambda_{\phi\pi}{\Lambda}'_{\pi\pi}/\Lambda_{\pi\pi}$
encodes all the information about the dynamics.

We then perform a second canonical transformation, in the form of the
Fourier expansion
\begin{align}
\label{eq:msfourier}
\hat{v}(\bm{x}) = \frac{1}{\left(2\pi\right)^{3/2}}
\int_{\setR^3}\dd^3\bm{k}\,  \ee^{-i\bm{k}\cdot\bm{x}} \hat{v}_{\bm{k}}
\end{align}
and a similar expression for $\hat{p}(\bm{x})$. The fact that this
defines a canonical transformation can be easily seen from combining
the inverse Fourier transform,
$\hat{v}_{\bm{k}}=(2\pi)^{-3/2}\int\dd^3\bm{x} \, \ee^{i \bm{k}\cdot
  \bm{x}} \hat{v}(\bm{x})$ (and a similar expression for
$\hat{p}_{\bm{k}}$) with \Eq{eq:Com:Rel:Fields} for the fields
$\hat{v}(\bm{x})$ and $\hat{p}(\bm{x})$, which yields
\begin{align}
  \left[\hat{v}_{\bm{k}},\hat{p}_{\bm{k}'}^\dagger\right]
  = i \delta(\bm{k}-\bm{k}')\, ,
\end{align} 
while $ [\hat{v}_{\bm{k}},\hat{v}_{\bm{k}'}^\dagger] =
[\hat{p}_{\bm{k}},\hat{p}_{\bm{k}'}^\dagger] = 0$.  Plugging the
Fourier expansions into the
Hamiltonian~\eqref{eq:Hamiltonian:realSpace}, one obtains
\begin{align}
\label{eq:Hsystem}
\hat{H} =\int _{\setR^{3+}}\dd ^3 \bm{k} \, \hat{\mathcal{H}}_{\bm{k}}
=\int_{\setR^{3+}} \dd ^3 \bm{k}\left[
  \hat{p}_{\bm{k}}\hat{p}_{\bm{k}}^{\dagger}
  +\omega^2\left(k, t \right)
  \hat{v}_{\bm{k}}\hat{v}_{\bm{k}}^{\dagger}\right],
\end{align}
which defines the Hamiltonian density in Fourier space
$\hat{\mathcal{H}}_{\bm{k}}$. In this expression, $\omega$ can depend
on $k$ since, as pointed out above, it may involve the gradient
operator. It is important to notice that the operators $\hat{v}_{\bm
  k}$ and $\hat{p}_{\bmk}$ are not Hermitian. Indeed, since $\hat{v}(
t ,{\bm x})$ is real, one has $\hat{v}_{\bm k}^{\dagger}=\hat{v}_{-\bm
  k}$ and a similar relation for the conjugate momentum. This shows
that independent degrees of freedom are labelled by half the Fourier
space only, and explains why the integral is performed over
$\setR^{3+}=\setR^2\times \setR^+$ in \Eq{eq:Hsystem}.  In the
  helicity basis, this also allows one to decompose the fields
  $\hat{v}_{\bm k}$ and $\hat{p}_{\bmk}$ onto creation and
  annihilation operators as
\begin{align}
\label{eq:ladder:operators}
\hat{v}_{\bm{k}} = \frac{1}{\sqrt{2k}}
\left(\hat{c}_{\bm{k}}+\hat{c}_{-\bm{k}}^\dagger \right)
\qquad \text{and}\qquad
\hat{p}_{\bm{k}} = -i\sqrt{\frac{k}{2}}
\left(\hat{c}_{\bm{k}}-\hat{c}_{-\bm{k}}^\dagger \right),
\end{align}
where $\hat{c}_{\bm{k}}$ and $\hat{c}_{\bm{k}'}^\dagger$ obey the
commutation relation
$[\hat{c}_{\bm{k}},\hat{c}_{\bm{k}'}^\dagger]=\delta(\bm{k}-\bm{k}')$. By
plugging the above into \Eq{eq:Hsystem}, one obtains
    \begin{align}
\label{eq:Hamiltonian:COperators}
\hat{H}=&\int_{\setR^{3+}} \dd ^3\bm k \biggl[  \frac{k}{2}
  \left( \frac{\omega^2}{k^2} +1 \right)
  \left( \hat{c}_{\bm k} \hat{c}_{\bm k}^\dagger
  + \hat{c}_{- \bm k}^{\dagger} \hat{c}_{- \bm k}
  \right)
 % \nonumber \\ &
  + \frac{k}{2} \left( \frac{\omega^2}{k^2} - 1 \right)
  \left( \hat{c}_{\bm k} \hat{c}_{-\bm k} + \hat{c}_{- \bm k}^{\dagger}
  \hat{c}_{\bm k}^{\dagger} \right)  \biggr] .
\end{align}
In this expression, the first term does not lead to net particle
creation and represents a collection of free oscillators while the
second term either creates or destroys a pair of particles with
momenta $\bm{k}$ and $-\bm{k}$, and can be seen as resulting from the
interaction with an exterior classical source.  Note that the four
combinations of ladder operators appearing in
\Eq{eq:Hamiltonian:COperators} are the only quadratic terms that are
allowed by statistical isotropy, \ie they are the only combinations
that ensure momentum conservation in the particle content. Let us also
notice that the form~(\ref{eq:Hamiltonian:COperators}) is not the one
commonly used in Cosmology [which is given by Eq.~(14)
  of~\Refa{Martin:2015qta}]. However, it is related to it by a simple
canonical transformation and is, therefore, equivalent to it.

Since $\hat{v}_{\bm k}$ and $\hat{p}_{\bmk}$ are not Hermitian, it is
convenient to perform a third and last canonical transformation, and
introduce the Hermitian operators corresponding to the Hermitian and
anti-Hermitian parts of $\hat{v}_{\bm{k}}$ and $\hat{p}_{\bm{k}}$,
\begin{align}
\label{eq:vRI:def}
\hat{v}_{\bm{k}}^{\mathrm{R}} =
\frac{\hat{v}_{\bm{k}}+\hat{v}_{\bm{k}}^\dagger}{\sqrt{2}} \, ,\qquad
\hat{v}_{\bm{k}}^{\mathrm{I}} =
\frac{\hat{v}_{\bm{k}}-\hat{v}_{\bm{k}}^\dagger}{\sqrt{2} i} \,
,\qquad \hat{p}_{\bm{k}}^{\mathrm{R}} =
\frac{\hat{p}_{\bm{k}}+\hat{p}_{\bm{k}}^\dagger}{\sqrt{2}} \, ,\qquad
\hat{p}_{\bm{k}}^{\mathrm{I}} =
\frac{\hat{p}_{\bm{k}}-\hat{p}_{\bm{k}}^\dagger}{\sqrt{2} i}\, .
\end{align}
The transformation~\eqref{eq:vRI:def} can be inverted according to
$\hat{v}_{\bm k}= \left(\hat{v}_{\bm k}^{\mathrm{R}}+i\hat{v}_{\bm
  k}^{\mathrm{I}}\right)/\sqrt{2}$ and $\hat{p}_{\bm k}=
\left(\hat{p}_{\bm k}^{\mathrm{R}}+i\hat{p}_{\bm
  k}^{\mathrm{I}}\right)/\sqrt{2}$, and one can readily check that
these relations define a canonical transformation, namely that
$[\hat{v}_{\bm{k}}^s,\hat{p}_{\bm{k}'}^{s'}]=i\delta(\bm{k}-\bm{k}')\delta_{s,s'}$
and that
$[\hat{v}_{\bm{k}}^s,\hat{v}_{\bm{k}'}^{s'}]
=[\hat{p}_{\bm{k}}^s,\hat{p}_{\bm{k}'}^{s'}]=0$
where $s=\mathrm{R}, \mathrm{I}$. It is also clear from these
expressions that $\hat{v}_{\bm{k}}^{s}$ and $\hat{p}_{\bm{k}}^{s}$ are
Hermitian operators, and that the Hamiltonian reads
\begin{align}
\label{eq:Hamiltonian:RI}
  \hat{H} &=\int_{\setR^{3+}} \dd ^3\bm k\sum_{s=\mathrm{R,I}}
  \hat{\mathcal{H}}_{\bm k}^s=\frac12\int_{\setR^{3+}} \dd ^3\bm k\sum_{s=\mathrm{R,I}}
\left[(\hat{p}_{\bm k}^s)^2+\omega^2(k, \tau )(\hat{v}_{\bm k}^s)^2\right] \, .
\end{align} 
%Let us note that, formally, the above set of transformations simply consists in working with a different parametrisation of the Hilbert space,
%\begin{align}
%\label{eq:HilbertSpace}
%{\mathcal E}=\bigotimes_{\bm{x}\in\setR^3} {\mathcal E}_{\bm{x}} 
%=\bigotimes_{\bm{k}\in\setR^{3+}}{\mathcal E}_{\bm{k}} \otimes {\mathcal E}_{-\bm{k}}
%=\bigotimes_{\bm{k}\in\setR^{3+}}\bigotimes_{s\in\{\mathrm{R},\mathrm{I}\}} {\mathcal E}_{\bm{k}}^s\, . 
%\end{align}
The advantage of this last parameterisation is that it makes the
Hamiltonian sum separable, see \Eq{eq:Hamiltonian:RI}. In other words,
it describes a collection of independent parametric oscillators. If
the initial quantum state is factorisable in that basis, which is for
instance the case for the vacuum state selected by that Hamiltonian,
it remains so at later time, and the dynamical evolution does not
generate correlation or entanglement between different subspaces.
\subsection{Partitions}
\label{sec:partitions}
As mentioned above, the system under consideration can be factorised
into independent Fourier subspaces, within which entangled pairs of
particles with opposite wave-momenta are created. Our goal is to
measure the amount of entanglement associated with this mechanism, and
to determine whether the resulting correlations have genuinely quantum
properties.

Experiments aimed at testing the quantum nature of a physical setup
usually rely on probing the properties of the correlations between two
of its subsystems. In Bell inequality experiments for instance, the
correlations between the spin states of two entangled particles are
tested against a possible local hidden-variables theory. In that case,
the way the physical system is split into two subsystems is obvious:
the two subsystems are simply the two space-like separated
particles. One may choose to parameterise phase space by means of other
combinations of the spin operators, but given that Bell experiments
test for locality, it is clear that the two particles constitute a
preferred partition.

The situation is however less clear for quantum fields. One may still
choose to work in real space, and probe the nature of the correlations
for two spatially-separated regions, see for instance
\Refs{Martin:2021xml, Martin:2021qkg}. However, in this approach, one
has to deal with mixed states, coming from the fact that when
observing the field at two distinct locations in real space one
implicitly traces over the configurations of the field in all other
locations, making the reduced state of interest a mixed one. This
problem does not occur in Fourier space since different Fourier
subspaces are uncoupled. Since we want to study the effect of
decoherence on the presence of quantum correlations, it seems
important to first isolate the decoherence associated with the
coupling to environmental degrees of freedom, from the one coming from
the effective mixing effect mentioned above. This is why in this work
we choose to study correlations within Fourier subspaces, leaving the
combination of both mixing effects (\ie the analysis of quantum
discord in real space in the presence of an environment) for future
work.

In Fourier space, there is no obvious way to split the system into two
subsystems. At the technical level, this implies that the construction
of Hermitian operators out of $\hat{v}_{\bm{k}}$ and
$\hat{p}_{\bm{k}}$ is not unique, and that \Eq{eq:vRI:def} is not the
only possibility. For instance, one can consider the set of operators
$\hat{q}_{\bm k}$ and $\hat{\pi}_{\bm k}$ involving ladder operators
of a single mode $\bm{k}$ (and excluding $-\bm{k}$),
namely~\cite{Martin:2015qta}
\begin{align}
\label{eq:defqpi}
\hat{q}_{\bm k}&=\frac{1}{\sqrt{2k}}
\left(\hat{c}_{\bm{k}}+\hat{c}_{\bm{k}}^\dagger \right)
  \qquad \text{and}\qquad
  \hat{\pi}_{\bm{k}} = -i\sqrt{\frac{k}{2}}
  \left(\hat{c}_{\bm{k}}-\hat{c}_{\bm{k}}^\dagger \right),
\end{align}
which are indeed Hermitian and satisfy $[\hat{q}_{\bm
    k},\hat{\pi}_{{\bm k}'}]=i\delta({\bm k}-{\bm k}')$. The
variables~\eqref{eq:vRI:def} and~\eqref{eq:defqpi} define two
partitions (namely a partition between the real and imaginary sector,
and between the $\bm{k}$ and $-\bm{k}$ sector, respectively), and
these two partitions feature different correlations of different
amount and nature.

Since there is no preferred partition, a generic approach is to probe
the nature of the correlations in all possible partitions. This is why
we now define the notion of quantum partitions at a more formal level,
and see how different partitions are related to each other (we refer
the reader to \App{ap:partition} for a more detailed analysis of
partitions). A partition of a Fourier subspace into two subsystems $1$
and $2$ is encoded in the phase-space vector
\begin{align}
  \hat{R}_{1/2}=\left(k^{1/2}\hat{q}_{\bm{k}}^{(1)},
  k^{-1/2}\hat{\pi}_{\bm{k}}^{(1)},k^{1/2}\hat{q}_{\bm{k}}^{(2)},
  k^{-1/2}\hat{\pi}_{\bm{k}}^{(2)}\right)^{\mathrm{T}}\,
,
\end{align} where the two first entries concern the first sector and the
two last entries describe the second sector (the prefactors $k^{1/2}$
and $k^{-1/2}$ are introduced to make all entries of the
$\hat{R}_{1/2}$ vector of the same dimension), and where the
commutators between the entries of $\hat{R}_{1/2}$ are canonical (\ie
the only non-vanishing commutators are between the first and the
second, and the third and the fourth, entries, and this commutator
equals $i$).  For instance, in the $\mathrm{R}/\mathrm{I}$ partition
corresponding to \Eq{eq:vRI:def}, one has
$\hat{R}_{\mathrm{R}/\mathrm{I}} =
(k^{1/2}\hat{v}_{\bm{k}}^{\mathrm{R}},k^{-1/2}\hat{p}_{\bm{k}}^{\mathrm{R}},
k^{1/2}\hat{v}_{\bm{k}}^{\mathrm{I}},k^{-1/2}\hat{p}_{\bm{k}}^{\mathrm{I}})^{\mathrm{T}}$,
while in the $\kmk$ partition corresponding to \Eq{eq:defqpi}, one has
$\hat{R}_{\kmk} =
(k^{1/2}\hat{q}_{\bm{k}},k^{-1/2}\hat{\pi}_{\bm{k}},
k^{1/2}\hat{q}_{-\bm{k}},k^{-1/2}\hat{\pi}_{-\bm{k}})^{\mathrm{T}}$.

Among all possible partitions, let us note that the
$\mathrm{R}/\mathrm{I}$ partition plays a specific role since it is
such that the Hamiltonian is sum separable [\ie \Eq{eq:Hamiltonian:RI}
  does not contain cross terms between the two subsectors]. In the
following, in order to preserve the quadratic nature of the
Hamiltonian density, we focus on partitions that are linearly related
to that reference partition,
\begin{align}
\hat{R}_{1/2}=T^{\mathrm{R}/\mathrm{I}\to 1/2}
\hat{R}_{\mathrm{R}/\mathrm{I}} \, ,
\end{align}
where $T^{\mathrm{R}/\mathrm{I}\to 1/2} $ is a four-by-four matrix
that encodes the change of partitions. This matrix must be such that
the commutator structure is preserved, \ie it must be a symplectic
matrix. Further imposing that different parameterisations share the
same vacuum state, \ie that $T^{\mathrm{R}/\mathrm{I}\to 1/2} $ does
not mix creation and annihilation operators, in \App{ap:partition} we
show that $T^{\mathrm{R}/\mathrm{I}\to 1/2} $ must be of the form
\begin{align}
\label{eq:T:RIto12}
T^{\mathrm{R}/\mathrm{I}\to 1/2} &  =
  \begin{pmatrix}
    \cos \alpha \cos \theta & -\sin \alpha \cos \theta
    & -\cos \delta \sin \theta & \sin \delta \sin \theta \\
    \sin \alpha \cos \theta & \cos \alpha \cos \theta
    & -\sin \delta \sin \theta & -\cos \delta \sin \theta \\
    \cos \beta \sin \theta & -\sin \beta \sin \theta &
    \cos(\alpha-\beta-\delta)\cos \theta &
    \sin(\alpha -\beta-\delta)\cos \theta \\
    \sin \beta \sin \theta & \cos \beta \sin \theta &
    -\sin(\alpha-\beta-\delta)\cos \theta &
    \cos(\alpha-\beta-\delta) \cos \theta
  \end{pmatrix} ,
\end{align}
where $\alpha$, $\beta$, $\delta$ and $\theta$ are four angles that
entirely characterise the partition. For instance, the
$\mathrm{R}/\mathrm{I}$ partition obviously corresponds to
$\alpha=\beta=\delta=\theta=0$, while the $\kmk$ partition corresponds
to $\alpha=0$, $\beta=-\pi$, $\delta=\pi/2$ and $\theta=-\pi/4$. It is
also worth mentioning that the one-parameter subset of partitions
studied in \Refa{Martin:2015qta} can be obtained by setting
$\alpha=0$, $\beta=3\pi/2+2\theta$ and $\delta =\pi/2$ in
\Eq{eq:T:RIto12}, leading to
\begin{align}
  \label{eq:matrixT}
  T^{\mathrm{R/I}\rightarrow 1/2}(\theta)=
  \begin{pmatrix}
    \cos \theta & 0 & 0 & \sin \theta \\
    0 & \cos \theta & -\sin \theta & 0 \\
    \sin \theta \sin (2\theta) & \sin \theta \cos (2\theta) & \cos \theta \cos(2\theta)
    & -\cos \theta \sin(2 \theta) \\
    -\sin \theta \cos(2\theta) & \sin \theta \sin(2\theta) &
    \cos \theta \sin(2\theta) & \cos \theta \cos (2\theta)
  \end{pmatrix}\, .
  \end{align}
This subset reaches the $\kmk$ partition since one can check that
$T^{\mathrm{R/I}\rightarrow \pm\bm{k}}=T^{\mathrm{R/I}\rightarrow
  1/2}(-\pi/4)$.  In what follows, we will focus on the
subclass~\eqref{eq:matrixT} of partitions for concrete applications of
our formalism, since it will be sufficient to study how the result may
depend on the choice of partitions, but the formalism will be kept
general enough to make it obvious how to apply it to the most generic
partitions~\eqref{eq:T:RIto12}.
\subsection{Covariance matrix}
\label{subsec:cov}
Since the Hamiltonian~\eqref{eq:Hamiltonian:generic} is quadratic, the
dynamics it generates is linear and admits Gaussian states as
solutions.\footnote{Note that this work is not restricted to pure
  states, so the quantum states we consider are in general represented
  by a density matrix $\hat{\rho}$, or equivalently by a Wigner
  function. Here, what ``Gaussian state'' means in practice is that
  the Wigner function is Gaussian.} Such states are entirely
characterised by their two-point correlation functions. The two-point
correlation functions are conventionally gathered in the real
symmetric covariance matrix $\gamma$ of the state defined by
\begin{align}
\label{eq:def:covariance}
\gamma_{ab}=\langle\{\hat{R}_a,\hat{R}_b\}\rangle\, ,
\end{align}
where $\{.\}$ denotes the anti-commutator, namely $\{a,b\}\equiv
ab+ba$. Upon a change of partition $\hat{R}\to \hat{R}'=T \hat{R}$,
the covariance matrix becomes
\begin{align}
  \label{eq:transcov}
  \gamma'=T\gamma T^\mathrm{T}\, .
\end{align}

As discussed around \Eq{eq:Hamiltonian:RI}, in the
$\mathrm{R}/\mathrm{I}$ partition, the two sectors decouple and have
the same reduced Hamiltonian. As a consequence, if the initial state
is uncorrelated and symmetric between the two sectors (which is the
case of the vacuum state selected by the Hamiltonian), it remains so
at any time, and the covariance matrix is of the form
\begin{align}
\label{eq:gamma:RI}
  \gamma^{\mathrm{R/I}} =
    \begin{pmatrix}
    \gamma_{11} & \gamma _{12} & 0 & 0 \\
    \gamma _{12} & \gamma_{22} & 0 & 0 \\
    0 & 0 & \gamma_{11} & \gamma_{12} \\
    0 & 0 & \gamma_{12} & \gamma_{22} 
  \end{pmatrix}\, ,
\end{align}
which depends on three parameters only, namely
\begin{align}
\label{eq:GammaijDef}
\gamma_{11}&=2k\left\langle
\left(\hat{v}_{\bm k}^\mathrm{R}\right)^2\right \rangle
  =2k\left \langle \left(\hat{v}_{\bm k}^\mathrm{I}\right)^2
  \right \rangle
  =k\left\langle \left\lbrace \hat{v}_{\bm{k}},
  \hat{v}_{\bm{k}}^\dagger  \right\rbrace\right\rangle
    \, , \\ \gamma_{12} & = \gamma_{21}
    =\left\langle \hat{v}_{\bm k}^\mathrm{R}\hat{p}_{\bm
  k}^\mathrm{R}+\hat{p}_{\bm
  k}^\mathrm{R}\hat{v}_{\bm k}^\mathrm{R}\right\rangle 
  =\left\langle \hat{v}_{\bm k}^\mathrm{I}\hat{p}_{\bm
  k}^\mathrm{I}+ \hat{p}_{\bm
  k}^\mathrm{I}\hat{v}_{\bm k}^\mathrm{I}\right\rangle 
  =\left\langle  \hat{v}_{\bm{k}} \hat{p}_{\bm{k}}^\dagger
  + \hat{p}_{\bm{k}} \hat{v}_{\bm{k}}^\dagger \right\rangle
  \, , \\ \gamma_{22}& =\frac{2}{k}
  \left \langle \left(\hat{p}_{\bm k}^\mathrm{R}\right)^2\right\rangle
  =\frac{2}{k}\left \langle
  \left(\hat{p}_{\bm k}^\mathrm{I}\right)^2\right\rangle
  =\frac{1}{k}\left\langle \left\lbrace \hat{p}_{\bm{k}},
  \hat{p}_{\bm{k}}^\dagger  \right\rbrace\right\rangle
  \, ,
  \label{eq:GammaijDef:end}
\end{align} 
where we have also related the entries of the covariance matrix to the
two-point function of the original $\hat{v}_{\bm{k}}$ and
$\hat{p}_{\bm{k}}$ operators (where one can also check that $\langle
\hat{v}_{\bm{k}} \hat{p}_{\bm{k}} + \hat{p}_{\bm{k}}^\dagger
\hat{v}_{\bm{k}}^\dagger \rangle = 0$).  Note that states represented
by a covariance matrix of the form~\eqref{eq:gamma:RI} are called
Gaussian and Homogeneous Density Matrices (GHDM) in
\Refa{CampoParentani}, where they are shown to yield the most general
partial reconstruction of the state using only the knowledge of the
two-point correlation function.

Making use of \Eq{eq:matrixT} and~\eqref{eq:transcov}, the covariance
matrix can then be written down in any partition, and we give the
result in \App{app:cov} for display convenience, where the specific
case of the $\kmk$ partition is also treated. Note that the
correlators of the ladder operators introduced in
\Eq{eq:ladder:operators} can also be expressed in terms of the entries
of the covariance matrix~\eqref{eq:gamma:RI}, and one obtains
\begin{align} 
\label{eq:nkck1}
\left \langle \left\{ \hat{c}_{\bm k} ,
\hat{c}^{\dagger}_{\bm k} \right\} \right \rangle 
&=
\left \langle \left\{ \hat{c}_{-{\bm k}} ,
\hat{c}^{\dagger}_{-{\bm k}} \right\} \right \rangle 
  = \frac{\gamma_{11}+\gamma_{22}}{2 }\equiv 2 {\cal N}_{ k} + 1 \, , 
\\
\label{eq:nkck2}
\left \langle \left\{ \hat{c}_{ \bm k} ,
\hat{c}_{- \bm k} \right\} \right \rangle &
= \frac{\gamma_{11} - \gamma_{22}}{2}
+ i \gamma_{12} \equiv  2 {\cal C}_{ k}  \, ,
\\
\label{eq:nkck3}
\left \langle \left\{ \hat{c}_{ \bm k} ^\dagger,
\hat{c}_{- \bm k}^\dagger \right\} \right \rangle &
 = \frac{\gamma_{11} - \gamma_{22}}{2} - i \gamma_{12} = 2 {\cal C}_{ k}^* \, .
\end{align}
where other correlators vanish. These expressions also define ${\cal
  N}_{ k}$, the number of particles in the modes $\bm k$ and $-\bm{k}$
(which are equal because of isotropy), and ${\cal C}_{ k}$, the
correlation between the modes $\pm \bm k$~\cite{Robertson:2017ysi}.

Note that the covariance matrix contains all information about the
quantum state, and any relevant quantity can be expressed in terms of
its entries. For instance, the purity of the state, $\mathrm{Tr} (
\hat{\rho}^2 ) $ is given by~\cite{Adesso_2014}
\begin{align}
\label{eq:purityGaussianStates}
\mathrm{Tr} \left( \hat{\rho}^2 \right) =  \frac{1}{\sqrt{\det \gamma  }}
=  \frac{1}{\gamma_{11}\gamma_{22}-\gamma_{12}^2} \, .
\end{align}
This quantity is comprised between $0$ and $1$ and measures the
deviation from a pure state, for which it equals $1$. In the
following, the purity will be thus used as a measure of
decoherence. Note that since symplectic matrices have unit
determinant, the purity is invariant under changes of partitions, and
more generally under any change of phase-space parameterisation.
\subsection{Quantum discord}
\label{subsec:quantumdiscGHDM}

The presence of quantum correlations between two subparts of a system
can be characterised by means of quantum discord~\cite{Henderson:2001,
  Zurek:2001}, which is briefly reviewed in Appendix
\ref{ap:discord}. The idea is to introduce two measures of correlation
that coincide for classically correlated setups because of Bayes theorem, but that may differ
for quantum systems. The first measure is the so-called mutual
information, which is defined as the sum between the von-Neumann
entropy of each reduced sub-systems (known as entanglement entropy), minus the entropy of the entire
system. The second measure evaluates the difference between the
entropy contained in the first subsystem, and the entropy contained in
that same subsystem when the second subsystem has been measured, where
an extremisation is performed over all possible ways to ``measure''
the second subsystem. Quantum discord is defined as the difference
between these two measures, and thus quantifies deviations from Bayes theorem.

It is worth mentioning that for pure states, the different measures of
correlations mentioned above (entanglement entropy, mutual
information and quantum discord) coincide up to
numerical prefactors. 
%Indeed, for a pure state, the von-Neumann entropy vanishes, hence mutual information is twice the entanglement entropy, and discord is half of mutual information. 
While this implies that correlated pure states necessarily feature quantum
correlations, it also means that quantum discord does not add particular insight in measuring them, since it contains the same information as entanglement entropy, which is easier to compute. 
However, quantum discord becomes more clearly useful when
considering mixed states, which is precisely the topic of this work. 
The reason is that there exist mixed states that feature classical correlations only. Contrary to pure states,
mixed states can thus possess classical \emph{and} quantum correlations, and the
role of discord is to isolate the part of the correlations that
is genuinely quantum.

In \App{ap:discord}, we show that for Gaussian homogeneous states such
as the ones introduced above, both measures depend only on the
symplectic eigenvalue of the reduced covariance matrix (\ie of the
diagonal $2$-by-$2$ blocks of the covariance matrix),
\begin{align}
\begin{split}
    \label{eq:sigma(theta)}
    \sigma(\theta) & = \sqrt{ \cos^2(2\theta)
      \left( \gamma_{11}\gamma_{22}-\gamma_{12}^2 \right) +
    \left( \frac{\gamma_{11}+\gamma_{22}}{2} \right)^2
    \sin^2(2\theta)} \, ,
\end{split}
\end{align}
and of the symplectic eigenvalue of the full covariance matrix, which
is nothing but $\sigma(0)$ [and which coincides with
  $(\det\gamma)^{1/4}$, see \Eq{eq:gamma:RI}]. This gives rise to the
following expression for quantum discord
\begin{align}
\label{eq:finaldiscord}
\mathcal{D}(\theta)
=f\left[\sigma(\theta)\right]
-2f\left[\sigma(0)\right]+
f\left[\frac{\sigma(\theta)+\sigma^2(0)}{\sigma(\theta)+1}\right]\, ,
\end{align}
where the function $f(x)$ is defined for $x\geq 1$ by
\begin{align}
\label{eq:deff}
  f(x)=\left(\frac{x+1}{2}\right)
  \log_2\left(\frac{x+1}{2}\right)
-\left(\frac{x-1}{2}\right)
\log_2\left(\frac{x-1}{2}\right),
\end{align}
One can check that, for the $\mathrm{R}/\mathrm{I}$ partition where
$\theta=0$, the above expressions give $\mathcal{D}=0$, in agreement
with the fact that the two subsystems are uncorrelated in this
partition.
\section{Discord in the absence of an environment}
\label{sec:timeevolutionwo}
In \Sec{sec:DiscordQField}, we have seen how the Fourier subspaces of
a real scalar field can be partitioned, and how the presence of
quantum correlations between its subparts can be characterised from
the knowledge of its covariance matrix. In this section, we treat the
situation where the field does not couple to any environmental degree
of freedom, and its quantum state remains pure. We describe its time
evolution using three different, though complementary, approaches: via
Bogoliubov coefficients in \Sec{subsec:bogo}, via squeezing parameters
in \Sec{subsec:visu} and via transport equations in
\Sec{subsec:transportwo}.  These three approaches are useful as they
will lead to different insights into the case with environmental
coupling, treated in \Sec{sec:caldleggeteom}. We finally analyse how
quantum discord evolves in time in \Sec{sec:evoldiscordnoenvi}.
\subsection{Bogoliubov coefficients}
\label{subsec:bogo}
In the Heisenberg picture, the equation of motion for the ladder
operators can be obtained from \Eq{eq:Hamiltonian:COperators}, and in
matricial form they are given by
\begin{align}
    \label{eq:evolcrea}
\frac{\dd }{\dd  {\tau} } \begin{pmatrix}
\hat{c}_{\bm k} \, \\
\hat{c}^{\dagger}_{- \bm k}
\end{pmatrix}
 = \begin{pmatrix}
 \displaystyle
-i \frac{k}{2} \left[ \frac{\omega^2(k, \tau )}{k^2} +1 \right] & 
\displaystyle - i \frac{k}{2}
\left[ \frac{\omega^2(k, \tau )}{k^2} - 1 \right] \\ \\
\displaystyle i \frac{k}{2}
\left[ \frac{\omega^2(k, \tau )}{k^2} - 1 \right]
& \displaystyle i \frac{k}{2} \left[ \frac{\omega^2(k, \tau )}{k^2} +1 \right]
\end{pmatrix} \begin{pmatrix}
\hat{c}_{\bm k} \, \\
\hat{c}^{\dagger}_{- \bm k}
\end{pmatrix} .
\end{align} 
This system being linear, it can be solved with a linear
transformation known as the Bogoliubov transformation
\begin{align}
\label{eq:BogoTransfo}
\begin{pmatrix}
\hat{c}_{\bm k} ( \tau ) \, \\
\hat{c}^{\dagger}_{- \bm k} ( \tau )
\end{pmatrix} 
= \begin{pmatrix}
u_{\bm k}( \tau ) & w_{\bm k} ( \tau ) \\
w^{*}_{- \bm k} ( \tau ) & u^{*}_{- \bm k} ( \tau )
\end{pmatrix} \begin{pmatrix}
\hat{c}_{\bm k} \left(  \tau _\uin \right) \, \\
\hat{c}^{\dagger}_{- \bm k} \left(  \tau _\uin \right)
\end{pmatrix} ,
\end{align}
where $ u_{\bm k}$ and $w_{\bm k}$ are the two complex Bogoliubov
coefficients satisfying
\begin{equation}
\label{eq:detBogoCoeff}
    \left| u_{\bm k} \right|^2 - \left| w_{- \bm k} \right|^2 = 1 \, .
\end{equation}
This condition ensures that the commutation relation
$[\hat{c}_{\bm{k}},\hat{c}_{\bm{k}'}^\dagger]=\delta(\bm{k}-\bm{k}')$
is preserved in time [which can be checked by differentiating this
  commutation relation with respect to time and using
  \Eq{eq:evolcrea}].  Solving the evolution of the system then boils
down to computing the Bogoliubov coefficients. They satisfy the same
differential system as the creation and annihilation operators, namely
\Eq{eq:evolcrea}, with initial conditions $u_{\pm \bm k} \left(
\tau_{\uin} \right) = 1$ and $w_{\pm \bm k} \left( \tau_{\uin} \right) =
0$. Note that because of statistical isotropy, the Bogoliubov
coefficients only depend on the norm of $\bm k$, so $u_{\bm k} = u_{-
  \bm k}\equiv u_k$ and $w_{\bm k} = w_{-\bm k}\equiv w_k$.

%For an arbitrary $\omega(k, \tau )$, the equations of evolution of the Bogoliubov coefficients cannot be solved in general. 

The two first-order differential equations for the Bogoliubov
coefficients can be combined into a single second-order equation for
the combination $u_{ k}+ w^{*}_{ k}$, namely
\begin{align}
\label{eq:SecondOrderDiffEquplusvstar}
\frac{\dd^2}{\dd{\tau}^2}\left(u_{k} + w^*_{k} \right)
+ \omega^2(k, \tau )\left(u_{k} + w^*_{k} \right) & = 0 \, .
\end{align}
This equation needs to be solved with the
initial conditions $(u_k+w_k^*)( \tau _\uin)=1$ and $(u_k+w_k^*)'( \tau _\uin)=-ik$,
where the latter comes from the relation
\begin{align}
    \label{eq:Relationuplusvuminusv}
\frac{\dd}{\dd{\tau}}\left(u_{k} + w^{*}_{k} \right) = - i k \left( u_{k} - w^{*}_{ k} \right)  ,
\end{align}
which itself follows from the fact that the Bogoliubov coefficients
satisfy the differential system~\eqref{eq:evolcrea}. Note that
\Eq{eq:Relationuplusvuminusv} also implies that $u_{k} - w^{*}_{ k}$
can be obtained from the solution of the second-order
equation~\eqref{eq:SecondOrderDiffEquplusvstar}, hence both $u_k$ and
$w_k$ can be reconstructed from that solution. In practice,
determining the full dynamics of the system thus boils down to solving
\Eq{eq:SecondOrderDiffEquplusvstar}.

The evolution can also be expressed in terms of the field variables,
since plugging \Eq{eq:BogoTransfo} into \Eq{eq:defqpi} leads to
\begin{align}
    \label{eq:EvolutionMatrixRI}
    \hat{R}_{\mathrm{R/I}} \left(  \tau  \right) =
    T_{\mathrm{R/I}} \left(\tau  \right)
    \hat{R}_{\mathrm{R/I}} \left(\tau_{\uin} \right)  ,
\end{align} 
where
\begin{align}
\label{eq:timeevoRI}
T_{ \mathrm{R/I}} \left(\tau  \right) & = \begin{pmatrix}
  \Rea \left( u_{ k} +  w_{ k} \right)
  & -\Ima \left( u_{ k} -  w_{ k} \right) & 0 & 0 \\
  \Ima \left( u_{ k} +  w_{ k} \right)
  & \Rea\left( u_{ k} -  w_{ k} \right) & 0 & 0  \\
  0  & 0 & \Rea \left( u_{ k} +  w_{ k} \right)
  & -\Ima \left( u_{ k} -  w_{ k} \right) \\
  0 & 0 & \Ima \left( u_{ k} +  w_{ k} \right)
  & \Rea \left( u_{ k} -  w_{ k} \right)
\end{pmatrix} \, .
\end{align}
The covariance matrix can then be evaluated by means of
\Eq{eq:transcov}, namely $\gamma \left( \tau \right) = T_{ \mathrm{R/I}}
\left( \tau \right) \gamma \left( \tau_{\uin} \right) T_{
  \mathrm{R/I}}^\mathrm{T} \left( \tau \right)$, which gives rise to
\begin{align}
\label{eq:EvolvedCovBogoCoeff}
\gamma_{11} \left(\tau  \right) & = \frac12
\left[\gamma_{11}( \tau _{\uin})+
  \gamma_{22}( \tau _{\uin})\right] \left| u_{k}( \tau )+w_{k}^*( \tau )\right| ^2
\nonumber \\ &
+\Rea \left\{ \left[u_{k}( \tau )+w_{k}^*( \tau )\right]^2
\left[\frac{\gamma_{11}( \tau _{\uin}) 
    - \gamma_{22}( \tau _{\uin})}{2} + i \gamma_{12}( \tau _{\uin}) \right]
\right\} \, , 
\\
\gamma_{22} \left(\tau  \right) & = \frac12\left[\gamma_{11}( \tau _{\uin})
  + \gamma_{22}( \tau _{\uin})\right] \left| u_{k}( \tau )-w_{k}^*( \tau )\right| ^2
\nonumber \\ &
-\Rea \left\{ \left[u_{k}( \tau )-w_{k}^*( \tau )\right]^2
\left[\frac{\gamma_{11}( \tau _{\uin}) 
    - \gamma_{22}( \tau _{\uin})}{2} + i \gamma_{12}( \tau _{\uin}) \right]
\right\} \, , 
\\
\gamma_{12} \left(\tau  \right) & = \left[\gamma_{11}( \tau _{\uin})
  + \gamma_{22}( \tau _{\uin})\right]\Ima \left[u_{k}( \tau ) w_{k}( \tau ) \right]  
\nonumber \\ &
-\Ima \left\{\left[u_{k}^*{}^2( \tau )-w_{k}^2( \tau )\right]
\left[\frac{\gamma_{11}( \tau _{\uin}) - \gamma_{22}( \tau _{\uin})}{2}
  -  i \gamma_{12}( \tau _{\uin}) \right]\right\} \, .
\end{align}
Notice that, using \Eqs{eq:nkck1}, (\ref{eq:nkck2})
and~(\ref{eq:nkck3}), the above relations can be rewritten in terms of
the initial number of particles ${\cal N}_k( \tau _\uin)$ and mode
correlation ${\cal C}_k( \tau _\uin)$, leading to
\begin{align} 
\label{eq:EvolvedCovBogoCoeff11}
\gamma_{11} \left(\tau  \right) & = \left[2 {\cal N}_k\left( \tau _\uin\right)
  +1\right] \left| u_{k}( \tau )+w_{k}^*( \tau )\right| ^2
+2\, \Rea \left\{ \left[u_{k}( \tau )+w_{k}^*( \tau )\right]^2
{\cal C}_k\left( \tau _\uin\right) \right\}  \\
\label{eq:EvolvedCovBogoCoeff22}
\gamma_{22} \left(\tau  \right) & =  \left[2 {\cal N}_k
  \left( \tau _\uin\right)+1\right]  \left| u_{k}( \tau )-w_{k}^*( \tau )\right| ^2
-2\, \Rea \left\{ \left[u_{k}( \tau )-w_{k}^*( \tau )\right]^2
{\cal C}_k\left( \tau _\uin\right) \right\}  \\
\label{eq:EvolvedCovBogoCoeff12}
\gamma_{12} \left(\tau  \right) & = 2 \left[2 {\cal N}_k\left( \tau _\uin\right)
  +1\right] \Ima \left[ u_{k}( \tau ) w_{k}( \tau ) \right]  -2\,
\Ima \left\{\left[u_{k}^*{}^2( \tau )-w_{k}^2( \tau )\right]
     {\cal C}_k^*\left( \tau _\uin\right) \right\} .
\end{align}
If the initial state is chosen as the vacuum state, ${\cal N}_k( \tau
_\uin)={\cal C}_k( \tau _\uin)=0$, the above expressions reduce to
\begin{align} 
\label{eq:EvolvedCovBogoCoeffVac}
\gamma_{11} \left(\tau  \right) & = \left| u_{k}( \tau )+w_{k}^*( \tau )\right| ^2  \, , 
\qquad
\gamma_{22} \left(\tau  \right) 
= \left|u_{k}( \tau )-w_{k}^*( \tau )\right| ^2  \, ,
 \\
\gamma_{12} \left(\tau  \right) & = 2\, \Ima \left[u_{k}( \tau ) w_{k}( \tau ) \right]  \, .
\label{eq:EvolvedCovBogoCoeffVac:end}
\end{align}
In that case, given the initial conditions $u_{ k} \left( \tau_{\uin}
\right) = 1$ and $w_{k} \left( \tau_{\uin} \right) = 0$, these
expressions also imply that $\gamma_{11} \left( \tau_{\uin} \right) =
\gamma_{22} \left( \tau_{\uin} \right) = 1$ and $\gamma_{12} \left(
t_{\uin} \right) = 0$.
\subsection{Squeezing parameters}
\label{subsec:visu}
An equivalent description of the dynamics is through the squeezing
parameters $(r_k,\varphi_k,\theta_k)$ (notice that the rotation
  angle $\theta_k$, which carries the index ``$k$'', should not be
  confused with the angle $\theta$ defining a partition), which are
defined in terms of the Bogoliubov coefficients as
\begin{align}
\label{eq:defrphiuv}
u_k( \tau )=e^{-i\theta_k}\cosh r_k, \qquad w_k( \tau )
=-e^{i\theta_k+2i\varphi_k}\sinh r_k\, ,
\end{align}
which ensures that the condition~\eqref{eq:detBogoCoeff} is
automatically satisfied. Given that the Bogoliubov coefficients
satisfy the differential system~\eqref{eq:BogoTransfo}, one can derive
equations of motion for the squeezing parameters, namely
\begin{align}
\label{eq:diffsqueezingr}
    \frac{\dd r_k}{\dd {\tau} } &=\frac{k}{2}\left(\frac{\omega^2}{k^2}-1\right)
    \sin\left(2\varphi_k\right), 
    \\
    \frac{\dd \varphi_k}{\dd {\tau} }&=
    -\frac{k}{2}\left(\frac{\omega^2}{k^2}+1\right)
    +\frac{k}{2}\left(\frac{\omega^2}{k^2}-1\right)
    \frac{\cos\left(2\varphi_k\right)}{\tanh(2 r_k)}\, , 
    \\
    \frac{\dd \theta_k}{\dd {\tau} }&=
    \frac{k}{2}\left(\frac{\omega^2}{k^2}+1\right)
    -\frac{k}{2}\left(\frac{\omega^2}{k^2}-1\right)
    \cos\left(2\varphi_k\right)
    \tanh r_k\, ,
    \end{align}
 where one can see that $\theta_k$ does not contribute to the time
 evolution of $r_k$ and $\varphi_k$.  Moreover, since we have already
 derived the relation between the covariance matrix elements and the
 functions $u_k$ and $v_k$, see \Eq{eq:EvolvedCovBogoCoeffVac}, one
 can also express the components $\gamma_{11}$, $\gamma_{22}$ and
 $\gamma_{12}$ in terms of the squeezing parameters. One
 finds\footnote{In the $\kmk$ partition, where the covariance matrix
   is given by \Eqs{eq:corr:qk2}-\eqref{eq:corr:qmkpik}, those
   expressions allow one to recover Eq.~(C28) of
   \Refa{Martin:2015qta}.}
\begin{align}
\label{eq:g11rphi}
    \gamma_{11} =& \cosh(2r_k)-\cos(2\varphi_k)\sinh(2 r_k),
    \\
    \label{eq:g22rphi}
    \gamma_{22} =& \cosh(2r_k)+\cos(2\varphi_k)\sinh(2 r_k),
    \\
    \label{eq:g12rphi}
    \gamma_{12} =& -\sin (2\varphi_k)\sinh(2 r_k) ,
\end{align}
where one can see that $\theta_k$ does not appear. 

The geometrical interpretation of the squeezing parameters becomes
clear when computing the Wigner function~\cite{PhysRevA.36.3868,
  SIMON1987223, doi:10.1119/1.2957889}, which is the Wigner-Weyl
transform of the density matrix. It can be seen as a quasi probability
distribution function, in the sense that the quantum expectation value
of any operator is given by the integral over phase-space of the
product between the Weyl transform of that operator and the Wigner
function. For a Gaussian state, it reads
\begin{align}
    W(R)=\frac{1}{\pi^2 \sqrt{\det \gamma}}
    \exp\left(-R^\mathrm{T}\gamma^{-1} R\right) \, .
\end{align}
In the $\mathrm{R}/\mathrm{I}$ partition, $\gamma^{\mathrm{R/I}}$ is
given by \Eq{eq:gamma:RI}, so
\begin{align}
    \left(\gamma^{\mathrm{R/I}}\right)^{-1}=
    \begin{pmatrix}
    \left(\gamma^\mathrm{R}\right)^{-1} & 0 \\
    0 & \left(\gamma^\mathrm{I}\right)^{-1}
    \end{pmatrix}
\end{align}
with 
\begin{align}
\label{eq:subblockinversecov}
   \left(\gamma^\mathrm{R}\right)^{-1}=\left(\gamma^\mathrm{I}\right)^{-1}
    =\begin{pmatrix}
    \cosh(2r_k)+\cos(2\varphi_k)\sinh(2 r_k) & \sin(2\varphi_k)\sinh(2r_k) \\
   \sin(2\varphi_k)\sinh(2r_k) & \cosh(2r_k)-\cos(2\varphi_k)\sinh(2 r_k)
    \end{pmatrix}.
\end{align}
Since $(\gamma^{\mathrm{R/I}})^{-1}$ is block diagonal, the Wigner
function factorises in that partition, \ie
$W(R)=W^\mathrm{R}(R^{\mathrm{R}}) W^\mathrm{I}(R^{\mathrm{I}})$,
where $R^{s}=(k^{1/2}v_{\bm k}^s,k^{-1/2}p_{\bm k}^s)^\mathrm{T}$ with
$s=\mathrm{R}$, $\mathrm{I}$. This translates the above remark that,
in the $\mathrm{R}/\mathrm{I}$ partition, the state is uncorrelated
and separable.

Owing to the Gaussian nature of $W^s$, the contours of the Wigner
function are ellipses in phase space, the geometrical parameters of
which can be derived as follows.  The quadratic form appearing in the
argument of the exponential in the Wigner function can be diagonalised
upon performing a phase-space rotation with angle $\varphi_k$
\begin{align}
\widetilde{R}^s = \mathcal{R}(-\varphi_k) R^s =
\begin{pmatrix}
\cos \varphi_k & \sin \varphi_k \\
- \sin\varphi_k & \cos \varphi_k 
\end{pmatrix}
R^s \, ,
\end{align}
along which the covariance matrix becomes
\begin{align}
\label{eq:gamma:m1:tilde}
(\widetilde{\gamma}^s)^{-1}=
\mathcal{R}(-\varphi_k)(\gamma^s)^{-1}\mathcal{R}^\mathrm{T}(-\varphi_k)=
\begin{pmatrix}
\ee^{2 r_k} & 0\\
0 & \ee^{-2 r_k}
\end{pmatrix} .
\end{align}
\begin{figure}
    \centering
    \includegraphics[width=0.7\textwidth]{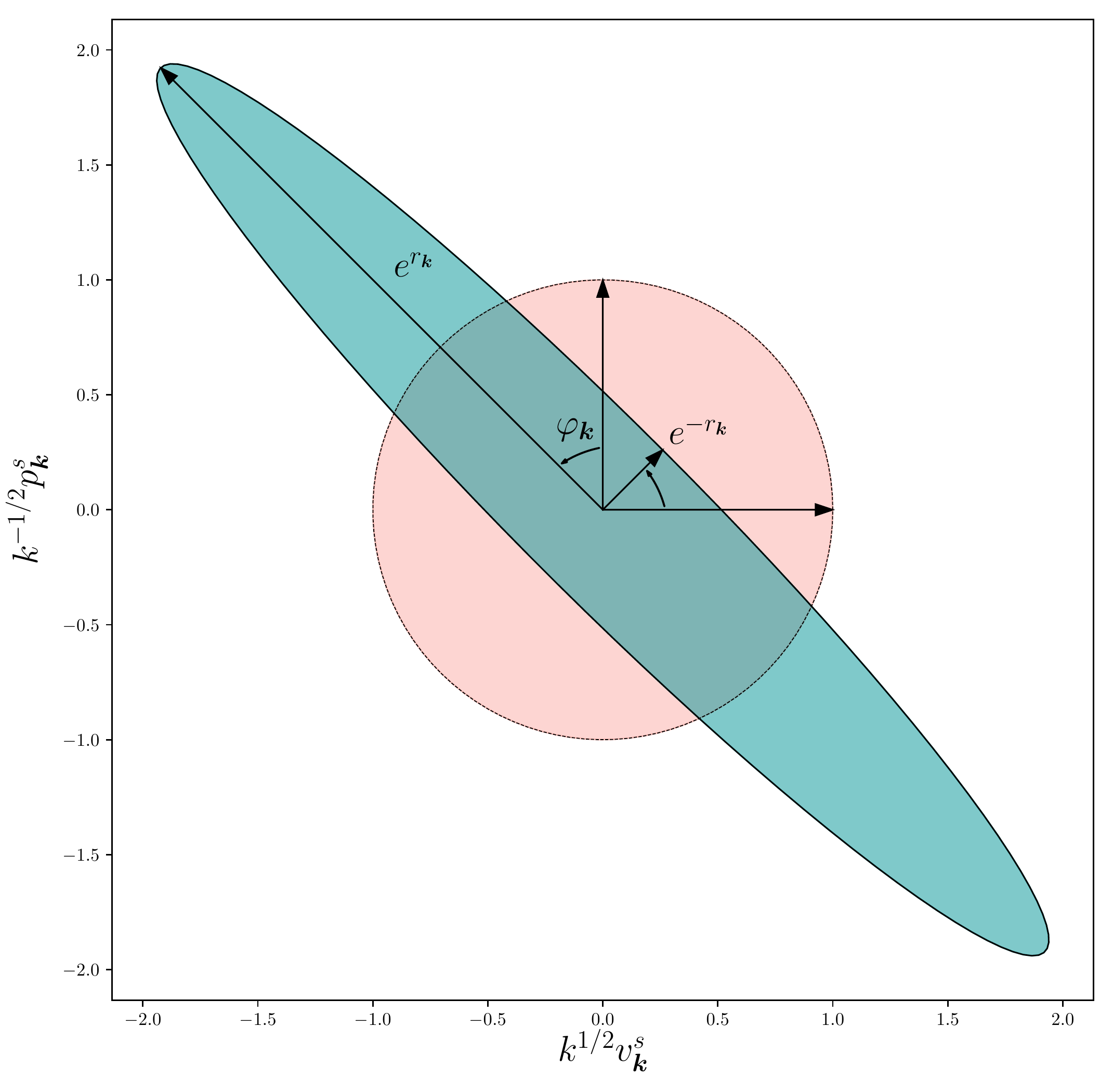}
    \caption{Phase space representation of the $\sqrt{2}$-$\sigma$
      contour level of the Wigner function $W^s$, for
      $\varphi_{\bm{k}} = \pi / 4 $, $r_{\bm{k}} = 1$ (green ellipse)
      compared to the pink circle corresponding to a vacuum state
      (coherent state) with vanishing squeezing.}
    \label{fig:ellipsesqueezingparameters}
\end{figure}
This implies that the semi-minor and the semi-major axes of the
above-mentioned ellipses are tilted by the angle $\varphi_k$ in phase
space, and that for the $\sqrt{2}$-$\sigma$ contour, their respective
lengths are given by $\ee^{r_k}$ and $\ee^{-r_k}$. Such an ellipse is
displayed in \Fig{fig:ellipsesqueezingparameters}, and fully describes
the quantum state of the system. This leads to a simple interpretation
of the squeezing parameters: $r_k$ controls the eccentricity of the
Wigner-function contours, and $\varphi_k$ its phase-space orientation.

Note that the area of the ellipse, which is proportional to the
product between the semi-major and the semi-minor axes lengths, is a
constant. This can be traced back to the fact that it is proportional
to the determinant of the covariance matrix, which is a constant given
that the time evolution is performed via a symplectic matrix in
\Eq{eq:EvolutionMatrixRI}. Alternatively, it can also be seen as a
consequence of \Eq{eq:detBogoCoeff}. Since the determinant of the
covariance matrix is related to the purity of the state via
\Eq{eq:purityGaussianStates}, it also simply translates the fact that
the state remains pure if it does not couple to an environment.
\subsection{Transport equations}
\label{subsec:transportwo}
The third method which allows us to follow the time evolution of the
system is to establish the differential equations obeyed by the
components of the covariance matrix, \ie by the two-point functions of
the system. In general, the quantum expectation value of any operator
$\hat{O}$ evolves according to the Heisenberg equation
\begin{align}
\label{eq:schrodingermean}
  \frac{\dd \langle \hat{O}\rangle}{\dd {\tau} }
  &=\left\langle \frac{\partial\hat{O}}{\partial {\tau}}
  \right\rangle-i\left\langle \left[\hat{O},\hat{H} \right]\right \rangle .
 \end{align}
For one-point correlation functions, using the
Hamiltonian~(\ref{eq:Hsystem}), this leads
\begin{align}
\label{eq:diffmean}
\frac{\dd \langle v_{\bm k}^s\rangle}{\dd {\tau} }
=\langle p_{\bm k}^s\rangle,\quad
\frac{\dd \langle p_{\bm k}^s\rangle}{\dd {\tau} }
=-\omega^2(k,\tau) \langle v_{\bm k}^s\rangle\, ,
\end{align}
which is nothing but Ehrenfest's theorem. Combined together, these two
equations lead to $\langle \hat{v}_{\bm k}^s\rangle''
+\omega^2(k,\tau)\langle \hat{v}_{\bm k}^s\rangle=0$, which coincides
with the equation satisfied by the combination $u_k+w_k^*$ of the
Bogoliubov coefficients, see \Eq{eq:SecondOrderDiffEquplusvstar}, and
which is nothing but the classical equation of motion.
  
For two-point correlation functions, one has
\begin{align}
\label{eq:diff2point1}
\frac{\dd}{\dd {\tau} }\langle \hat{v}_{{\bm k}_1}^s
\hat{v}_{{\bm k}_2}^s\rangle &=
\langle \hat{v}_{{\bm k}_1}^s\hat{p}_{{\bm k}_2}^s
+\hat{p}_{{\bm k}_1}^s\hat{v}_{{\bm k}_2}^s
  \rangle, \\
  \label{eq:diff2point2}
\frac{\dd}{\dd {\tau} }\langle \hat{p}_{{\bm k}_1}^s\hat{v}_{{\bm k}_2}^s\rangle &=
\langle \hat{p}_{{\bm k}_1}^s\hat{p}_{{\bm k}_2}^s\rangle
-\omega^2(k_1,\tau)\langle\hat{v}_{{\bm k}_1}^s\hat{v}_{{\bm k}_2}^s
  \rangle, \\
  \label{eq:diff2point3}
\frac{\dd}{\dd {\tau} }\langle \hat{v}_{{\bm k}_1}^s\hat{p}_{{\bm k}_2}^s\rangle &=
\langle \hat{p}_{{\bm k}_1}^s\hat{p}_{{\bm k}_2}^s\rangle
-\omega^2(k_2,\tau)\langle\hat{v}_{{\bm k}_1}^s\hat{v}_{{\bm k}_2}^s
\rangle, \\
\label{eq:diff2point4}
\frac{\dd}{\dd {\tau} }\langle \hat{p}_{{\bm k}_1}^s\hat{p}_{{\bm k}_2}^s\rangle &=
-\omega^2(k_2,\tau)\langle \hat{p}_{{\bm k}_1}^s\hat{v}_{{\bm k}_2}^s\rangle
-\omega^2(k_1,\tau)\langle\hat{v}_{{\bm k}_1}^s\hat{p}_{{\bm k}_2}^s
\rangle ,
\end{align}
where, as expected, the time derivative of correlators mixing R and I
quantities vanish, \ie $\dd {\langle R^s_i R^{\tilde{s}}_j\rangle}/(\dd{\tau})
\propto \delta(s-\tilde{s})$.  Making use of
\Eqs{eq:GammaijDef}-\eqref{eq:GammaijDef:end}, this leads to the
following differential system for the entries of the covariance
matrix,
\begin{align}
\label{eq:diffcov1}
\frac{1}{k}\frac{\dd \gamma_{11}}{\dd {\tau} }&=\gamma_{12}+\gamma_{21}, \\
\label{eq:diffcov2}
\frac{1}{k}\frac{\dd}{\dd {\tau} }(\gamma_{12}+\gamma_{21})
&=2\gamma_{22}  -2\frac{\omega^2}{k^2}\gamma_{11}, \\
\label{eq:diffcov3}
\frac{1}{k}\frac{\dd \gamma_{22}}{\dd {\tau} }&=
-\frac{\omega^2}{k^2}(\gamma_{12}+\gamma_{21}) .
\end{align}
Let us recall that, as pointed out below
\Eq{eq:EvolvedCovBogoCoeffVac:end}, if the initial state is chosen to
be the vacuum state, these equations must be solved with initial
conditions $\gamma_{11}( \tau _\uin)=\gamma_{22}( \tau _\uin)=1$ and
$(\gamma_{12}+\gamma_{21})( \tau _\uin)=0$. One can check that, in
agreement with the remark made at the end of \Sec{subsec:visu}, these
equations imply that $\det
\gamma^s=\gamma_{11}\gamma_{22}-\gamma_{12}\gamma_{21}$ is preserved
in time. One may also note that the above three first-order
differential equations lead to a single, third-order, differential
equation for $\gamma_{11}$, namely
\begin{align}
\label{eq:thirddiffstandard}
  \frac{1}{k^3}\frac{{\dd}^3\gamma_{11}}{{\dd} {\tau} ^3}
  &+4\frac{\omega^2}{k^2}\frac{1}{k}\frac{{\dd}\gamma_{11}}{{\dd} {\tau} }
  +\frac{2}{k}\frac{{\dd}}{{\dd} {\tau} }\left(\frac{\omega^2}{k^2}\right)
  \gamma_{11}=0.
\end{align}
The order of that differential equation can however be reduced upon
introducing the (complex) change of variable $\gamma_{11}=v_{ k}v_{
  k}^*$. Indeed, one can show that \Eq{eq:thirddiffstandard} is
satisfied if $v_{ k}''+\omega^2v_{ k}=0$. One recovers again the same
second-order differential equation as the one satisfied by the
combination of Bogoliubov coefficients $u_k+w_k^*$, see
\Eq{eq:SecondOrderDiffEquplusvstar} [note that the initial conditions
  also match, \ie the initial conditions give above for $\gamma_{11}$,
  $\gamma_{12}$ and $\gamma_{22}$ lead to $v_k({\tau}_\uin)=1$ and
  $v_k'({\tau}_\uin)=-i k$], which also coincides with the classical
equation of motion as pointed out above. This shows that the evolution
of Gaussian quantum states can be entirely described by the dynamics
of its classical counterpart, and that the three approaches introduced
above to solve the dynamics are technically equivalent.
\subsection{Quantum discord}
\label{sec:evoldiscordnoenvi}
As explained in \Sec{subsec:quantumdiscGHDM}, the computation of
quantum discord boils down to the computation of the symplectic
eigenvalue $\sigma(\theta)$. Setting $\theta=0$ in
\Eq{eq:sigma(theta)} leads to $\sigma(0)=\sqrt{\det \gamma^s}$, so for
a pure state, one has $\sigma(0)=1$. Since \Eq{eq:deff} leads to
$f(1)=0$, the expression~\eqref{eq:finaldiscord} for quantum discord
reduces to
\begin{align}
  \label{eq:discordredf}
  \mathcal{D} = f\left[\sigma(\theta)\right]\, .
  \end{align}
The symplectic eigenvalue $\sigma(\theta)$ can be expressed in terms
of the Bogoliubov coefficients by plugging
\Eq{eq:EvolvedCovBogoCoeffVac} into \Eq{eq:sigma(theta)}, and one
finds $\sigma(\theta) = \sqrt{1 + 4 \vert u_k\vert^2 \vert w_k\vert^2
  \sin^2(2\theta)}$.  Making use of \Eq{eq:defrphiuv}, it can also be
written in terms of the squeezing parameters as
\begin{align}
\label{eq:sigmasqueezdS}
\sigma(\theta) = \sqrt{1 + \sinh^2(2r_k) \sin^2(2\theta)} \, ,
\end{align}
where only the squeezing amplitude $r_k$ enters the expression. This
shows that, when $\theta\neq 0$, the discord increases with the
squeezing amplitude but does not depend on the squeezing angle. Let us
also mention that, plugging \Eq{eq:nkck1} into \Eq{eq:sigma(theta)},
one obtains an expression that only involves the number of particles
$\mathcal{N}_k$, namely
\begin{align}
\label{eq:sigma:Nk}
\sigma(\theta) = \sqrt{1 + 4 \sin^2(2\theta)\mathcal{N}_k
  \left(\mathcal{N}_k+1\right)} \, .
\end{align}
This shows that discord increases with the number of entangled
particles created between the sectors $\bm{k}$ and $-\bm{k}$, as
expected. This also indicates that discord is maximal when
$\theta=-\pi/4$, \ie in the partition $\kmk$. These considerations are
in agreement with the results found in \Refa{Martin:2015qta}.

\section{Discord in the presence of an environment}
\label{sec:caldleggeteom}
In \Sec{sec:timeevolutionwo}, we have described the evolution of the
system and its quantum discord in the case where it is placed in a
pure state, without interactions with environmental degrees of
freedom. We now study how these considerations generalise to the
situation where an environment is present and couples to the
system. Formally, we write down the total Hamiltonian as the sum of a
term acting on the system, $\hat{H}$, a term acting on the
environment, $\hat{H}_\mathrm{env}$, and an interaction term,
$\hat{H}_\mathrm{int}$,
\begin{align}
\label{eq:Htot}
  \hat{H}_{\mathrm{tot}}=\hat{H} \otimes \hat{\setI}_\mathrm{env}
  +\hat{\setI}\otimes \hat{H}_\mathrm{env}
  +g\hat{H}_\mathrm{int},
\end{align}
where $g$ is a coupling constant that controls the interaction
strength and we recall that $\hat{H}$ is given in \Eq{eq:Hsystem}. Our
goal is to analyse the state of the system, which is described by the
reduced density matrix
\begin{align}
\label{eq:reduced:rho:def}
\hat{\rho} = \mathrm{Tr}_{\mathrm{env}}\left(\hat{\rho}_{\mathrm{tot}}\right)\, .
\end{align}
In practice, we assume that the interaction term is local, so it can
be written as
\begin{align}
  \hat{H}_\mathrm{int}( \tau )=\int \dd ^3 {\bm x}\, \hat{A}( \tau ,{\bm x})
  \otimes \hat{E}( \tau ,{\bm x}),
  \end{align}
where $\hat{A}$ is an operator acting in the Hilbert space of the
system and $\hat{E}$ an operator acting in the Hilbert space of the
environment. 
\subsection{Caldeira-Leggett model}
\label{subsec:inter}
Under the assumption that the auto-correlation time of $\hat{E}$ in the
environment, which we denote $\tau_\uc$, is much shorter than the time
scale over which the system evolves, one can show that the reduced
density matrix~\eqref{eq:reduced:rho:def} obeys the Lindblad
equation~\cite{CohenTannoudji:1992, LeBellac:2006, Burgess:2006jn,
  Pearle:2012, Brasil:2012}
\begin{align}
  \label{eq:Lindblad}
  \frac{\dd \hat{\rho}}{\dd{\tau}}
  &=-i\left[\hat{H} ,\hat{\rho}\right]
 % \nonumber \\ &
  -\frac{\Gamma}{2}\int \dd ^3{\bm x}\, \dd ^3 {\bm y}
  \, C_E(\tau;{\bm x},{\bm y})\left[\hat{A}({\bm x}),
    \left[\hat{A}({\bm y}),\hat{\rho}\right]\right],
\end{align}
where $C_{E}({\tau}; {\bm x},{\bm y})=\langle \hat{E}({\tau},{\bm
  x})\hat{E}({\tau},{\bm y})\rangle$ is the equal-time correlation function
of $\hat{E}$, and $\Gamma\equiv 2g^2 \tau_\uc$. Let us note that the
Lindblad equation generates all quantum dynamical
semigroups~\cite{Lindblad:1975ef}, and that even though it is derived
at leading order in $g$, it allows for efficient late-time
re-summation~\cite{Kaplanek:2019dqu}.

Similarly to \Eq{eq:schrodingermean}, the equation controlling the
quantum expectation value of a given operator $\hat{O}$, namely
$\langle \hat{O}\rangle =\mathrm{Tr}(\hat{\rho}\, \hat{O})$, can be
obtained from the Lindblad equation, and one finds
\begin{align}
\label{eq:Lindblad:mean}
  \frac{\dd \langle \hat{O}\rangle}{\dd{\tau}}
  &=\left\langle \frac{\partial\hat{O}}{\partial {\tau}}\right\rangle
  -i\left\langle \left[\hat{O},\hat{H} \right]\right \rangle
 % \nonumber \\
  & -\frac{\Gamma}{2}(2\pi)^{3/2}
  \int _{\setR^3}\dd ^3 {\bm k} \, \tilde{C}_{E}({\tau},{\bm k})
  \left\langle \left[\left[\hat{O},\hat{A}_{\bm k}\right],
    \hat{A}_{-{\bm k}}\right]\right\rangle,
  \end{align}
where $\tilde{C}_{E}({\tau},{\bm k})$ is the Fourier transform of the
correlation function [assuming statistical homogeneity,
  $C_{{E}}(\bm{x},\bm{y})$ depends only on $\bm{x}-\bm{y}$, so we mean
  the Fourier transform with respect to $\bm{x}-\bm{y}$].

These equations are difficult to solve in general, but they greatly
simplify under the assumption that $\hat{A}$ is linear in the
phase-space variables. The reason is that, in that case, all
interactions involving the system are linear, so the state of the
system remains Gaussian (although it becomes a mixed state). This
allows one to still fully describe it in terms of a covariance matrix,
and to generalise most of the considerations presented in
\Sec{sec:timeevolutionwo}. Such a setup is called the Caldeira-Leggett
model~\cite{Caldeira:1981rx, Caldeira:1982uj, Caldeira:1982iu}, and in
what follows we will use it to understand how decoherence may affect
the presence of quantum discord within the system. For simplicity, we
will consider the case where $\hat{A}=\hat{v}$, but the more generic
situation where $\hat{A}$ is a linear combination of $\hat{v}$ and
$\hat{p}$ can be dealt with along very similar lines, see
\Refa{Martin:2018zbe}.
\subsection{Transport equations}
Let us first follow the approach presented in \Sec{subsec:transportwo}
and derive transport equations from \Eq{eq:Lindblad:mean}. For
one-point correlation functions, one still obtains \Eq{eq:diffmean},
\ie the classical equations of motion. For two-point correlation
functions, one finds
\begin{align}
\label{eq:diff2point1deco}
  \frac{\dd}{\dd{\tau}}\langle \hat{v}_{{\bm k}_1}^s\hat{v}_{{\bm k}_2}^s\rangle &=
  \langle \hat{v}_{{\bm k}_1}^s\hat{p}_{{\bm k}_2}^s
  +\hat{p}_{{\bm k}_1}^s\hat{v}_{{\bm k}_2}^s
  \rangle, \\
  \label{eq:diff2point2deco}
\frac{\dd}{\dd{\tau}}\langle \hat{p}_{{\bm k}_1}^s\hat{v}_{{\bm k}_2}^s\rangle &=
\langle \hat{p}_{{\bm k}_1}^s\hat{p}_{{\bm k}_2}^s\rangle
-\omega^2({ k}_1,\tau)\langle\hat{v}_{{\bm k}_1}^s\hat{v}_{{\bm k}_2}^s
  \rangle, \\
  \label{eq:diff2point3deco}
\frac{\dd}{\dd{\tau}}\langle \hat{v}_{{\bm k}_1}^s\hat{p}_{{\bm k}_2}^s\rangle &=
\langle \hat{p}_{{\bm k}_1}^s\hat{p}_{{\bm k}_2}^s\rangle
-\omega^2({k}_2,\tau)\langle\hat{v}_{{\bm k}_1}^s\hat{v}_{{\bm k}_2}^s
\rangle, \\
\label{eq:diff2point4deco}
\frac{\dd}{\dd{\tau}}\langle \hat{p}_{{\bm k}_1}^s\hat{p}_{{\bm k}_2}^s\rangle &=
-\omega^2({k}_2,\tau)\langle \hat{p}_{{\bm k}_1}^s\hat{v}_{{\bm k}_2}^s\rangle
-\omega^2({k}_1,\tau)\langle\hat{v}_{{\bm k}_1}^s\hat{p}_{{\bm k}_2}^s
\rangle
%\nonumber \\ &
+\Gamma(2\pi)^{3/2}\tilde{C}_{E}(\tau,{\bm k}_1)
\delta({\bm k}_2-{\bm k}_1),
\end{align}
where correlators mixing R and I quantities still vanish, \ie
${\langle R^s_i R^{\tilde{s}}_j\rangle}\propto
\delta(s-\tilde{s})$. Compared to
\Eqs{eq:diff2point1}-\eqref{eq:diff2point4}, one can see that only the
last equation gets modified, and receives an additional contribution
proportional to $\Gamma$. Making use of
\Eqs{eq:GammaijDef}-\eqref{eq:GammaijDef:end}, this leads to the
following differential system for the entries of the covariance
matrix,
\begin{align}
\label{eq:diffcov1deco}
\frac{1}{k}\frac{\dd \gamma_{11}}{\dd{\tau}}&=\gamma_{12}+\gamma_{21}, \\
\label{eq:diffcov2deco}
\frac{1}{k}\frac{\dd}{\dd{\tau}}(\gamma_{12}+\gamma_{21})
&=2\gamma_{22}  -2\frac{\omega^2}{k^2}\gamma_{11}, \\
\label{eq:diffcov3deco}
\frac{1}{k}\frac{\dd \gamma_{22}}{\dd{\tau}}&=
-\frac{\omega^2}{k^2}(\gamma_{12}+\gamma_{21})+2\Gamma (2\pi)^{3/2}
\frac{\tilde{C}_{E}}{k^2}\, ,
\end{align}
which should be compared with
\Eqs{eq:diffcov1}-\eqref{eq:diffcov3}. Let us recall that, under the
assumption that the system is initially in the vacuum state, these
equations should be solved with initial conditions
$\gamma_{11}({\tau}_\uin)=1$, $(\gamma_{12}+\gamma_{21})({\tau}_\uin)=0$ and
$\gamma_{22}({\tau}_\uin)=1$.

Another important consequence of these transport equations is that
they lead to the following evolution for $\det( \gamma^s)
=\gamma_{11}\gamma_{22}-\gamma_{12}^2$:
\begin{align}
\label{eq:detgammadeco}
\frac{\dd}{\dd{\tau}}\det \left(\gamma^s\right)=2\Gamma  \gamma_{11} (2\pi)^{3/2}
\frac{\tilde{C}_{{E}}}{k}.
  \end{align}
 When $\Gamma=0$, \ie in the absence of an environment, one recovers
 the fact that this determinant is preserved, hence the system remains
 in a pure state. Otherwise, \Eq{eq:detgammadeco} indicates that the
 interaction with the environment induces decoherence of the system,
 since it makes the purity decrease away from one, see
 \Eq{eq:purityGaussianStates}.
  
Finally, similarly to \Eq{eq:thirddiffstandard}, one can derive a
single, third-order differential equation for $\gamma_{11}$, which
reads
\begin{align}
\label{eq:thirddiffstandarddeco}
  \frac{1}{k^3}\frac{{\dd}^3\gamma_{11}}{{\dd}{\tau}^3}
  &+4\frac{\omega^2}{k^2}\frac{1}{k}\frac{{\dd}\gamma_{11}}{{\dd}{\tau}}
  +\frac{2}{k}\frac{{\dd}}{{\dd}\tau}\left(\frac{\omega^2}{k^2}\right)
  \gamma_{11}=4\Gamma (2\pi)^{3/2}
\frac{\tilde{C}_{E}}{k^2}\, .
\end{align}
As pointed out below \Eq{eq:thirddiffstandard}, in the absence of a
source term in the right-hand side, the solution to this equation
reads $\gamma_{11} = v_k v_k^*$, where $v_k$ satisfies the classical
equation of motion $ v_{ k}''+\omega^2v_{ k}=0$. Using Green's
function method, the solution in the presence of a source term is thus
given by
\begin{align}
\label{eq:thirdorderdeco}
  &\gamma_{11}(\tau)=v_{ k}(\tau)v_{ k}^*(\tau)
  -\frac{8k}{W^2}(2\pi)^{3/2}
  \int_{\tau_\uin}^{\tau}
 \Gamma({\tau}') \, \tilde{C}_{E}({\tau}',k)\, 
\Ima^2\left[v_{ k}(\tau)
  v_{k}^*({\tau}')\right]{\dd}{\tau} ',
\end{align}
where $W\equiv v_{ k}v_{ k}^*{}'-v_{ k}^*v_{ k}'$ is the Wronksian of
the $v_k$ mode function. Given the equation of motion that $v_k$
satisfies, one can readily show that $W$ is preserved in time. It can
therefore be evaluated at initial time, where the initial conditions
derived for $v_k$ below \Eq{eq:thirddiffstandard} lead to
$W=2ik$. Using \Eqs{eq:diffcov1deco} and~\eqref{eq:diffcov2deco}
again, one thus obtains the following expressions for the entries of
the covariance matrix,
\begin{align}
\label{eq:exactgammadeco23}
\gamma_{11} &= \left\vert v_k\right\vert^2 + \mathcal{I}_k,
\qquad
\gamma_{12}  = \frac{\Rea \left( v_{k}  v^{* \, \prime}_{k}  \right)}{k}
+ \mathcal{J}_k ,
\qquad
\gamma_{22}  =  \frac{\left| v_{ k}^{\prime} \right|^2 }{k^2}  + \mathcal{K}_k\, ,
\end{align}
where
\begin{align}
\label{eq:I:integrals:def}
\mathcal{I}_k(\tau) &= \frac{2}{k}(2\pi)^{3/2}
  \int_{{\tau}_\uin}^{\tau}
 \Gamma({\tau}') \, \tilde{C}_{E}({\tau}',k)\, 
\Ima^2\left[v_{ k}({\tau}')
  v_{k}^*(\tau)\right]{\dd}{\tau} '\, ,\\
\label{eq:J:integrals:def}
\mathcal{J}_k(\tau)&=\frac{2}{k^2}(2\pi)^{3/2}
  \int_{{\tau}_\uin}^{\tau}
 \Gamma({\tau}') \, \tilde{C}_{E}({\tau}',k)\, 
\Ima\left[v_{ k}({\tau}')
  v_{k}^*(\tau)\right]\Ima\left[v_{ k}({\tau}')
  v_{k}^{*\prime}(\tau)\right]{\dd}{\tau} '\, ,\\
\label{eq:K:integrals:def}
\mathcal{K}_k(\tau)&=\frac{2}{k^3}(2\pi)^{3/2}
  \int_{{\tau}_\uin}^{\tau}
 \Gamma({\tau}') \, \tilde{C}_{E}({\tau}',k)
 \Ima^2\left[v_{ k}({\tau}')
  v_{k}^{*\prime}(\tau)\right]\dd{\tau}'\, .
\end{align}
These formula provide an explicit expression for the covariance
matrix, hence for the full quantum state of the reduced system.
\subsection{Generalised squeezing parameters}
\label{subsec:generalized}
We now follow the approach presented in \Sec{subsec:visu} where the
quantum state of the system is described in terms of squeezing
parameters. Note that the Bogoliubov coefficients introduced in
\Sec{subsec:bogo} cannot be directly generalised to the case where an
environment is present, since they are related to the unitary
evolution of the system. Therefore, one cannot use \Eq{eq:defrphiuv}
to define squeezing parameters in the present situation. However, the
geometrical interpretation developed around
\Fig{fig:ellipsesqueezingparameters} can still be used to introduce
generalised squeezing parameters. This will be particularly useful to
understand the state purity, and quantum discord, from a phase-space
geometrical perspective.

In the Caldeira-Leggett model indeed, the state is still described by
a covariance matrix, that can still be diagonalised as in
\Eq{eq:gamma:m1:tilde}. The only difference is that, as mentioned
around \Eq{eq:detgammadeco}, the determinant of the covariance matrix
does not remain equal to one. This introduces a new ``squeezing
parameter'', denoted $\lambda_k \equiv \det(\gamma^s)$, such that
\Eq{eq:gamma:m1:tilde} becomes
\begin{align}
\label{eq:gamma:m1:tilde:extended}
(\widetilde{\gamma}^s)^{-1}=\lambda_k^{-1/2}
\begin{pmatrix}
\ee^{2 r_k} & 0\\
0 & \ee^{-2 r_k}
\end{pmatrix} .
\end{align}
Performing the phase-space rotation of angle $\varphi_k$ introduced in
\Eq{eq:gamma:m1:tilde}, this leads to the following expression for the
covariance matrix in the $\mathrm{R}/\mathrm{I}$ partition,
\begin{align}
\label{eq:covgeneralizedsqueez}
\gamma^{s} =  \sqrt{\lambda_k} \begin{pmatrix}
  \cosh \left( 2 r_k \right) - \cos \left( 2 \varphi_k \right)
  \sinh\left( 2 r_k \right) & -\sin \left( 2 \varphi_k \right)
  \sinh\left( 2 r_k \right) \\
  -\sin \left( 2 \varphi_k \right) \sinh \left( 2 r_k \right) &
  \cosh\left( 2 r _k\right) + \cos \left( 2 \varphi_k \right)
  \sinh \left( 2 r_k \right)
\end{pmatrix} \, ,
\end{align}
which generalises \Eqs{eq:g11rphi}-\eqref{eq:g12rphi}. This shows
that, in the Caldeira-Leggett model, the quantum state of the system
can still be described with an ellipse in phase space, where $r_k$
describes the eccentricity of the ellipse, $\varphi_k$ its
orientation, and $\lambda_k$ its area (which is given by $\pi
\lambda_k$).

Note that equations of motion for the generalised squeezing parameters
$r_k$, $\varphi_k$ and $\lambda_k$ can also be derived, by plugging
\Eq{eq:covgeneralizedsqueez} into
\Eqs{eq:diffcov1deco}-\eqref{eq:diffcov3deco}. This leads to 
\begin{align}
\label{eq:eom:lambdak}
\frac{{\dd}\lambda_k}{{\dd} {\tau}} &=
2\Gamma (2\pi)^{3/2}\frac{\tilde C_{E}}{k}\lambda_k^{1/2}
\left[\cosh\left(2r_k\right)
  -\cos\left(2\varphi_k\right)\sinh \left(2r_k\right)\right],
\\
\frac{{\dd}r_k}{{\dd}{\tau}} &=
\frac{k}{2}\left(\frac{\omega^2}{k^2}-1\right)\sin(2\varphi_k)
-\frac{\Gamma}{\sqrt{\lambda_k}} \frac{(2\pi)^{3/2}}{2}\frac{\tilde C_{E}}{k}
\left[\sinh\left(2r_k\right)-\cos\left(2\varphi_k\right)
  \cosh \left(2r_k\right)\right],
\\
\frac{{\dd}\varphi_k}{{\dd}{\tau}} &=
-\frac{k}{2}\left(\frac{\omega^2}{k^2}+1\right)
+\frac{k}{2}\left(\frac{\omega^2}{k^2}-1\right)
\frac{\cos(2\varphi_k)}{\tanh(2r_k)}
- \Gamma\frac{\sin(2\varphi_k)}{2 \sinh(2r_k)\sqrt{\lambda_k}}
(2\pi)^{3/2}\frac{\tilde C_{E}}{k}.
\end{align}
One can check that, in the limit $\Gamma\to 0$,
\Eqs{eq:diffsqueezingr} are recovered, and that \Eq{eq:eom:lambdak} is
essentially a rewriting of \Eq{eq:detgammadeco} for the
non-conservation of the determinant.
\subsection{Quantum discord}
\label{sec:evoldiscordenvi}
\begin{figure} 
    \centering
    \includegraphics[width=0.49\textwidth]{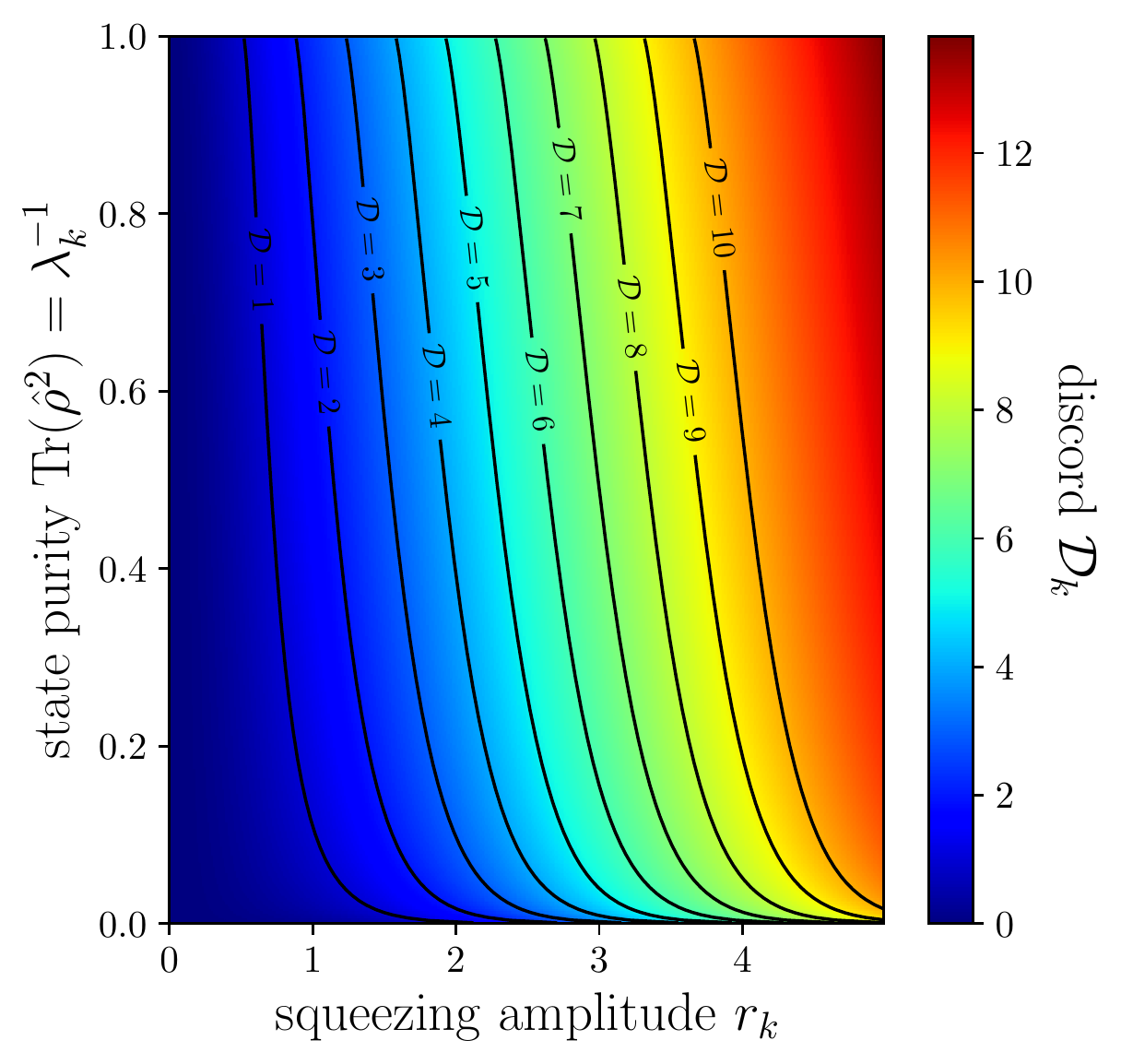}
    \includegraphics[width=0.49\textwidth]{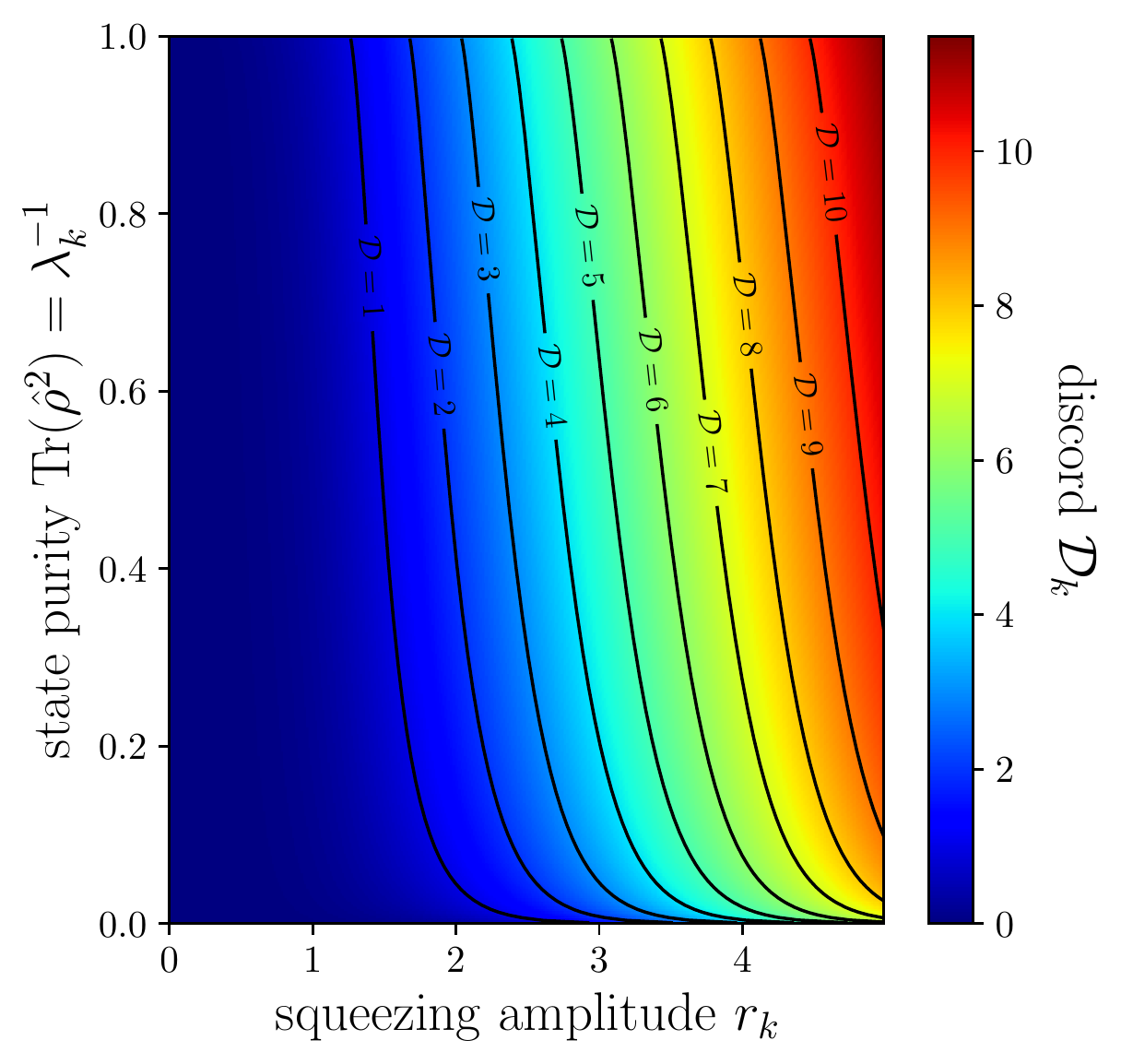}
    \caption{Quantum discord $\mathcal{D}_k$ in terms of the
      generalised squeezing amplitude $r_k$ and the state purity
      $\mathrm{Tr}(\hat{\rho}^2)=1/\lambda_k$, for $\theta=-\pi/4$
      (left panel) and $\theta=0.1$ (right panel). The black solid
      lines show a few contour lines of $\mathcal{D}_k$.}
    \label{fig:mapdiscordsqueezing}
\end{figure}

Let us now turn to the main goal of this article, namely the
calculation of the quantum discord in the presence of an
environment. As explained in \Sec{subsec:quantumdiscGHDM}, a single
quantity needs to be computed, namely $\sigma(\theta)$ given in
\Eq{eq:sigma(theta)}, which in terms of the generalised squeezing
parameters reads
\begin{align}
    \sigma(\theta)=\lambda_k^{1/2}\sqrt{1+\sinh^2(2r_k)\sin^2(2\theta)}\, .
\end{align}
Plugging this expression into \Eq{eq:finaldiscord}, one obtains an
explicit formula for quantum discord in terms of three parameters
only: the squeezing amplitude $r_k$, the state purity
$\mathrm{Tr}(\hat{\rho}^2)=1/\lambda_k$, and the partition angle
$\theta$. This formula is displayed in \Fig{fig:mapdiscordsqueezing}
for $\theta=-\pi/4$ (left panel, corresponding to the $\kmk$ partition)
and $\theta=0.1$ (right panel).
One can see that, as the squeezing amplitude increases, quantum
discord increases, as in the case where no environment was
present. When the state purity decreases, quantum discord decreases,
which means that interactions with an environment tend to reduce the
amount of quantum correlations. This is in agreement with the common
lore that decoherence is associated with the emergence of classical
properties. One can also check that quantum discord increases with the
partition angle as it varies between $0$ and $\pi/4$, \ie as it
interpolates between the $\mathrm{R}/\mathrm{I}$ partition which is
separable (hence uncorrelated) and the $\kmk$ partition where it is
maximally correlated and discordant.

In order to gain more analytical insight in the behaviour of quantum
discord, let us consider the large-squeezing limit $r_k\gg 1$ (this
limit is particularly relevant to the cosmological setting considered
in \Sec{sec:CosmoPerturbations}). From \Eq{eq:finaldiscord}, it is
clear that different behaviours are obtained depending on whether
$\sigma(\theta)/\sigma^2(0)$ is small or large. In the large-squeezing
limit, this ratio is given by
\begin{align}
\label{eq:ratio:crit:large:squeezing}
\frac{\sigma(\theta)}{\sigma^2(0)} \simeq
\lambda_k^{-1/2} \ee^{2 r_k} \frac{\left\vert \sin(2\theta) \right\vert}{2}\, .
\end{align}
In general, as the time evolution proceeds, $r_k$ increases (denoting
particle creation) and $\lambda_k$ increases too (as an effect of
decoherence). The two effects therefore compete in
\Eq{eq:ratio:crit:large:squeezing}, and whether the ratio
$\sigma(\theta)/\sigma^2(0)$ is small or large depends on the details
of the dynamics. Expanding \Eq{eq:finaldiscord} in these two limits,
one obtains
\begin{align}
\mathcal{D}_k \simeq
\begin{cases}
  \dfrac{2 r_k}{\ln 2 } \quad \text{if}\quad
  \ee^{2 r_k} \left\vert \sin(2\theta) \right\vert \gg \sqrt{\lambda_k}\\
  \lambda_k^{-1/2} \ee^{2 r_k}
  \dfrac{\left\vert \sin(2\theta) \right\vert}{2\ln 2}\quad
  \text{if}\quad \ee^{2 r_k} \left\vert \sin(2\theta)\right\vert
  \ll \sqrt{\lambda_k}
\end{cases}\, .
\end{align}
This shows that quantum discord is large in the first case and small
in the second case. As a consequence, whether quantum discord is large
or small depends on which of squeezing or decoherence wins in the
ratio~\eqref{eq:ratio:crit:large:squeezing}. This provides a simple
criterion for assessing when decoherence substantially reduces the
amount of quantum correlations, namely it happens when
\begin{align}
\mathrm{Tr}(\hat{\rho}^2) \ll \ee^{-4 r_k}\, .
\end{align}
A useful geometrical interpretation of
\Eq{eq:ratio:crit:large:squeezing} is that, according to
\Eq{eq:gamma:m1:tilde:extended}, the combination
$\sqrt{\lambda_k}\ee^{-2r}$ happens to be the length of the semi-minor
axis of the phase-space ellipse. The semi-minor axis increases as an
effect of decoherence, which increases the overall area of the
ellipse, and decreases because of quantum squeezing: this competition
determines whether the semi-minor axis increases or decreases, hence
it determines the fate of quantum discord.

Finally, in the same way as we have introduced generalised squeezing
parameters, one can extend the definition of the number of particles
$\mathcal{N}_k$ and the correlation $\mathcal{C}_k$, by plugging
\Eq{eq:covgeneralizedsqueez} into \Eqs{eq:nkck1}-\eqref{eq:nkck3},
leading to
\begin{align}
\label{eq:nkckeffectivesqueezing}
2 {\cal N}_{ k} + 1 = \sqrt{\lambda_k} \cosh \left( 2 r_k \right) \, , \quad
2 {\cal C}_{ k} =  -\sqrt{\lambda_k} e^{i 2 \varphi_k} 
\sinh \left( 2 r_k \right) \, .
\end{align}
In terms of these parameters, one has $\sigma(\theta) =
\sqrt{4(\mathcal{N}_k+1/2)^2 - 4 \vert \mathcal{C}_k\vert^2
  \cos^2(2\theta)}$, which allows one to express quantum discord as a
function of $\mathcal{N}_k$ and $\vert \mathcal{C}_k\vert$ only.
\section{Application : Cosmological perturbations}
\label{sec:CosmoPerturbations}
In this section, we apply the formalism developed so far to the case
of cosmological perturbations. The goal is twofold. First, this will
allow us to exemplify in a concrete situation how the tools introduced
above work in practice. Second, as explained in \Sec{sec:intro}, the
presence of quantum correlations in the primordial field of
cosmological perturbations, and how decoherence might partly remove
them, is of great importance regarding our understanding of the origin
of cosmic structures as emerging from a quantum-mechanical mechanism,
and for our ability to test this aspect of the cosmological scenario.

When the universe is dominated by a single scalar-field, there is a
single scalar gauge-invariant perturbation known as the
Mukhanov-Sasaki variable~\cite{Mukhanov:1981xt,Kodama:1985bj}. If time
is parameterised by conformal time $\eta$,\footnote{So far the time
  variable was left unspecified and denoted with the generic variable
  ``$\tau$''. The above considerations can thus be applied to any time
  variable, provided the Hamiltonian is adapted accordingly. In what
  follows, we therefore make the identification $\tau=\eta$ and
  $\prime = \partial/\partial\eta$.} which is related to cosmic time
$t$ by $\dd{t}=a \dd \eta$, where $a$ is the Friedman-Lema\^ \i
tre-Robertson-Walker scale factor, then the Hamiltonian for the
Mukhanov-Sasaki variable $v$ is by \Eq{eq:Hsystem} where
\begin{align}
\label{eq:defomega}
\omega^2\left(k,\eta\right)=
k^2-\frac{\left(a\sqrt{\epsilon_1}\right)^{\prime\prime}}
{a\sqrt{\epsilon_1}}\, .
\end{align}
In this expression, a prime denotes derivation with respect to $\eta$,
$\epsilon_1=1-{\cal H}'/{\cal H}^2$ is the first slow-roll parameter
and ${\cal H}=a'/a$.

In practice, we will consider the case of a de-Sitter expansion where
$a(\eta)=-1/(H\eta)$ with $H$ the Hubble parameter, since cosmological
observations indicate that it is a good proxy for the dynamics of the
universe expansion during the inflationary phase. In that case,
$\omega^2=k^2-2/\eta^2$, where $\eta$ varies between $-\infty$ to $0$.
\subsection{Inflationary perturbations in the absence of an environment}
\label{sec:cosmo:free}
Following the approach presented in \Sec{subsec:bogo}, let us first
derive the Bogoliubov coefficients. The solution of the equation of
motion~\eqref{eq:SecondOrderDiffEquplusvstar} satisfied by
$v_k=u_k+w_k^*$, \ie $v_k''+\omega^2 v_k=0$, is given by
\begin{align}
\label{eq:desittermodefunction}
v_k\left( \eta \right) = \left(1-\frac{i}{k\eta}\right)\ee^{-i k \eta} \, ,
\end{align}
where we have made use of the initial conditions derived for $v_k$
below \Eq{eq:thirddiffstandard}, namely $v_k(\eta_\uin)=1$ and
$v_k'(\eta_\uin)=-ik$, at initial time $\eta_\uin$ set to the infinite
past, $\eta_\uin=-\infty$. As explained below
\Eq{eq:Relationuplusvuminusv}, this allows one to derive both
Bogoliubov coefficients, and one finds
\begin{align}
\label{eq:soluvds}
    u_k=\left(1-\frac{i}{k\eta}-\frac{1}{2k^2\eta^2}\right)e^{-ik\eta}\, ,
    \qquad\qquad
    w_k^* =\frac{1}{2k^2\eta^2}e^{-ik\eta}\, .
\end{align}
One can easily check that these solutions satisfy
\Eq{eq:detBogoCoeff}, \ie $\vert u_k\vert^2-\vert w_k\vert^2=1$. Using
\Eq{eq:EvolvedCovBogoCoeffVac}, one can then calculate the
covariance-matrix element and one obtains
\begin{align}
\label{eq:solcovds}
\gamma_{11}(\eta)&=1+\frac{1}{k^2\eta^2},
\qquad \gamma_{12}(\eta)=-\frac{1}{k^3\eta^3}, 
    \qquad
    \gamma_{22}(\eta)=1-\frac{1}{k^2\eta^2}+\frac{1}{k^4\eta^4}.
\end{align}
As a consistency check, one can verify that these expressions are
solutions of the differential
system~\eqref{eq:diffcov1}-\eqref{eq:diffcov3}, and that they satisfy
the initial conditions that were given for it. Finally, the squeezing
parameters can be derived from \Eqs{eq:g11rphi}-\eqref{eq:g12rphi},
which lead to
\begin{align}
  \cosh ^2\left(r^\udS_k \right) &=1+\frac{1}{4 (k\eta)^4}, \quad
  \label{eq:tanphiketa}
  \tan\left(2\varphi^{\udS}_k\right)= \frac{2k\eta}{1-2k^2\eta^2}.
  \end{align}
These expressions can be inverted, and one finds\footnote{The
  inversion for $\varphi_k$ should be done noting that from
  \Eq{eq:diffsqueezingr} and the fact that $r_k$ grows during
  inflation, one has $\sin(2\varphi_k^{\udS})<0$. Moreover,
  \Eq{eq:tanphiketa} implies that $\tan(2\varphi_k^{\udS})>0$ if
  $k\eta<-1/\sqrt{2}$ and $\tan(2\varphi_k^{\udS})<0$ if
  $-1/\sqrt{2}<k\eta<0$. This implies that
  $2\varphi_k\in[-\pi,-\pi/2]$ when $k\eta<-1/\sqrt{2}$, and that
  $2\varphi_k\in[-\pi/2,0]$ when $k\eta<-1/\sqrt{2}$ (modulo
  $2\pi$). The Heaviside function in \Eq{eq:deSittersqueezinphi}
  ensures that $\varphi_k$ is continuous when $k\eta=-1/\sqrt{2}$.}
\begin{align}
\label{eq:deSittersqueezingr}
r^{\udS}_k(\eta) &= \frac{1}{2} \mathrm{arccosh} 
\left[ 1+\frac{1}{2 (k\eta)^4} \right] \, ,\\
\label{eq:deSittersqueezinphi}
\varphi_k^{\udS} & = \frac{1}{2}
\arctan \left(\frac{2k\eta}{1-2k^2\eta^2} \right)   
-\frac{\pi}{2} {\mathrm H}\left(-k\eta-\frac{1}{\sqrt{2}}\right)+\ell \pi\, ,
\end{align}
where $\ell$ is an integer number and $\mathrm{H}$ is the Heaviside
step function defined as $\mathrm{H}(x)=1$ when $x>0$ and $0$
otherwise. One can see that the squeezing amplitude increases as
inflation proceeds, and at late time (\ie when the wavelength of the
mode under consideration is much larger than the Hubble radius, $k\ll
\calH=-1/\eta$), $r_k\simeq -2\ln(-k\eta)= 2 \ln[a/a(k)]$. In this
expression, $a(k)$ denotes the value of the scale factor when $k$
crosses out the Hubble scale, \ie when $k\eta=-1$. For the modes
observed in the cosmic microwave background, $\ln[a/a(k)]\simeq 50$ at
the end of inflation, hence the squeezing amplitude is of order
$100$. The squeezing angle starts out from $\varphi_k=-\pi/2$ in the
asymptotic past and approaches $0$ at late time, where
$\varphi_k\simeq k\eta \simeq -a(k)/a \simeq -\ee^{-r_k/2}$ [here we set
  $\ell=0$ in \Eq{eq:deSittersqueezinphi}].
\begin{figure}[!t]
\centering
\includegraphics[width=14.0cm]{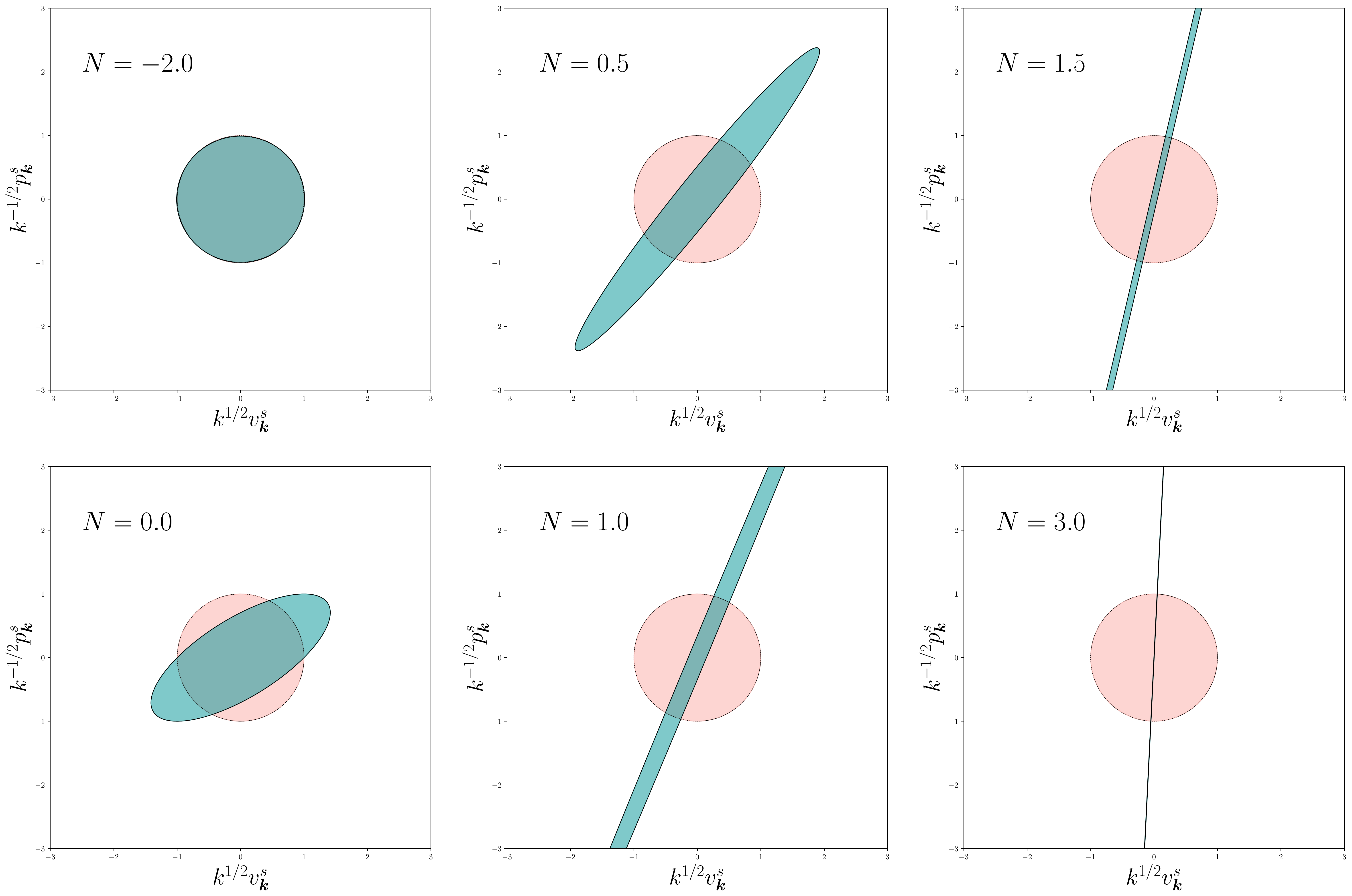}
\caption{Phase-space ellipse in the plane
  $(k^{1/2}v_{\bm{k}}^s,k^{-1/2}p_{\bm{k}}^s)$ (see the discussion
  around \Fig{fig:ellipsesqueezingparameters}) at different instants
  during inflation, labelled by $N=\ln[a/a(k)]$, \ie the number of
  \efolds~measured from the Hubble-crossing time of the mode under
  consideration. On sub-Hubble scales, the ellipse remains a circle,
  while it gets squeezed and rotates in the super-Hubble regime.}
\label{fig:wigner}
\end{figure}
The evolution of the squeezing parameters is displayed at the level of
the phase-space ellipse in \Fig{fig:wigner}. While the mode under
consideration remains in the vacuum state, \ie on sub-Hubble scales
when $a\ll a(k)$, the squeezing amplitude is small and the ellipse is
close to a circle. When the mode crosses out the Hubble radius, the
ellipse gets squeezed ($r_k>0$) and rotates. In the asymptotic future,
it gets infinitely squeezed and its semi-minor axis becomes aligned
with the horizontal axis ($\varphi_k\to 0$, in agreement with
\Fig{fig:ellipsesqueezingparameters}).

Regarding quantum discord, plugging \Eq{eq:deSittersqueezingr} into
\Eq{eq:sigmasqueezdS} leads to
\begin{align}
  \sigma(\theta) = \sqrt{1+\frac{1}{k^4\eta^4}
    \left(1+\frac{1}{4 k^4\eta^4}\right) \sin^2(2\theta) }\, ,
 \end{align}
 which together with \Eq{eq:discordredf} leads to an explicit
 expression for quantum discord. In the super-Hubble regime, \ie at
 late time when the squeezing amplitude is large, it can be expanded
 according to
\begin{align}
\label{eq:discordds}
\mathcal{D}_k \simeq
\log_2 \left(\frac{\vert \sin(2\theta)\vert}{4 k^4 \eta^4}\right)
+ \frac{1}{\ln 2}  \sim \frac{4}{\ln 2}
\ln\left[\frac{a}{a(k)}\right] \, .
\end{align}
Recalling that modes of astrophysical
interest are such that $r_k\sim 50$ at the end of inflation, their
discord is of order $300$, which makes the cosmic microwave background
an extremely discordant state by laboratory-experiment
standard~\cite{Martin:2015qta}.
\subsection{Inflationary perturbations in the presence of an environment}
Let us now generalise the above considerations to the case where
cosmological perturbations interact with environmental degrees of
freedom. In practice, we will use the Caldeira-Leggett model and the
approach laid out in \Sec{sec:caldleggeteom}.
Note that this framework relies on the assumption that interactions are linear in the system's variables, which is indeed the case at leading order in cosmological perturbation theory. In principle, higher-order coupling terms may also be present, which would lead to non-Gaussian states. While Lindblad equations can still be derived in that case, and the environmental imprint on the power spectrum and higher-order correlation functions can be investigated, see \Refs{Martin:2018zbe,Martin:2018lin},  this does not allow for a straightforward calculation of quantum discord. Nevertheless, these effects are parametrically suppressed by the amplitude of primordial fluctuations, which are constrained to be small, so they can be safely neglected as a first approximation.

In practice, the relevant environmental degrees of freedom during inflation can be additional fields, (since most physical setups that have been proposed to embed inflation contain extra fields), to which the inflaton couples at least gravitationally. Because of the non-linearities of General Relativity, unobserved scales also couple to the ones of observable interest, and they may constitute another ``environment''. The advantage of the present formalism is that the microphysical details of the environment do not need to be further specified.

The calculation of the integrals derived in \Eqs{eq:I:integrals:def},
(\ref{eq:J:integrals:def}) and~(\ref{eq:K:integrals:def}) require to
specify the function $\tilde{C}_E(k)$, \ie the Fourier transform of
the equal-time environment correlator $C_E(\bm{x}-\bm{y})$. In
practice, we will consider that the environment is correlated on
length scales $\ell_E$, \ie that $C_E(\bm{x}-\bm{y})$ is suppressed
when $a \vert \bm{x}-\bm{y}\vert \gg \ell_E$ (here $\bm{x}$ and
$\bm{y}$ are comoving coordinates, which explains why the scale factor
has been introduced). This implies that the Fourier transform
$\tilde{C}_E(k)$ is suppressed at scales $k\ll a/\ell_E$, which in
practice we model via a simple Heaviside function
\begin{align}
\label{eq:kG:def}
\Gamma \tilde{C}_E(k) = (2\pi)^{-3/2} k_\Gamma^2
\left(\frac{a}{a_*}\right)^{p-3}
\mathrm{H}\left(1-\frac{k\ell_E}{a}\right) .
\end{align}
In this expression, $k_\Gamma$ sets the strength of the environmental
effects and has the same dimension as a comoving wavenumber, hence the
notation [the prefactor $(2\pi)^{-3/2}$ is introduced for later
  convenience, and in order to match the notations of
  \Refs{Martin:2018zbe, Martin:2018lin}]. The possible time dependence
of $\Gamma$ is accounted for in the factor $(a/a_*)^{p-3}$, where
$a_*$ denotes the scale factor at some reference time, \ie we assume
that $\Gamma \tilde{C}_E$ evolves as a power of the scale factor. The
model has therefore two free parameters, namely $k_\Gamma$ and $p$.

\begin{figure}[!t]
\centering
\includegraphics[width=10.0cm]{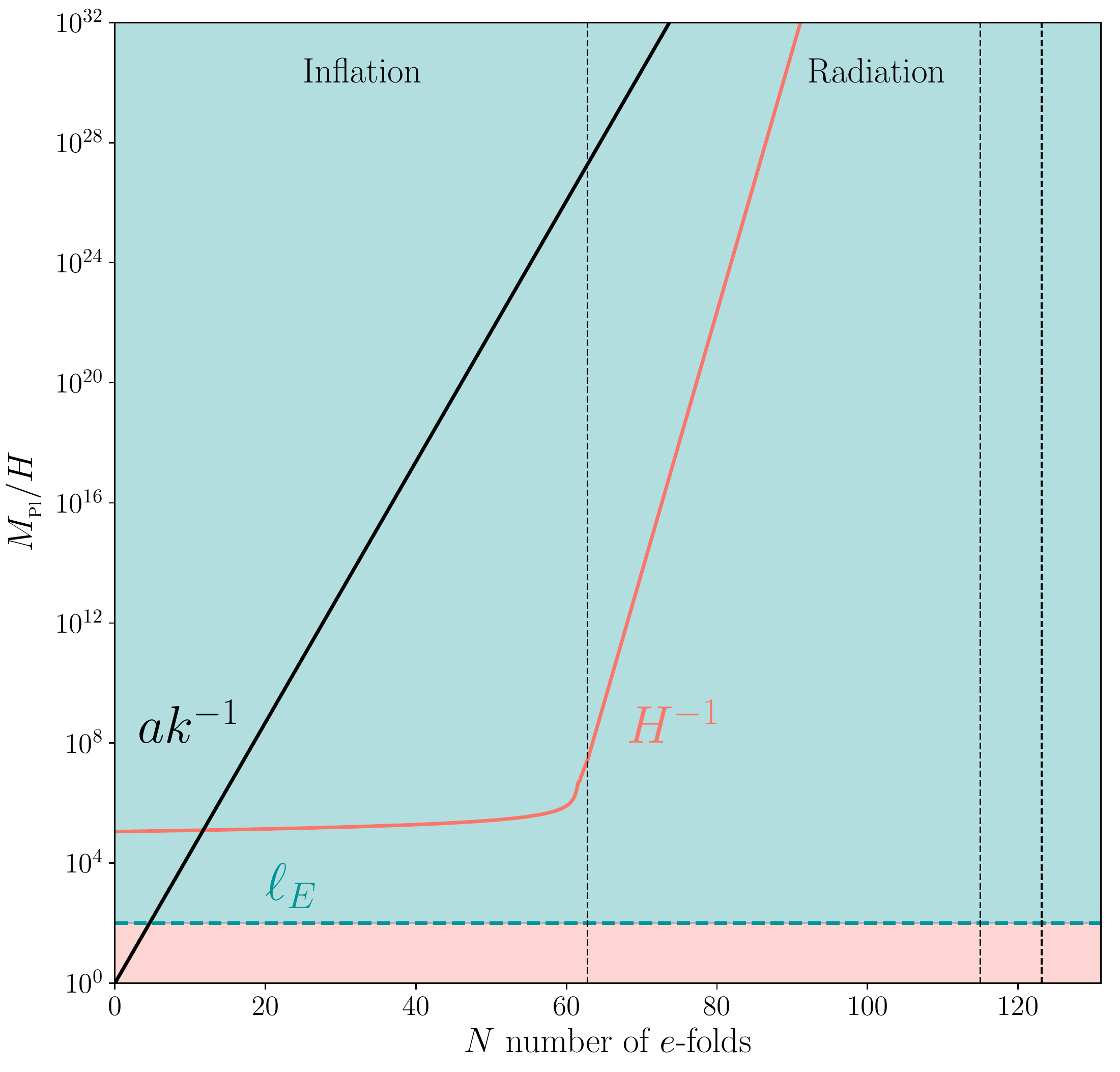}
\caption{Schematic representation of the evolution of the different
  physical scales. $a/k$ (black) represents the physical wavelength of
  the mode $\bm{k}$ under consideration. When $a/k$ crosses $\ell_{E}$
  (green dashed), the coherence length of the environment, decoherence
  starts to be effective. When $a/k$ crosses out $H^{-1}$ (pink), the
  Hubble radius, the perturbation starts to be amplified.}
\label{fig:crossing}
\end{figure}

In realistic situations, one may want to consider smoother kernels
than Heaviside functions, but this only slightly affects the boundary
terms in the integrals of \Eqs{eq:I:integrals:def},
(\ref{eq:J:integrals:def}) and~(\ref{eq:K:integrals:def}) and does not
lead to substantial modifications of the following considerations. In
practice, \Eq{eq:kG:def} indicates that small-scales fluctuations are
immune to environmental effects (which, conveniently, leave the vacuum
state unaffected at early time). The effect of the interaction with
the environment becomes relevant when the mode under consideration
crosses out the correlation length $\ell_E$. In practice, if the
environment is comprised of heavy (\ie with respect to the Hubble
scale) degrees of freedom, one has $\ell_E < a/\mathcal{H}$, hence a
given mode first crosses the environment correlation length before
crossing the Hubble radius. The situation is depicted in
\Fig{fig:crossing}.

\subsubsection{Covariance matrix}
\label{sec:cov:cosmo}
With the ansatz~\eqref{eq:kG:def}, the integrals appearing in
\Eqs{eq:I:integrals:def}, (\ref{eq:J:integrals:def})
and~(\ref{eq:K:integrals:def}) can be performed exactly, and a
detailed calculation is presented in \App{ap:sec:covmatcaldlegget}. In
the late-time limit, \ie on super-Hubble scales, $-k\eta \rightarrow
0$, they can be approximated by
\begin{align}
  \label{eq:expansiong11}
   \gamma_{11}
   \simeq &
   \frac{1}{(-k\eta)^2}\left\lbrace 1 -
     2  \left(\frac{k_\Gamma}{k}\right)^2\left[
     B_{11}\left(\frac{k}{k_*},p,\ell_EH\right)+
      A_{11}\left(\frac{k}{k_*},p\right)(-k\eta)^{8-p}\right]\right\rbrace
%   + \mathcal{O} \left[\left(-k\eta\right)^0,\left(-k\eta\right)^{7-p}\right]
   , \\
   \label{eq:expansiong12}
   \gamma_{12} \simeq &
   \frac{1}{(-k\eta)^3}
   \left\lbrace 1 - 2  \left(\frac{k_\Gamma}{k}\right)^2\left[
     B_{12}\left(\frac{k}{k_*},p,\ell_EH\right)+A_{12}\left(\frac{k}{k_*},p\right)\left(-k\eta\right)^{8-p} \right]\right\rbrace
%   + {\cal O}\left[\left(-k\eta\right)^0,\left(-k\eta\right)^{7-p} \right]
  ,
   \\
   \label{eq:expansiong22}
\gamma_{22} \simeq & 
\frac{1}{(-k\eta)^4} \left\lbrace 1
  - 2  \left(\frac{k_\Gamma}{k}\right)^2 \left[
  B_{22}\left(\frac{k}{k_*},p,\ell_EH\right)+A_{22}\left(\frac{k}{k_*},p\right)\left(-k\eta\right)^{8-p} \right]\right\rbrace
%+{\cal O}\left( \frac{1}{\left(-k\eta\right)^2},\left(-k\eta\right)^{6-p} \right)
,
  \end{align}
  see \Eqs{eq:g11approx}, (\ref{eq:g12approx})
  and~(\ref{eq:g22approx}), where $k_*$ denotes the comoving scale
  that crosses out the Hubble radius at the reference time $\eta_*$.
  One can check that, in the limit $k_\Gamma \rightarrow 0$, one
  recovers the covariance matrix calculated in the absence of an
  environment, namely \Eqs{eq:solcovds}. The coefficients $A_{11}$,
  $B_{11}$, $A_{12}$, $B_{12}$, $A_{22}$ and $B_{22}$ are functions of
  the parameters $p$ and $\ell_EH$ and their explicit expressions can
  be found in~\App{ap:sec:covmatcaldlegget}.

Let us note that the expression for $\gamma_{11}$ is of observational
interest as it gives the relative correction to the power spectrum of
the Mukhanov-Sasaki variable, see \Eq{eq:GammaijDef}, or of any
quantity proportional to the Mukhanov-Sasaki variable such as the
curvature perturbation $\zeta$ that is measured on the cosmic
microwave background. Upon expanding the expression given for $B_{11}$
in \App{ap:sec:covmatcaldlegget} in the regime $\ell_E H\ll 1$, one
obtains
\begin{align}
  B_{11}& \simeq
  \frac{1}{2}\left(\frac{k}{k_*}\right)^{p-3}
  \biggl(\frac{\left(\ell_EH\right)^{p-4}}{p-4}
  \left\{1-\frac{p-4}{2}\left(\ell_EH\right)\sin \left(\frac{2}{\ell_EH}\right)
    +{\cal O}\left[\left(\ell_EH\right)^2\right]\right\}
\nonumber \\ &
    -\frac{(p-3)(p-6)}{2^{4-p}}\Gamma(2-p)
    \cos\left(\frac{\pi}{2}p\right)\biggr).
\end{align}
It is interesting to notice that, at next-to-leading order, this
expression contains non-analytical terms. However, it is likely that
this non-analytical behaviour would be smoothed out if a non-sharp
window function were used. On the other hand, we also have
$A_{11}=-2(k/k_*)^{p-3}/[(p-8)(p-5)(p-2)]$, see \Eq{eq:A11}. Which
term dominates in $B_{11}$ depends on the relative position of $p$
with respect to $4$, while which of the corrections in
\Eq{eq:expansiong11} dominates depends on whether $p<8$ or $p>8$. There
are therefore three cases to distinguish, and one finds
\begin{align}
\frac{\Delta\calP_\zeta}{\calP_\zeta}\simeq
\begin{cases}
\displaystyle   \frac{\left(\ell_E H\right)^{p-4}}{4-p}
  \left(\frac{k_\Gamma}{k_*}\right)^2 \left(\frac{k}{k_*}\right)^{p-5}
  \quad &\text{if}\quad p<4,\\
  \displaystyle 2^{p-4}(3-p)(6-p)\Gamma(2-p)\cos\left(\frac{\pi p}{2}\right)
  \left(\frac{k_\Gamma}{k_*}\right)^2 \left(\frac{k}{k_*}\right)^{p-5}
  \quad &\text{if}\quad 4<p<8,\\
  \displaystyle \frac{4}{(p-8)(p-5)(p-2)}\left(\frac{k_\Gamma}{k_*}\right)^2
  \left(\frac{k}{k_*}\right)^3\left(\frac{\eta}{\eta_*}\right)^{8-p}
  \quad &\text{if}\quad p>8.
\end{cases} 
\end{align}
One can check that these expressions coincide with the result obtained
in \Refa{Martin:2018zbe}.\footnote{More precisely, they should be
  compared to Eqs.~(3.35), (3.32) and (3.29) of that reference (when
  setting $\epsilon_{1*}=0$ and $\nu=3/2$ in those expressions), to
  which they agree up to a factor $2$ that corresponds to a factor $2$
  difference in the definition of $\Gamma$.} In particular, when
$p<8$, the correction to the power spectrum freezes on large scale and
is scale invariant when $p=5$, while it continues to increase on large
scales for $p>8$. These formulas allow one to set upper bound on
$\Gamma$ such that the modifications to observables remain negligible,
see the white dotted line in \Fig{fig:mapdiscordcosmo} below.
\subsubsection{State purity}
\label{sec:purity:cosmo}
Endowed with the above expressions of the components $\gamma_{11}$,
$\gamma_{12}$ and $\gamma_{22}$ of the covariance matrix, we are now
in a position to calculate $\sigma(0)$ and the state purity. However,
upon evaluating $\sigma(0)$ by plugging
\Eqs{eq:expansiong11}-\eqref{eq:expansiong22} into
\Eq{eq:sigma(theta)}, one can see that the terms controlled by
$k_\Gamma$ all cancel out when $\theta=0$, which implies that
\Eqs{eq:expansiong11}-\eqref{eq:expansiong22} must be expanded to
higher order in order to get the first correction to
$\sigma(0)$. Before following that route, let us note that such an
expansion can be avoided by using \Eq{eq:detgammadeco} directly. The
right hand side of this equation is proportional to $k_\Gamma^2$, so
it is enough to evaluate it by using the solution~\eqref{eq:solcovds}
for $\gamma_{11}$ in the free theory, which leads to
\begin{align}
\sigma^2(0)=
  \det \gamma ^s\simeq 1 -2\left(\frac{k_\Gamma}{k_*}\right)^2
  \left(\frac{k}{k_*}\right)^{p-5}\int _{1/(\ell_EH)}^{-k\eta}
  \left(y^{3-p}+y^{1-p}\right)\dd y,
  \end{align}
namely
\begin{align}
\label{eq:sigma0:exp:cosmo}
  \sigma^2(0)\simeq 1+2\left(\frac{k_\Gamma}{k_*}\right)^2\left(\frac{k}{k_*}\right)^{p-5}
  \left[\frac{1}{p-2}\left(\frac{k}{k_*}\right)^{2-p}
    \left(\frac{a_*}{a}\right)^{2-p}-\frac{\left(\ell_EH\right)^{p-4}}{p-4}
   \right] ,
\end{align}
where we have kept the leading terms in $\ell_E H$ and in $-k\eta$
only. This again coincides with the result found in
\Refa{Martin:2018zbe}, see Eq.~(4.6) of that reference, and it implies
that decoherence occurs at the pivot scale $k_*$ when
\begin{align}
\label{eq:decoherence:param:cosmo:simplified}
\frac{k_\Gamma}{k_*}\gg
\begin{cases}
\displaystyle \left(\ell_E H \right)^{2-\frac{p}{2}}
\qquad& \text{if}\qquad p<2,\\
\displaystyle \left(\frac{a}{a_*}\right)^{1-\frac{p}{2}}
\qquad& \text{if}\qquad p>2,
\end{cases} 
\end{align}
where we recall that the state purity is related to $\sigma(0)$ via
\Eq{eq:purityGaussianStates}. This domain is delineated by the white
dashed line in \Fig{fig:mapdiscordcosmo}.

Before moving on and addressing the calculation of quantum discord,
let us note that the above result can be recovered from a higher-order
expansion of the covariance matrix. This is done in detail
in~\App{ap:sec:covmatcaldlegget}, see~\Eqs{eq:expansioncorr11},
(\ref{eq:expansioncorr12}) and~(\ref{eq:expansioncorr22}). This leads
to expressions for $\gamma_{11}$, $\gamma_{12}$ and $\gamma_{22}$
which, compared to \Eqs{eq:expansiong11}, (\ref{eq:expansiong12})
and~(\ref{eq:expansiong22}), contain extra coefficients (\ie beyond
$A_{11}$, $B_{11}$, $A_{12}$, $B_{12}$, $A_{22}$ and $B_{22}$), namely
$C_{11}$, $D_{11}$, \dots, $C_{12}$, $D_{12}$, \dots and $C_{22}$,
$D_{22}$, \dots, the explicit expressions of which are given
in~\App{ap:sec:covmatcaldlegget}. Of course, these coefficients are
also functions of the parameters $p$ and $\ell_EH$. Plugging the
result into \Eq{eq:sigma(theta)}, one obtains that, on super-Hubble
scales,
\begin{align}
\label{eq:finalsig0}
\sigma^2(0)&=\Sigma_{-6}\left(-k\eta\right)^{-6}
+\Sigma_{-5}\left(-k\eta\right)^{-5}
+\Sigma_{-4}\left(-k\eta\right)^{-4}
+\Sigma_{-3}\left(-k\eta\right)^{-3}
+\Sigma_{-2}\left(-k\eta\right)^{-2}
\nonumber \\ &
+\Sigma_{-1}\left(-k\eta\right)^{-1}
+1+\Sigma_0+\Sigma_1 \left(-k\eta\right)+\cdots 
+\Sigma_{2-p}\left(-k\eta\right)^{2-p}+
\Sigma_{4-p}\left(-k\eta\right)^{4-p}
\nonumber \\ &
+\Sigma_{5-p}\left(-k\eta\right)^{5-p}+\cdots .
\end{align}
Each coefficient $\Sigma_i$ is a combination of the coefficients
$A_{11}$, $B_{11}$, \dots, $A_{12}$, $B_{12}$, \dots and $A_{22}$,
$B_{22}$ \dots . As a consequence, the $\Sigma_i$'s are also functions
of the parameters $p$ and $\ell_EH$ and are proportional to
$k_\Gamma^2$ or $k_\Gamma^4$, which guarantees that, without an
environment (\ie when $k_\Gamma\rightarrow 0$), one recovers
$\sigma(0)=1$. Then, using the explicit expressions of $A_{11}$,
$B_{11}$, \dots, $A_{12}$, $B_{12}$, \dots and $A_{22}$, $B_{22}$
\dots given in~\App{ap:sec:covmatcaldlegget}, one can show that
$\Sigma_{-6}=\Sigma_{-5}=\Sigma_{-4}=\Sigma_{-3}
=\Sigma_{-2}=\Sigma_{-1}=\Sigma_1=\cdots
=0$. These relationships are direct consequences of the cancellations
mentioned before, and indicate that the expansion has to be performed
to very high order indeed. Therefore, at leading order in the
super-Hubble limit, one has
\begin{align}
  \label{eq:superHsigzero}
  \sigma^2(0)\simeq 1+\Sigma_0+\Sigma_{2-p}\left(-k\eta\right)^{2-p}+\cdots,
  \end{align}
with
  \begin{align}
    \Sigma_0&= \left(\frac{k_\Gamma}{k}\right)^2\left(-2C_{11}+4E_{12}
    -2E_{22}-2F_{11}-2G_{22}\right)
\nonumber \\ &
    +\left(\frac{k_\Gamma}{k}\right)^4\left(
    -4C_{12}^2+4D_{11}D_{22}-8B_{12}E_{12}+4C_{11}E_{22}
    +4B_{22}F_{11}+4B_{11}G_{22}\right),
      \\
      \Sigma_{2-p}&=\left(\frac{k_\Gamma}{k}\right)^2\left(-2A_{11} +4A_{12}
      -2A_{22}\right)
      + \left(\frac{k_\Gamma}{k}\right)^4\left(4A_{22}B_{11}-8A_{12}B_{12}
      +4A_{11}B_{22}\right) \, .
  \end{align}
As mentioned above, the coefficients appearing in the expansions of
$\gamma_{11}$, $\gamma_{12}$ and $\gamma_{22}$ are functions of $p$
and $\ell_EH$. However, the coefficients $A_{11}$, $A_{12}$ and
$A_{22}$ only depend on $p$, see the explicit expressions
in~\App{ap:sec:covmatcaldlegget}, \Eqs{eq:A11}, (\ref{eq:A12})
and~(\ref{eq:A22}). It follows that the term proportional to
$k_\Gamma^2$ in the expression of $\Sigma_{2-p}$ is also a function of
$p$ only. Explicitly, one has
\begin{align}
  \label{eq:kg21}
  -2A_{11}+4A_{12}-2A_{22}=\frac{2}{p-2}\left(\frac{k}{k_*}\right)^{p-3}\, .
  \end{align}
By contrast, the term proportional to $k_\Gamma^2$ in the coefficient
$\Sigma_0$ contains the terms $C_{11}$, $E_{12}$, $E_{22}$, $F_{11}$
and $G_{22}$ and, as consequence, depends on $p$ but also on
$\ell_EH$. Explicitly, one finds
\begin{align}
  \label{eq:kg22}
-2C_{11}+4E_{12}
-2E_{22}-2F_{11}-2G_{22}=-2\left(\frac{k}{k_*}\right)^{p-3}
\left[\frac{\left(\ell_EH\right)^{p-4}}{p-4}
  +\frac{\left(\ell_EH\right)^{p-2}}{p-2} \right].
\end{align}
Combining the above results, one recovers \Eq{eq:sigma0:exp:cosmo},
which is an important consistency check of our calculations.

\begin{figure*}[t]
\begin{center}
\includegraphics[width=0.49\textwidth,clip=true]{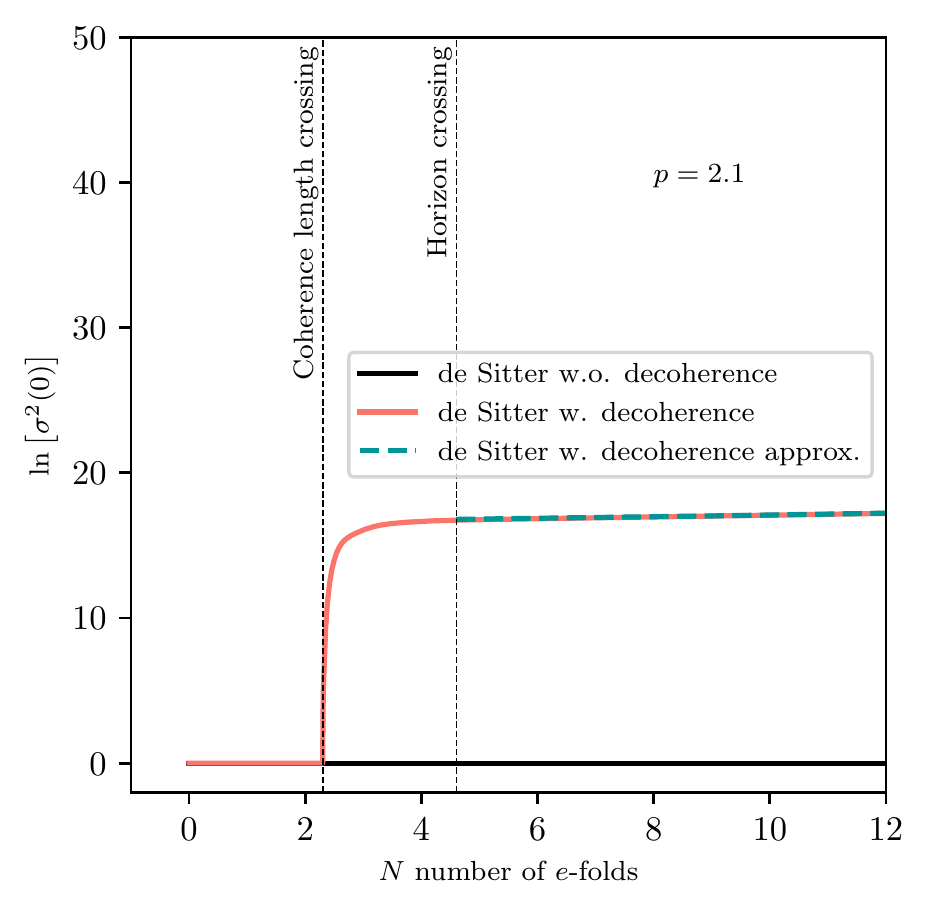}
\includegraphics[width=0.49\textwidth,clip=true]{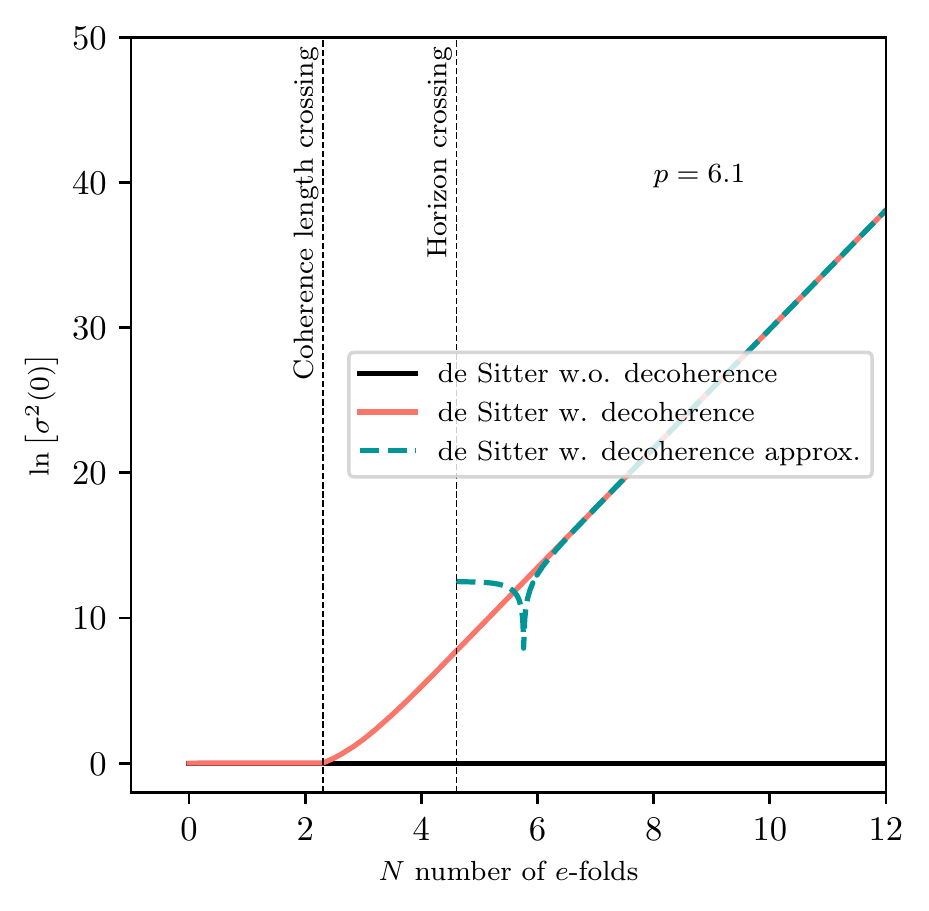}
\caption{$\ln \left[\sigma^2 (0) \right]$ with (pink) and without
  (black) decoherence for the de Sitter case. The approximated version
  (green dashed) is obtained using \Eq{eq:superHsigzero} for $\sigma^2
  (0)$. The first vertical dashed line shows the time when the mode
  $\bm{k}$ starts to decohere $\ell_{E} a / k = 1$, the second the
  time when the mode $\bm{k}$ exits the Hubble radius. The parameters
  are $\ell_{E} H=0.1 $, $x_{\star} = 1$, $p = 2.1$ (left) or $p = 6.1$
  (right), and $k_{\Gamma}/k = 10$.}
\label{fig:sigzerosquare}
\end{center}
\end{figure*}

\subsubsection{Quantum discord}
\label{subsubsec:discord_decoherence_infl}
The final step is to calculate $\sigma^2(\theta)$ and extract quantum
discord. Given that we already have computed $\sigma^2(0)$, see
\Eq{eq:sigma0:exp:cosmo}, and since \Eq{eq:sigma(theta)} can be
rewritten as
\begin{align}
  \label{eq:siggammatheta}
  \sigma(\theta)=\sqrt{\sigma^2(0)
    +\frac14\left[\left(\gamma_{11}-\gamma_{22}\right)^2+4\gamma_{12}^2\right]
    \sin^2\left(2\theta\right)},
  \end{align}
we see that we only need to estimate the second term, \ie the one
proportional to $\sin^2(2\theta)$. This is easier since no
cancellation occurs in that term. In \App{ap:sec:covmatcaldlegget}, we
find that the dominant contribution comes from $\gamma_{22}$, and that
$ (\gamma_{11}-\gamma_{22})^2+4\gamma_{12}^2 \simeq (-k\eta)^{-8}
[1-({k_\Gamma}/{k})^2B_{11}/2]^2 $, see \Eqs{eq:g11approx},
\eqref{eq:g12approx} and \eqref{eq:g22approx}. This leads to
\begin{align}
\label{eq:finalsigtheta}
\sigma^2(\theta) & =\left[1-\frac12\left(\frac{k_\Gamma}{k}\right)^2B_{11}\right]^2
\frac{ \sin^2(2\theta)}{4\left(-k\eta\right)^8}
 + \mathcal{O} \left[\left(-k\eta\right)^{-6}\right] \, .
 \end{align}
 Recalling that $B_{11}$ does not depend on time, one can see that the
 effect of the environment is only to change the prefactor in
 $\sigma(\theta)$, while it does not affect its time behaviour
 $\sigma(\theta) \propto (k\eta)^{-4}$ on super-scales.

\begin{figure} 
    \centering
    \includegraphics[width=0.6\textwidth]{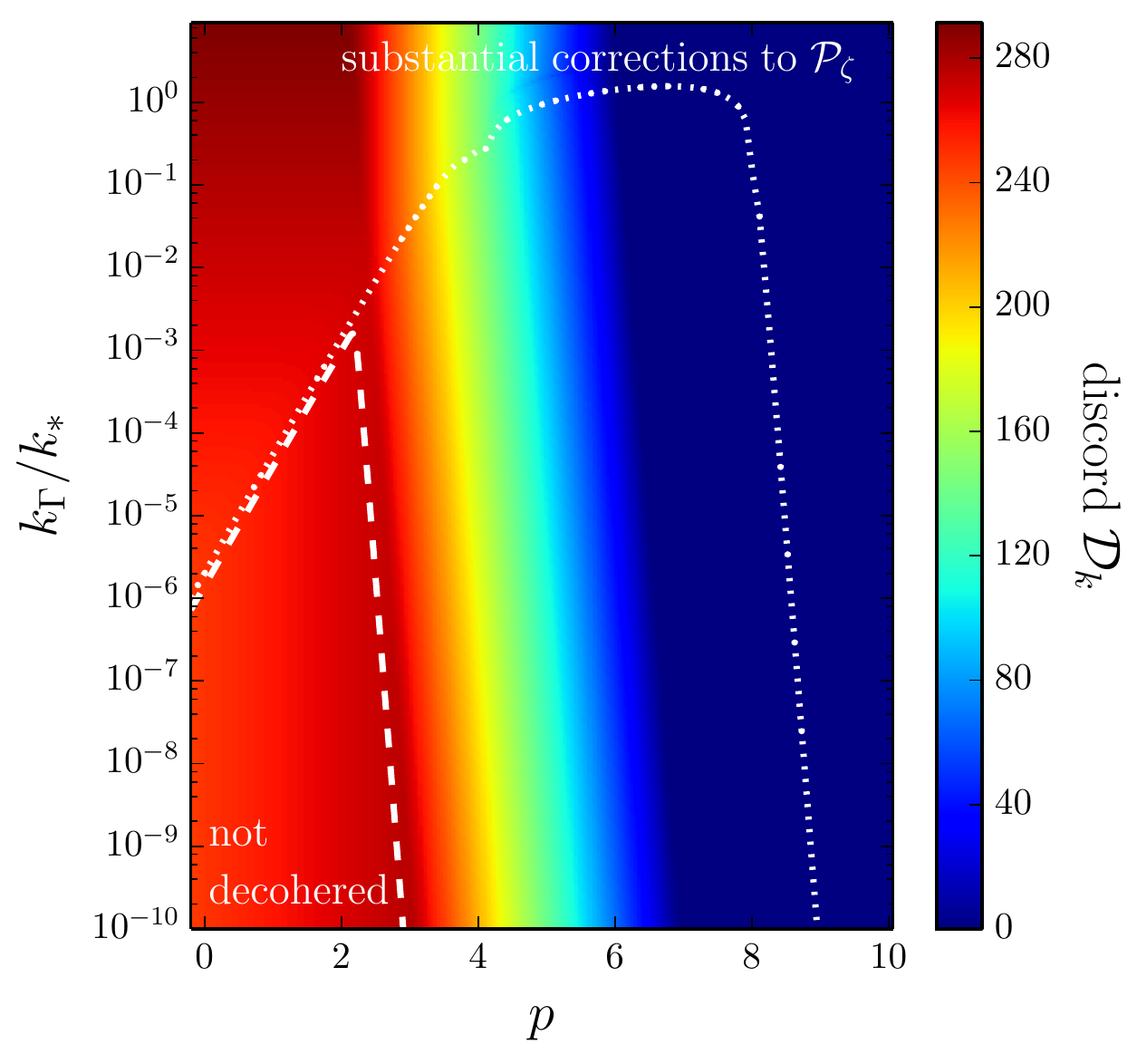}
    \caption{Quantum discord $\mathcal{D}_k$ as a function of $p$ and
      $k_\Gamma/k_*$, for $k=k_*$, $-k\eta=\ee^{-50}$ (corresponding
      to the scales probed in the cosmic microwave background at the
      end of inflation), $\ell_E H=10^{-3}$ and $\theta=-\pi/4$. The
      white dashed line is the contour line of
      $\mathrm{Tr}(\rho^2)=1/2$, and above that line the quantum state
      is decohered. The white dotted line is the contour line of
      $\Delta\calP_\zeta/\calP_\zeta=1$, above which the power
      spectrum is spoilt by environmental effects. These two contours
      essentially correspond to Fig.~6 of \Refa{Martin:2018zbe}; but,
      in Fig.~6 of \Refa{Martin:2018zbe}, the region where there is a
      substantial change of the spectral index is displayed, as
      opposed to the region where there is a substantial change of
      $\calP_\zeta$ in the above figure. This is the reason why, in
      Fig.~6 of \Refa{Martin:2018zbe}, there is a feature at $p=5$,
      for which the corrections are scale-invariant, which does not
      appear in the above figure. One of the main result of the
      present paper is the value of the quantum discord in the
      parameter space $(p,k_\Gamma/k_*)$. Let us also notice that the
      ``not decohered'' region invades the whole figure for
      sufficiently small values of $k_\Gamma$. Here, it looks bounded
      because, due to the logarithmic scale used, $k_\Gamma$ is
      ``cut'' at $k_\Gamma/k_*=10^{-10}$.}
    \label{fig:mapdiscordcosmo}
\end{figure}
\begin{figure*}[t]
\begin{center}
\includegraphics[width=0.49\textwidth,clip=true]{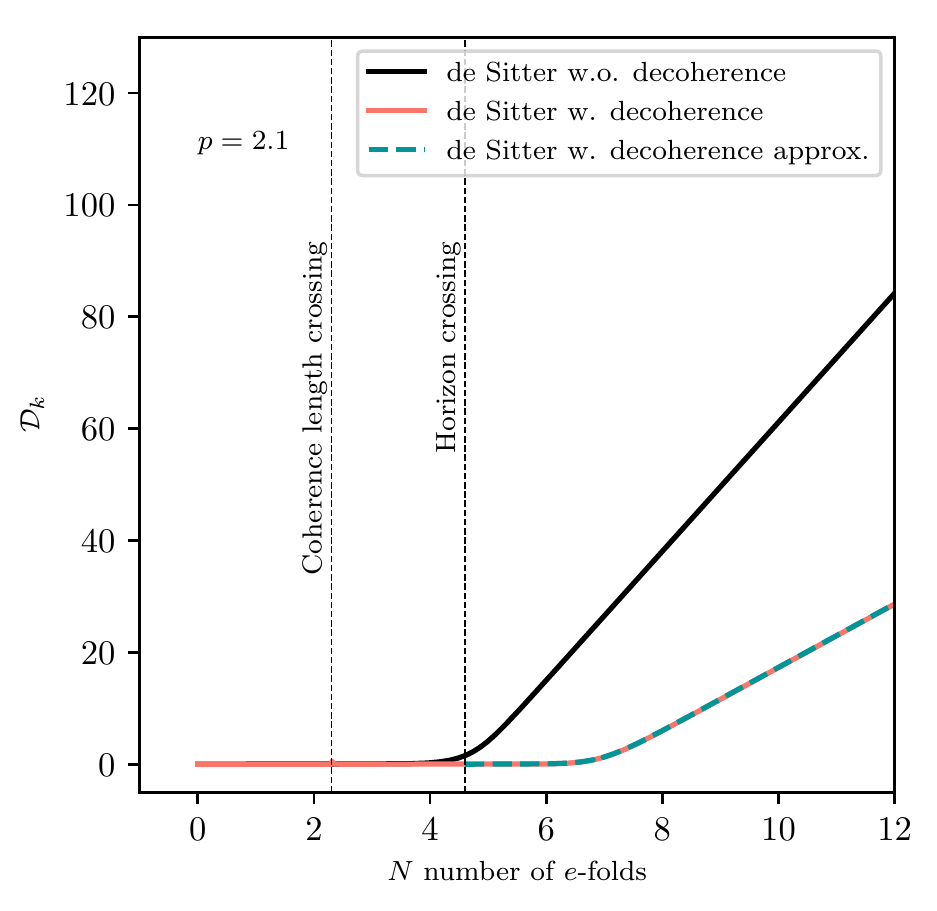}
\includegraphics[width=0.49\textwidth,clip=true]{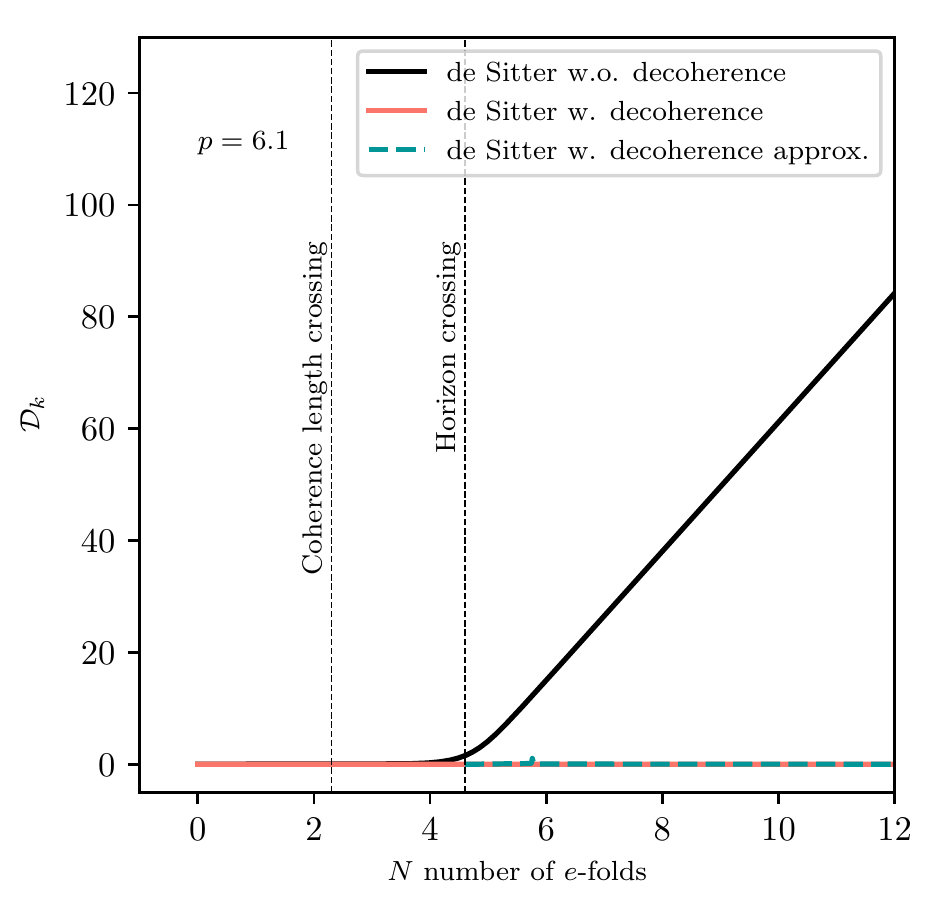}
\caption{Quantum discord $\mathcal{D}_{k}$ with (pink) and without
  (black) decoherence for the de Sitter case. The approximated version
  (green dashed) is obtained using the first order
  approximations~(\ref{eq:expansiong11}), (\ref{eq:expansiong12})
  and~(\ref{eq:expansiong22}) for $\gamma_{ij}$ and the
  expression~(\ref{eq:superHsigzero}) for $\sigma^2 (0)$ to obtain
  $\sigma^2 (\theta)$ in~(\ref{eq:siggammatheta}). The quantity
  $\sigma^2 (0)$ and $\sigma^2 (\theta)$ are then used in the
  expression~(\ref{eq:finaldiscord}) for the discord. The first
  vertical dashed line shows the time when the mode $\bm{k}$ starts to
  decohere $\ell_{E} a / k = 1$, the second the time when the mode
  $\bm{k}$ exits the Hubble radius. The parameters are $\ell_{E}H =0.1
  $, $x_{\star} = 1$, $p = 2.1$ (left) or $p = 6.1$ (right), and
  $k_{\Gamma}/k = 10$.}
\label{fig:plotdiscord}
\end{center}
\end{figure*}

 Let us now consider the ratio $\sigma^2(0)/\sigma(\theta)$ which, as
 explained in \Sec{sec:evoldiscordenvi}, determines the fate of
 quantum discord. If $p<2$, the second term dominates in
 \Eq{eq:sigma0:exp:cosmo}, hence $\sigma^2(0)$ reaches a constant on
 large scales. One thus has $\sigma^2(0)/\sigma(\theta)\propto
 a^{-4}$, which is highly suppressed on super-Hubble scales, and which
 gives rise to $\mathcal{D}_k \propto 4 \log_2[a/a(k)]$. This shows
 that quantum discord is large in that case, and one recovers the
 result obtained in \Eq{eq:discordds}. If $p\geq 2$, the first term
 dominates in \Eq{eq:sigma0:exp:cosmo}, hence $\sigma^2(0)\propto
 a^{p-2} $ on super-Hubble scales. As a consequence,
 $\sigma^2(0)/\sigma(\theta)\propto a^{p-6}$, the time behaviour of
 which depends on whether $p<6$ or $p>6$ as illustrated by \Fig{fig:sigzerosquare} . If $p<6$,
 $\sigma^2(0)/\sigma(\theta)$ decays, and one has $\mathcal{D}_k
 \propto (6-p) \log_2[a/a(k)]$, so the discord remains large. If
 $p>6$, $\sigma^2(0)/\sigma(\theta)$ increases on super-Hubble scales,
 and upon expanding \Eq{eq:finaldiscord} in that regime one finds that
 $\mathcal{D}_k\propto [a/a(k)]^{6-p}$, so it becomes highly
 suppressed. The behaviour of $\mathcal{D}_k$ on each side of the threshold is illustrated by \Fig{fig:plotdiscord}

These considerations can be checked in \Fig{fig:mapdiscordcosmo},
where quantum discord is displayed as a function of $p$ and
$k_\Gamma/k_*$. It confirms that the pivotal value of $p$ from the
point of view of quantum discord is $p=6$: discord remains large on
super-Hubble scales when $p<6$, and is highly suppressed
otherwise. Note however that the above formulas only describe the time
behaviour at large scales, and do not incorporate the constant
prefactors that can otherwise be readily established from combining
the above results. These prefactors depend on both $\ell_E H$ and
$\Gamma$ (through $k_\Gamma$), and they explain why the discord shown
in \Fig{fig:mapdiscordcosmo} does not depend only on $p$ (for
instance, even for $p>6$, one can get a substantial discord by
considering extremely low values of $k_\Gamma$, in agreement with the
fact that in the limit $\Gamma\to 0$, one recovers the results of
\Sec{sec:cosmo:free}). Nonetheless, for reasonable values of the
coupling constant $\Gamma$, the result is mostly determined by the
value of $p$.
 
In \Fig{fig:mapdiscordcosmo}, we have also displayed the region in
parameter space where decoherence does not occur (below the white
dashed line, see \Sec{sec:purity:cosmo}) and the region where
substantial corrections to the power spectrum are obtained (above the
white dotted line, see \Sec{sec:cov:cosmo}). One can see that when the
coupling to the environment is not strong enough to make the system
decohere, quantum discord is always large on super-Hubble scales. This
corresponds to the bottom left corner in \Fig{fig:mapdiscordcosmo}. In
the opposite corner, namely the top-right region in
\Fig{fig:mapdiscordcosmo}, the coupling with the environment is so
strong that it both decoheres the system very efficiently, while it
prevents its quantum discord from growing. An important remark is
that, between these two regimes, there are regions where quantum
discord remains large even though the system decoheres and even when
imposing that the observed power spectrum is unaffected by
environmental effects.
\section{Conclusions}
\label{sec:conclusions}

In this work, we have studied how quantum discord behaves in the
presence of an environment.  When a quantum system couples to
environmental degrees of freedom, the entanglement between the open
system and the environment leads to decoherence of the system, which
is usually associated with the loss of certain quantum features
displayed by the system. Since discord characterises how ``genuinely
quantum'' the correlations between subparts of the system are, it is
naturally expected that decoherence leads to a suppression of
discord. The goal of this work was to study this effect on generic
grounds, since any practical experiment aiming at revealing the
presence of quantum effects is a priori subject to such environmental
limitations.

For simplicity, we have considered the case of a quantum scalar field
with (homogeneous and isotropic) quadratic Hamiltonian, which boils
down to a collection of independent pairs of quantum parametric
oscillators, one for each pair of opposite Fourier modes. We have
shown that, in general, quantum discord depends on the precise way
these systems are partitioned into two subsystems. We have found a
generic parameterisation that describes all possible partitions, which
has allowed us to derive quantum discord for any partitioning. Note
that the way a given physical system should be split into two
subsystems is sometimes obvious: for instance, for two particles
located in each polariser of a Bell's inequality experiment, the two
sub-systems are clearly the two space-like separated
particles. However, the Fourier sub-sector of a quantum field does not
feature such a clear preferred partitioning, and it is therefore
necessary to study how the result depends on the partition in
general.\footnote{An alternative approach is to study correlations
  between the field configuration at two separated positions in real
  space (as opposed to between opposite Fourier modes). In that case,
  a natural partitioning is available (namely the two real-space
  locations), and the relevant bipartite system is mixed even if the
  full quantum field is in a pure state. This is because, when
  considering the field configuration at two locations, one implicitly
  traces over the configuration of the field at every other location,
  to which the bipartite system is a priori entangled. The formalism
  developed in this work is still relevant for that case, since it
  merely boils down to the calculation of quantum discord in a mixed
  Gaussian state. This is the topic of \Refs{Martin:2021xml,
    Martin:2021qkg}.}

In the absence of interactions with an environment, the system is
placed in a Gaussian state known as the two-mode squeezed state, which
can be equivalently described in terms of Bogoliubov coefficients,
squeezing parameters, or covariance matrix. In that case, we recovered
the formula derived in \Refa{Martin:2015qta} for quantum discord,
which we nonetheless extended to any partitioning. 

In the case where an environment is present and couples to the system,
for explicitness, we have assumed that the coupling is linear in the
phase-space variables describing the system, and that it can be cast
in terms of a Lindblad equation (this is the so-called
Caldeira-Leggett model). In this context, the state remains Gaussian
(though not pure anymore), so it can still be described in terms of a
covariance matrix, for which we have derived the modified evolution
equation. We have shown that generalised squeezing parameters can also
still be defined, from their geometrical phase-space interpretation. A
third parameter describing the area of the elliptic contours of the
Wigner function should be added to the usual squeezing amplitude and
squeezing angle, which respectively correspond to the eccentricity and
orientation of these ellipses.  This ``third squeezing parameter''
equals one for a pure state and is larger than one otherwise, and it
is directly related to the inverse purity of the state. We have thus
derived the modified evolution equations for these three ``squeezing
parameters'', which represent an alternative description of the state
with an intuitive geometrical interpretation~\cite{Chen:2021vwq}.

We have then computed quantum discord in this model, both in terms of
the covariance matrix and in terms of the generalised squeezing
parameters. As in the case of pure states, we have found that quantum
discord does not depend on the squeezing angle, but only involves the
squeezing amplitude, the state purity and the partition
parameters. More precisely, in a given partitioning, quantum discord
increases as the length of the semi-minor axis of the phase-space
ellipse decreases, which provides a simple geometrical interpretation
of discord. This implies that discord increases with both the
squeezing amplitude and the state purity. In general, as the time
evolution proceeds, the squeezing amplitude increases (denoting
particle creation) and the state purity decreases (because of
decoherence), so the two effects compete. The details of how quantum
discord is affected by an environment thus depend on the rate at which
these two parameters vary, which has to be discussed on a
model-by-model basis. Those findings are consistent with the work of \Refa{Hollowood:2017bil} where the authors considered a model corresponding to p=5 in our parametrisation. They computed an upper bound on the discord which despite decoherence grows in time, albeit at a reduced rate, in line with the discussion of \Sec{subsubsec:discord_decoherence_infl}.

To go beyond those generic considerations, we have finally applied our
framework to the case of primordial cosmological perturbations. This
case study is of particular interest not only because it provides a
useful illustration of the tools introduced in this work, but also
since the possible presence of quantum correlations in cosmic
structures, and the potential of decoherence to make them
undetectable, is of great importance for our understanding of their
origin. Assuming that the coupling parameter $\Gamma$ between the
system (here cosmological perturbations) and the environment (possibly
heavier fields, smaller-scale degrees of freedom, etc.) grows as a
power of the scale factor $a$ of the universe, $\Gamma\propto a^p$,
whether or not decoherence leads to a suppression of discord (\ie
whether or not the phase-space semi-minor axis increases) crucially
depends on $p$. More precisely, if $p<6$, discord remains large on
large scales, and is strongly suppressed otherwise. Let us also note
that for $p<2$, decoherence cannot proceed without substantially
affecting the observed power spectrum of the cosmological density
field, so in the region of parameter space that is in agreement with
current observations, environmental effects are mostly irrelevant. For
$2<p<6$, there exists a regime where the state decoheres but remains
strongly discordant, while preserving its power spectrum.

Those considerations imply that there is no simple relationship
between decoherence and discord: one can find situations where the
state of the system becomes decohered and non-discordant, where it
becomes decohered but remains discordant, where it remains pure and
discordant, or where it is pure and non discordant (this case does not
appear in cosmology but may be encountered in other contexts, see the
discussion around \Fig{fig:mapdiscordsqueezing}). As a consequence,
although decoherence may affect our ability to reveal the presence of
quantum correlations within a given quantum system, this effect cannot
be simply assessed by considering the amount of decoherence (\ie the
state purity). One alternative criterion may be the quantum discord
discussed in this work, although its relationship with concrete
observables is not clear, in particular in the context of mixed
states~\cite{2010PhRvL.105b0503G}. This is why a next natural step
would be to apply the present framework to investigate violations of
Bell-inequalities in the presence of environmental effects, using the
techniques developed in \Refs{Martin:2016tbd, Martin:2016nrr,
  Choudhury:2016cso, Martin:2017zxs, Ando:2020kdz}. In this way, one
may be able to better understand the relationship between discord,
decoherence, Bell inequalities violation, and maybe other criteria
such as Peres-Horodecki separability~\cite{Simon2000}, in a broad
context.

\begin{acknowledgments}
A.~Micheli is supported by the French National Research Agency under
the Grant No. ANR-20-CE47-0001 associated with the project COSQUA.
We thank Ashley Wilkins for pointing out a typo in the labels of \Fig{fig:wigner} in a previous version of the manuscript. 
\end{acknowledgments}
%                                                                                          
%\newpage                                                                                  

\appendix
\addtocontents{toc}{\protect\setcounter{tocdepth}{1}}

\section{Partitions}
\label{ap:partition}
As explained in \Sec{sec:partitions}, when studying the nature of the
correlations present within a given (classical or quantum) system, one
first has to split this system into two (or more) sub-systems, and
then to analyse how these sub-systems are correlated. This way to
divide the system into several sub-systems is called a ``partition'',
and in this appendix we formally study how partitions can be defined
on generic grounds, and how different partitions are related to each
other.

\subsection{Quantum phase space}
\label{app:description}

In this article, we consider continuous-variable systems, \ie systems
described by Hermitian operators satisfying canonical commutation
relation. It can be, for instance, the positions $\hat{q}_i$ and
momenta $\hat{\pi}_i$ of $n$ particles (with $i=1\cdots n$), with
$[\hat{q}_i,\hat{\pi}_j]=i\delta_{ij}$. This can also correspond to
the Fourier modes of a quantum field, see
\Sec{subsec:description}. The quantum state of the system is an
element of the Hilbert space
\begin{align}
\label{eq:HilbertSpace:Discrete}
{\mathcal E}=\bigotimes_{i=1\, \cdots \, n} {\mathcal E}_i\, ,
\end{align}
where ${\mathcal E}_i$ is the Hilbert space associated to the
$i^{\mathrm{th}}$ particle. It can be described by the vector
\begin{align}
\label{eq:def:VectorCCO}
\hat{R}=(\hat{q}_1,\hat{\pi}_1,\cdots, \hat{q}_i,
\hat{\pi}_i,\cdots , \hat{q}_n,\hat{\pi}_n)^{\mathrm{T}}\, .
\end{align}
In terms of the components of the vector $\hat{R}$, the commutation
relations can be written as\footnote{Hereafter, the indices
  $a,b,c,\cdots $ label the components of the vectors $\hat{R}$, while
  the indices $i,j,k,\cdots$ label the degrees of freedom of the
  system. For instance, for a two-``particle'' system, one has
  $\hat{R}=(\hat{q}_1,\hat{\pi}_1,\hat{q}_2,\hat{\pi}_2)^\mathrm{T}$,
  so $\hat{R}_1=\hat{q}_1$, $\hat{R}_2=\hat{\pi}_1$,
  $\hat{R}_3=\hat{q}_2$ and $\hat{R}_4=\hat{\pi}_2$.}
\begin{align}
\label{eq:commutator:R}
\left[\hat{R}_a,\hat{R}_b\right]=iJ_{ab}^{(n)}\, ,
\end{align}
where $J^{(n)}$ is the $2n \times 2n$ block-diagonal matrix
\begin{align} 
\label{eq:def:SymplecticFormJ}
J^{(n)} = 
\left( 
\begin{array}{ccc}
J^{(1)} & & \\
& \ddots & \\
& &  J^{(1)}
\end{array}
\right)
\quad\text{with}\quad
 J^{(1)}=
  \begin{pmatrix}
    0 & 1 \\
    -1 & 0
  \end{pmatrix}
  \, .
\end{align}

An alternative description of the system is by means of the creation
and annihilation operators $\hat{c}_i$ and $\hat{c}_i^{\dagger}$,
defined by
\begin{align}
\label{eq:defqp}
  \hat{q}_i=\frac{1}{\sqrt{2}}\left(\hat{c}_i+\hat{c}_i^{\dagger}\right),
  \quad
    \hat{\pi}_i=-\frac{i}{\sqrt{2}}\left(\hat{c}_i-\hat{c}_i^{\dagger}\right).
      \end{align}
They can be assembled into the vector 
\begin{align}
\label{eq:def:C:def}
\hat{C}=(\hat{c}_1,\cdots,
\hat{c}_i,
\cdots,\hat{c}_n,\hat{c}_1^\dagger,\cdots,\hat{c}_i^\dagger,
\cdots,\hat{c}_n^\dagger)^\mathrm{T}\, .
\end{align}
Contrary to the vector $\hat{R}$, notice that $\hat{C}$ is not
arranged such that the variables describing the subsystem $i$ directly
follow each other, and the reason for this choice will be made clear
below.\footnote{In practice, one may also consider the vector
  $\hat{\overline{C}}=(\hat{c}_1,\hat{c}_1^\dagger,\cdots,\hat{c}_n,\hat{c}_n^\dagger)^\mathrm{T}$
  which is related to $\hat{C}$ through $\hat{\overline{C}} =
  P^{(n)}\cdot \hat{C}$, where $P^{(n)}$ is a permutation matrix that
  can be readily written down.
\label{footnote:Cbar}
}

The relation between $\hat{R}$ and $\hat{C}$ is linear and can thus be
written in matricial form as $\hat{R}=M^{(n)} \cdot \hat{C}$, where
$M^{(n)}$ is an unitary matrix that can be obtained from
\Eq{eq:defqp}.\footnote{In general, the matrix $M^{(n)}$ can be
  computed as follows. One first writes
  $\hat{R}=\overline{M}^{(n)}\hat{\overline{C}}$, where
  $\hat{\overline{C}}$ was introduced in footnote~\ref{footnote:Cbar}
  and where $\overline{M}^{(n)}$ is a simple block-diagonal matrix :
\begin{align}
\overline{M}^{(n)}=\left(
\begin{array}{ccc}
\overline{M}^{(1)} & & \\
& \ddots & \\
& & \overline{M}^{(1)}
\end{array}
\right)\, ,
\quad \text{where} \quad
\overline{M}^{(1)}  = \frac{1}{\sqrt{2}}\left(
\begin{array}{cc}
1 & 1\\
-i & i
\end{array}
\right)\, .
\end{align}
Since $\overline{M}^{(n)} = M^{(n)}\cdot P^{(n)}$, one has $M^{(n)} =
\overline{M}^{(n)} P^{(n),\mathrm{T}}$ since permutation matrices are
orthogonal. This allows one to compute $M^{(n)}$ from the above
expression for $\overline{M}^{(n)}$. This also allows one to show that
$M^{(n)}$ is unitary: one can check that $\overline{M}^{(1)}
\overline{M}^{(1),\dagger} = \setI_2$ and so $\overline{M}^{(n)}\cdot
\overline{M}^{(n),\dagger} = \setI_{2n}$, which in turns implies that
${M}^{(n)}\cdot {M}^{(n),\dagger} = \setI_{2n}$, using the fact that
$P^{(n)}$ is orthogonal.\label{footnote:M:unitary}} For instance, with
$n=2$, one has
\begin{align}
  \label{eq:defmatrixM}
  M^{(2)}=\frac{1}{\sqrt{2}}
  \begin{pmatrix}
    1 & 0 & 1 & 0 \\
    -i & 0 & i & 0 \\
    0 & 1 & 0 & 1 \\
    0 & -i & 0 & i
  \end{pmatrix}.
  \end{align}
The commutation relations can be expressed as 
\begin{align}
\label{eq:commutator:C}
\left[\hat{C}_a,\hat{C}_b\right]=\Omega_{ab}^{(n)}\, ,
\end{align}
with $\Omega^{(n)}=i M^{(n),-1}\cdot J^{(n)}\cdot M^{(n),-1,\mathrm{T}}$, \ie
\begin{align}
  \Omega^{(n)}=
  \begin{pmatrix}
    0 & \setI_n\\
    -\setI_n & 0 
  \end{pmatrix},
  \end{align}
where $\setI_n$ is the identity matrix of size $n$.

\subsection{Partitioning}
\label{subsec:defpartition}

For $n\geq 2$, the degrees of freedom can always be split into two
subsets $A$ and $B$. This defines a partition and allows us to see the
whole system as a bipartite system. An important point is that one can
define several partitions for the same system. As an introductory (and
elementary) example, let us consider the case $n=4$ where
$\hat{R}=(\hat{q}_1,\hat{\pi}_1,\hat{q}_2,\hat{\pi}_2,\hat{q}_3,
\hat{\pi}_3,\hat{q}_4,\hat{\pi}_4)^{\mathrm{T}}$. For instance, one
can choose the subsystem $A$ to be made of the first two degrees of
freedom and to be described by
$\hat{R}^{(A)}=(\hat{q}_1,\hat{\pi}_1,\hat{q}_2,\hat{\pi}_2)^{{\mathrm{
      T}}}$, and the subsystem $B$ to contain the third and fourth
degrees of freedom,
$\hat{R}^{(B)}=(\hat{q}_3,\hat{\pi}_3,\hat{q}_4,\hat{\pi}_4)^{{\mathrm{
      T}}}$.  Then, the vector $\hat{R}$ can be written as
$\hat{R}=(\hat{R}^{(A)},\hat{R}^{(B)})^{{\mathrm T}}$. This definition
of $\hat{R}$, namely the way we order its components, is, implicitly,
a definition of a partition. Obviously, other partitions are possible,
for instance the one defined by
$\hat{R}^{(A)}{}'=(\hat{q}_1,\hat{\pi}_1,\hat{q}_3,\hat{\pi}_3)^{{\mathrm{
      T}}}$ and
$\hat{R}^{(B)}{}'=(\hat{q}_2,\hat{\pi}_2,\hat{q}_4,\hat{\pi}_4)^{{\mathrm{
      T}}}$ with
$\hat{R}'=(\hat{R}^{(A)}{}',\hat{R}^{(B)}{}')^{{\mathrm T}}$.

More generally, changing the partition can be viewed as performing a
canonical transformation on the system (\ie a transformation that
preserves the commutator structure). A linear canonical transformation
is a transformation $\hat{R}\rightarrow \hat{R}' = T \hat{R}$, where
$T$ is a real matrix since $\hat{R}$ and $\hat{R}'$ are Hermitian,
which preserves the commutators, \ie such that
$[\hat{R}'_a,\hat{R}'_b] = [\hat{R}_a,\hat{R}_b]$ (hereafter we drop
the index $n$ for notational convenience).  This leads to the
condition
\begin{align}
\label{eq:def:symplectic}
TJ T^{{\mathrm{T}}}=J \, ,
\end{align}
which defines the group of symplectic matrices
$T$~\cite{goldstein2002classical, Grain:2019vnq,Colas:2021llj}. In
particular, any symplectic matrix has determinant $1$
\cite{goldstein2002classical},
\begin{equation}
\label{eq:det1canonicaltransfo}
    \det T  = 1 \, .
\end{equation}
In the simple example mentioned above, one can check that the
transformation matrix $T$ is indeed symplectic.  Let us note that
canonical transformations can also be defined at the level of the
vectors $\hat{C}$, since $\hat{R}\rightarrow T \hat{R}$ leads to
$\hat{C}\rightarrow S \hat{C}$ with
\begin{align}
\label{eq:link:S:T}
S=M^{-1} T M=M^\dagger TM
\end{align}
(where, in the last equation, we have used that $M $ is unitary, see
footnote~\ref{footnote:M:unitary}). Using the definition of $\Omega$
given below \Eq{eq:commutator:C}, \Eqs{eq:def:symplectic}
and~\eqref{eq:link:S:T} lead to the condition
\begin{align}
\label{eq:def:symplectic:c}
S \Omega S^\mathrm{T} = \Omega\, ,
\end{align}
which also implies that
\begin{equation}
\label{eq:det1canonicaltransfo:c}
    \det S  = 1 \, .
\end{equation}
Another useful property for the matrix $S$ comes from the ordering of
the entries of the vector $\hat{C}$ in \Eq{eq:def:C:def}, which leads
to $\hat{C}^\dagger = A\cdot \hat{C}$, where $\hat{C}^\dagger$ is
defined as $(\hat{C}^\dagger)_a=(\hat{C}_a)^\dagger$ and where
$A=\left(\begin{array}{cc} 0 & \setI_n \\ \setI_n &
  0\end{array}\right)$. Since $\hat{R}^\dagger=\hat{R}$, the adjoint
  of the relation $\hat{R}=M \hat{C}$ gives rise to $M^* A =
  M$. Evaluating the complex conjugate of $S=M^{-1} T M$, this leads
  to
\begin{align}
\label{eq:Sstar}
S^* = A S A\, ,
\end{align}
where we have used that $A^2=\setI_n$ and that $T^*=T$ (since
$\hat{R}$ and $\hat{R}'$ are both Hermitian).

It is worth pointing out that there are canonical transformations that
do not mix the two subsystems (they are called ``local''
transformations) and hence do not change the partition. A simple
example is
$\hat{R}'=(\hat{q}_2,\hat{\pi}_2,\hat{q}_1,\hat{\pi}_1,\hat{q}_3,
\hat{\pi}_3,\hat{q}_4,\hat{\pi}_4)^\mathrm{T}$ where we have simply
flipped the ordering of the two first degrees of freedom. As a less
trivial example, let us define
$\hat{Q}_{ij}^{\pm}=(\hat{q}_i\pm\hat{q}_j)/\sqrt{2}$ and
$\hat{\Pi}_{ij}^{\pm}=(\hat{\pi}_i\pm\hat{\pi}_j)/\sqrt{2}$ so that
$[\hat{Q}_{ij}^{\pm},\Pi_{ij}^{\pm}]=i$, thus ensuring that the
corresponding transformation can be described by a symplectic
matrix. Clearly,
$\hat{R}'=(\hat{Q}_{12}^+,\hat{\Pi}_{12}^+,\hat{Q}_{12}^-,\hat{\Pi}_{12}^-,
\hat{q}_3,\hat{\pi}_3,\hat{q}_4,\hat{\pi}_4)^{{\mathrm T}}$
corresponds to the same partition as
$\hat{R}=(\hat{q}_1,\hat{\pi}_1,\hat{q}_2,\hat{\pi}_2,\hat{q}_3,
\hat{\pi}_3,\hat{q}_4,\hat{\pi}_4)^{\mathrm{T}}$ since we still have
$1$ and $2$ in subsystem $A$ and $3$ and $4$ in subsystem
$B$. Therefore, partition changes are described by only a subclass of
symplectic matrices (namely those that are not $n$$\times$$n$-block
diagonal).

In order for two parameterisations of the system to share the same
vacuum state, let us first impose that $S$ does not mix creation and
annihilation operators. This implies that $S$ is block diagonal (which
is the reason why the ordering made in \Eq{eq:def:C:def} was indeed
convenient). The condition~\eqref{eq:Sstar} then imposes that the two
blocks are complex conjugate to each other, so $S$ can be written as
\begin{align} 
\label{eq:S:RI:to:12}
  S
  =
  \begin{pmatrix}
    s^{(n)} &0 \\
    0 & s^{(n)}{}^*
  \end{pmatrix}\, .
\end{align}
The symplectic condition~\eqref{eq:def:symplectic:c} leads to
$s^{(n)}{}^\dagger s^{(n)}=\setI_n$, \ie the matrices $s^{(n)}$ belong
to the unitary group $U(n)$. This discussion shows that the space of
partitions is essentially the group $U(n)$, so any parameterisation of
that group, which has dimension $n^2$, leads to a parameterisation of
all possible partitions, by means of the above formulas. For instance,
with $n=2$, matrices of $U(2)$ can be written in the form
\begin{align}
  s^{(2)}=
  \begin{pmatrix}
    e^{i\alpha} \cos \theta & -e^{i\delta} \sin \theta \\
    e^{i\beta}\sin \theta & e^{i(\delta+\beta-\alpha)}\cos \theta
  \end{pmatrix},
\end{align}
where $\alpha$, $\beta$, $\delta $ and $\theta$ are four arbitrary
real numbers, which thus parameterise all possible partitions. The
matrix $T$ can be also written in terms of these parameters by making
use of \Eq{eq:link:S:T} together with \Eqs{eq:defmatrixM}
and~\eqref{eq:S:RI:to:12}, leading to
\begin{align}
\label{eq:T:n_eq_2:gen}
  T&  =
  \begin{pmatrix}
    \cos \alpha \cos \theta & -\sin \alpha \cos \theta
    & -\cos \delta \sin \theta & \sin \delta \sin \theta \\
    \sin \alpha \cos \theta & \cos \alpha \cos \theta
    & -\sin \delta \sin \theta & -\cos \delta \sin \theta \\
    \cos \beta \sin \theta & -\sin \beta \sin \theta &
    \cos(\alpha-\beta-\delta)\cos \theta &
    \sin(\alpha -\beta-\delta)\cos \theta \\
    \sin \beta \sin \theta & \cos \beta \sin \theta &
    -\sin(\alpha-\beta-\delta)\cos \theta &
    \cos(\alpha-\beta-\delta) \cos \theta
  \end{pmatrix}.
\end{align}
As a consistency check, one can verify that such a matrix is indeed
symplectic, namely that it satisfies \Eq{eq:def:symplectic}. However,
note that the group of real $4$$\times$$4$ symplectic matrices,
usually denoted $\mathrm{Sp}(4,\setR)$, is of dimension
10~\cite{Colas:2021llj}. Therefore, partition changes, that are
described by 4 parameters, only correspond to a subgroup (isomorphic
to $\mathrm{U}(2)$, and corresponding to the ``rotation'' generators
of table 2 in \Refa{Colas:2021llj}) of the symplectic group.

Note that, in agreement with the group structure of $U(n)$, changes of
partitions can be composed according to
\begin{align}
T^{1/2\to 1'/2'} = T^{1/2\to 1''/2''} \cdot T^{1''/2''\to 1'/2'} \, ,
\end{align}
and that 
\begin{align}
T^{1/2\to 1'/2'} = \left(T^{1'/2'\to 1/2}\right)^{-1} \, ,
\end{align}
with similar expressions for $S^{1/2\to 1'/2'} $.
\section{Covariance matrix in arbitrary partition}
\label{app:cov}
The covariance matrix $\gamma$ in a given phase-space parameterisation
$\hat{R}$ is defined by \Eq{eq:def:covariance}.  Since $\hat{R}_a
\hat{R}_b =(\{\hat{R}_a,\hat{R}_b\} + [\hat{R}_a,\hat{R}_b])/2 $,
\Eqs{eq:commutator:R} and~\eqref{eq:def:covariance}, give rise to
\begin{align}
  \langle \hat{R}_a\hat{R}_b\rangle=\frac12 \gamma_{ab}
  +\frac{i}{2}J_{ab}^{(n)}\, ,
\end{align} 
where the matrix $J^{(n)}$ has been defined in
\Eq{eq:def:SymplecticFormJ}.  Furthermore, since $\hat{R}_a$ and
$\hat{R}_b$ are Hermitian, $\{\hat{R}_a,\hat{R}_b\}$ is also
Hermitian, and \Eq{eq:def:covariance} implies that $\gamma$ is a real
symmetric matrix. Note that the correlators of the ladder operators
introduced in \Eq{eq:def:C:def}, and arranged into the vector
$\hat{C}=M^{(n),-1} \hat{R}$, can also be expressed in terms of
the covariance matrix:
\begin{align}
  \left\langle \left\lbrace \hat{C}_a ,
  \hat{C}_b \right\rbrace\right\rangle
  = M^{(n),-1}_{ac} \gamma_{cd} M^{(n),*}_{db} 
\end{align}
where we have used that $M^{(n)}$ is unitary, see
footnote~\ref{footnote:M:unitary}.

In the Fourier subspaces of a real scalar field, the covariance matrix
is given by \Eq{eq:gamma:RI} in the $\mathrm{R}/\mathrm{I}$
partition. Making use of \Eq{eq:matrixT} and~\eqref{eq:transcov}, the
covariance matrix can then be computed in all partitions, and one
finds
\begin{align}
  \gamma=
  \begin{pmatrix}
    \gamma_A & \gamma_C \\
    \gamma_C & \gamma_B
  \end{pmatrix},
\end{align}
with
\begin{align}
  \label{eq:def:gammaA}
  \gamma_A&=
  \begin{pmatrix}
    \displaystyle
    \gamma_{11} \cos^2 \theta+\gamma_{22}\sin^2 \theta & 
    \displaystyle
    \gamma_{12} \cos(2\theta) \\
    \displaystyle
    \gamma_{12}\cos(2\theta) &
    \displaystyle
    \gamma_{22} \cos^2 \theta+\gamma_{11}\sin^2 \theta
  \end{pmatrix},
  \\
    \label{eq:def:gammaB}
  \gamma_B&=
  \begin{pmatrix}
    \displaystyle
\gamma_B\vert_{11} & \gamma_B\vert_{12} \\
     \displaystyle
     \gamma_B\vert_{21} & \gamma_B\vert_{22}
     \end{pmatrix},
  \\
  \label{eq:def:gammaC}
  \gamma_C&=\left(
  \begin{array}{cccc}
    \displaystyle
    \frac12(\gamma_{11}-\gamma_{22})\sin^2(2\theta)+\frac12 \gamma_{12}\sin(4\theta) & & &
    \displaystyle
    -\frac14(\gamma_{11}-\gamma_{22})\sin(4\theta)+\gamma_{12}\sin^2(2\theta) \\
    \displaystyle
    -\frac14(\gamma_{11}-\gamma_{22})\sin(4\theta)+\gamma_{12}\sin^2(2\theta)& & &
    \displaystyle
    -\frac12(\gamma_{11}-\gamma_{22})\sin^2(2\theta)-\frac12 \gamma_{12}\sin(4\theta)
  \end{array}\right),
\end{align}
where the components of $\gamma_B$ are given by
\begin{align}
\gamma_B\vert_{11}&=
\frac{1}{2}\gamma_{11}+\frac{1}{2}\gamma_{22}+\frac12(\gamma_{11}-\gamma_{22})
\cos(2\theta)\cos(4\theta)-\gamma_{12}\cos(2\theta)\sin(4\theta),
\\
\gamma_B\vert_{12}&=\gamma_B\vert_{21}=
\gamma_{12}\cos(2\theta)\cos(4\theta)
+\frac12(\gamma_{11}-\gamma_{22})\cos(2\theta)\sin(4\theta),
\\
\gamma_B\vert_{22}&=
\frac{1}{2}\gamma_{11}+\frac{1}{2}\gamma_{22}
-\frac12(\gamma_{11}-\gamma_{22})\cos(2\theta)\cos(4\theta)
+\gamma_{12}\cos(2\theta)\sin(4\theta).  
\end{align}

For instance, recalling that the $\kmk$ partition is reached by
setting $\theta=\pi/4$, the correlation functions in the ``${\bm
  k}$''-sector are given by
\begin{align}
\label{eq:corr:qk2}
k\left\langle \hat{q}_{\bm{k}}^2\right\rangle &=
\frac12 \gamma_A\vert _{11}+\frac{i}{2}J_{11}^{(2)}
=\frac{\gamma_{11}+\gamma_{22}}{4}
\, ,\\
\frac{\left\langle  \hat{\pi}_{\bm{k}}^2\right\rangle}{k}&=
\frac12 \gamma_A\vert _{22}+\frac{i}{2}
J_{22}^{(2)}=\frac{\gamma_{11}+\gamma_{22}}{4}
\, \\
\left \langle \hat{q}_{\bm k}\hat{\pi}_{\bm k}\right\rangle&=
\gamma _A\vert_{12}+\frac{i}{2}J_{12}^{(2)}=\frac{i}{2}, \quad 
\left \langle \hat{\pi}_{\bm k}\hat{q}_{\bm k}\right\rangle=
\gamma _A\vert_{21}+\frac{i}{2}J_{21}^{(2)}=-\frac{i}{2},
\end{align}
and we have the same results in the ``$-{\bm k}$''-sector since, for
$\theta=-\pi/4$, one has $\gamma_A=\gamma_B=(\gamma_{11}+\gamma_{22})
\mathrm{diag}(1,1)/2$. The correlation functions mixing ${\bm k}$ and
$-{\bm k}$ modes depend on the matrix $\gamma_C$, and are given by
\begin{align}
k\left\langle \hat{q}_{\bm k}\hat{q}_{-{\bm k}}
\right \rangle &= \frac{1}{2}\gamma_C\vert_{11}+\frac{i}{2}
J_{13}^{(2)}=\frac{1}{4}(\gamma_{11}-\gamma_{22})
\, 
\\
\frac{1}{k}\left\langle \hat{\pi}_{\bm k}\hat{\pi}_{-{\bm k}}
\right \rangle &= \frac{1}{2}\gamma_C\vert_{22}+\frac{i}{2}
J_{24}^{(2)}=-\frac{1}{4}(\gamma_{11}-\gamma_{22})
\,
\\
\left\langle \hat{q}_{\bm k}\hat{\pi}_{-{\bm k}}
\right \rangle &= \frac{1}{2}\gamma_C\vert_{12}+\frac{i}{2}
J_{14}^{(2)}=\frac12 \gamma_{12}, 
\quad
\left\langle \hat{\pi}_{-{\bm k}}\hat{q}_{{\bm k}}
\right \rangle = \frac{1}{2}\gamma_C\vert_{21}+\frac{i}{2}
J_{41}^{(2)}=\frac12\gamma_{12}, 
\\
\left\langle \hat{q}_{-{\bm k}}\hat{\pi}_{{\bm k}}
\right \rangle &= \frac{1}{2}\gamma_C\vert_{12}+\frac{i}{2}
J_{32}^{(2)}=\frac12 \gamma_{12}, 
\quad
\left\langle \hat{\pi}_{{\bm k}}\hat{q}_{-{\bm k}}
\right \rangle = \frac{1}{2}\gamma_C\vert_{21}+\frac{i}{2}
J_{23}^{(2)}=\frac12\gamma_{12}
\, .
\label{eq:corr:qmkpik}
\end{align}
\section{Quantum discord for Gaussian homogeneous states}
\label{ap:discord}
In this appendix we explain how the quantum discord can be used as a
tool to assess the presence of quantum correlations between two
subsystems. In \Secs{ap:subsec:classCorr} and
\ref{ap:subsec:quantCorr}, we first present a brief introduction to
the main ideas behind quantum discord and give its mathematical
definition. In \Secs{ap:subsec:calcI} and \ref{ap:subsec:calcJ}, we
then derive the expression of quantum discord for Gaussian homogeneous
states we use in the main text, \ie \Eq{eq:finaldiscord} .

\subsection{Classical correlations}
\label{ap:subsec:classCorr}
Let us consider two systems $A$ and $B$, and denote by $\{a_i\}$ and
$\{ b_j\}$ their possible respective configurations. The probability
to find the system $A$ in the configuration $a_i$ is noted $p_i$, and
similarly for $p_j$. How uncertain the state of system $A$ is can be
characterised by the von Neumann entropy which is defined by the
following expression
\begin{align}
S(A) = -\sum_{i} p_i \log_2(p_i)\, .
\end{align}
One can check that $S(A)=0$ corresponds indeed to the situation where
all $a_i$ vanish but one (so the state of A is certain), and that, in
general, $S(A)\geq 0$. A similar expression for $S(B)$ can be
introduced, and this can also be done for the joint system
\begin{align}
S(A,B) = -\sum_{i,j} p_{ij} \log_2(p_{ij}),
\end{align}
where $p_{ij}$ denotes the joint probability to find the system $A$ in
configuration $a_i$ and the system $B$ in configuration $b_j$. Then,
the mutual information between $A$ and $B$ can be measured by
\begin{align}
\label{eq:calI:def}
\mathcal{I}(A,B) = S(A)+S(B)-S(A,B)\, .
\end{align}
The fact that $\mathcal{I}(A,B)$ measures the presence of correlations
between $A$ and $B$ can be seen by noting that if $A$ and $B$ are
uncorrelated, then the mutual information vanishes. Indeed, if $p_{ij}
= p_i p_j$, then $\mathcal{I} = -\sum_i p_i \log_2(p_i)-\sum_j p_j
\log_2(p_j) + \sum_{i,j} p_i p_j [\log_2(p_i) + \log_2(p_j)]=0$, where
we have used that $\sum_i p_i = \sum_j p_j = 1$. More generally,
$p_{ij}$ can be expressed by means of Baye's theorem
\begin{align}
\label{eq:Bayes}
p_{i,j} = p_j p_{i\vert j}\, ,
\end{align}
where $p_{i\vert j}$ denotes the conditional probability to find $A$
in configuration $a_i$ knowing that $B$ is in configuration
$b_j$. Plugging \Eq{eq:Bayes} into the definition~\eqref{eq:calI:def},
one obtains $\mathcal{I} = -\sum_i p_i \log_2(p_i)-\sum_j p_j
\log_2(p_j) + \sum_{i,j} p_jp_{i\vert j} [\log_2(p_j)+\log_2(p_{i\vert
    j})] = -\sum_i p_i \log_2(p_i) + \sum_{i,j} p_jp_{i\vert j}
\log_2(p_{i\vert j})$ where we have used that $\sum_i p_{i\vert
  j}=1$. This justifies the introduction of the following quantity
\begin{align}
\label{eq:conditional:entropy}
S(A\vert B) = - \sum_{j }p_j  \sum_i p_{i\vert j} \log_2(p_{i\vert j})\, ,
\end{align}
which stands for the conditional entropy contained in $A$ after
finding the system $B$ in configuration $b_j$, averaged over all
possible configurations for $B$. The above calculation thus suggests
an alternative expression for mutual information, namely
\begin{align}
\label{eq:def:J}
\mathcal{J}(A,B) = S(A)-S(A\vert B)\, .
\end{align}
It also shows that, in classical systems,
$\mathcal{I}=\mathcal{J}$.
\subsection{Quantum correlations}
\label{ap:subsec:quantCorr}
We now want to reproduce the above discussion in a quantum-mechanical
context. This implies to construct quantum analogues of $\mathcal{I}$
and $\mathcal{J}$. The full quantum system can be described by its
density matrix $\hat{\rho}_{A,B}$, where information about the
subsystem $A$ is obtained by tracing over the degrees of freedom
contained in $B$, \ie
\begin{align}
\hat{\rho}_A = \mathrm{Tr}_B \left(\hat{\rho}_{A,B}\right)\, ,
\end{align}
and similarly for $\hat{\rho}_B$. The von Neumann entropy can then be
written as
\begin{align}
  S(A) = - \mathrm{Tr} \left[\hat{\rho}_A
    \log_2\left(\hat{\rho}_A\right)\right]\, ,
\end{align}
with similar expressions for $S(B)$ and $S(A,B)$. This allows us to
evaluate $\mathcal{I}(A,B)$ with \Eq{eq:calI:def}. In order to
evaluate $\mathcal{J}(A,B)$, one needs to introduce the notion of
entropy after performing a (quantum) measurement, $S(A\vert B)$. To
this end, let us introduce $\hat{\Pi}_j$, a complete set of projectors
on subsystem $B$, and denote by $\vert b_j\rangle$ the quantum states
on which they project. One thus has $\hat{\Pi}_j =
\hat{\setI}_A\otimes \vert b_j \rangle \langle b_j \vert$. It is
important to notice that such complete sets of projectors
$\hat{\Pi}_j$ (or equivalently, of states $\vert b_j \rangle$) are not
unique. For instance, for a spin particle, one can consider $\vert +
\rangle_{\vec{e}}$ and $\vert - \rangle_{\vec{e}}$ along any unit
vector $\vec{e}$.  We will come back to this point below. The
probability to find $B$ in the state $b_j$ is given by $p_j =
\mathrm{Tr}(\hat{\rho} \hat{\Pi}_j)$, and a measurement of $B$ that
returns the result $b_j$ projects the state into $\hat{\rho} \to
\hat{\Pi}_j\hat{\rho} \hat{\Pi}_j/p_j $. This leads us to introduce
\begin{align}
  \hat{\rho}_{A\vert \hat{\Pi}_i} = \mathrm{Tr}_B
  \left(\frac{ \hat{\Pi}_j\hat{\rho} \hat{\Pi}_j}{p_j}\right),
\end{align}
which describes the state of $A$ after measuring $B$ and finding $b_j$
as a result of the measurement, and in terms of which the conditional
entropy is given by
\begin{align}
S(A\vert B) = \sum_j p_j S\left(\hat{\rho}_{A\vert \hat{\Pi}_i} \right) .
\end{align}
This is the analogue of \Eq{eq:conditional:entropy}, and these
formulas then allow one to evaluate $\mathcal{J}(A,B)$ with
\Eq{eq:def:J}. Quantum discord is finally defined as
\begin{align}
\label{eq:discord:def}
  \delta(A,B)= \min_{\{\hat{\Pi}_i\}}\left[{\cal I}(A,B)-{\cal J}(A,B)\right]\, .
\end{align}
In this expression, we have minimised over all possible complete sets of
projectors. This ensures that a non-vanishing discord signals
genuine quantum correlations, for any projection basis.
\subsection{Mutual information $\mathcal{I}$}
\label{ap:subsec:calcI}
For a Gaussian state, which is entirely characterised by its
covariance matrix $\gamma$, the von Neumann entropy is given
by~\cite{1999quant.ph.12067H}
\begin{align}
  S(\hat{\rho})=\sum_{i=1}^nf(\sigma_i),
\end{align}
where the function $f(x)$ is defined for $x\geq 1$ by \Eq{eq:deff} in
the text and $\sigma_i$ are the symplectic eigenvalues of the
covariance matrix, that is to say the quantities $\sigma_i$ such that
$\mathrm{Sp}(J^{(n)}\gamma)=\{i\sigma_1,-i\sigma_1,\cdots,
i\sigma_n,-i\sigma_n\}$.

For a partition of the kind~\eqref{eq:matrixT}, the full covariance
matrix can be obtained from~\eqref{eq:transcov}
and~\Eq{eq:gamma:RI}. One can show that a Gaussian state remains
Gaussian after partial tracing over, with a covariance matrix given by
the relevant entries of the full covariance matrix. In other words,
$\hat{\rho}_A$ is still a Gaussian state with covariance matrix
$\gamma_A$ given in \Eq{eq:def:gammaA}, and $\hat{\rho}_B$ is still a
Gaussian state with covariance matrix $\gamma_B$ given in
\Eq{eq:def:gammaB}. This leads to
$\mathrm{Sp}(J^{(1)}\gamma_A)=\mathrm{Sp}(J^{(1)}\gamma_B)=\{i\sigma(\theta),
-i\sigma(\theta)\}$ where $\sigma(\theta)$ has been defined in the
text \Eq{eq:sigma(theta)}.

Moreover, for the full system, one obtains
$\mathrm{Sp}(J^{(2)}\gamma)=\{i\sigma(0),-i\sigma(0),i\sigma(0),-i\sigma(0)\}$
with $\sigma^2(0)=\gamma_{11}\gamma_{22}-\gamma_{12}^2$. Combining the
above results, one obtains
\begin{align}
  \label{eq:finalI}
  {\cal I}=2f\left[\sigma(\theta)\right]-2f\left[\sigma(0)\right].
  \end{align}
 Note that because of Heisenberg's uncertainty principle,
 $\sigma(0)\geq 1$ since $\sigma^2(0)$ is the determinant of the
 covariance matrix written in the $\mathrm{R}$ or $\mathrm{I}$
 subspace, see \Eq{eq:gamma:RI}. This also guarantees that
 $\sigma(\theta)\geq 1$ since \Eq{eq:sigma(theta)} implies that
 $\sigma(\theta)\geq \sigma^2(0)$. This ensures that the function $f$
 can be safely applied to $\sigma(0)$ and $\sigma(\theta)$.
\subsection{Mutual information $\mathcal{J}$}
\label{ap:subsec:calcJ}
The calculation of $\mathcal{J}$ is less straightforward and we will
follow the approach presented in \Refa{2010PhRvL.105c0501A}. It relies
on noting that the mutual information $\mathcal{J}$ (like
$\mathcal{I}$) is invariant under local canonical transformations,
which means that correlation measures do not depend on the way each
subsystem is parameterised internally. As a consequence, it is
convenient to first perform local canonical transformations that bring
the covariance matrix into the simple form
\begin{align}
  \label{eq:covform}
  \gamma=
  \begin{pmatrix}
    \mathfrak{A} & \mathfrak{C} \\
    \mathfrak{C}^\mathrm{T} & \mathfrak{B}
  \end{pmatrix},
  \end{align}
with $\mathfrak{A}=\mathfrak{a}\setI_2$,
$\mathfrak{B}=\mathfrak{b}\setI_2$ and
$\mathfrak{C}=\mathrm{diag}\left(\mathfrak{c},\mathfrak{d}\right)$.

This can be achieved by performing two transformations. The first
local transformation is realised by
\begin{align}
  T=
  \begin{pmatrix}
    T_A & 0 \\
    0 & T_B
  \end{pmatrix},
  \end{align}
with $T_A=\sqrt{\sigma(\theta)}\setI_2 \gamma_A^{-1/2}$ and
$T_B=\sqrt{\sigma(\theta)}\setI_2 \gamma_B^{-1/2}$.  One can check
that $T_A$ and $T_B$, hence $T$, satisfy \Eq{eq:def:symplectic}, so
they generate symplectic transformations. Making use of
\Eq{eq:transcov}, the covariance matrix becomes
\begin{align}
  \label{eq:covgammap}
  \gamma'=
\left(
\begin{array}{cc}
\sigma(\theta)\setI_2 & \gamma_C\\
\gamma_C &  \sigma(\theta)\setI_2
\end{array}
\right)\, ,
\end{align}
where we used the fact that $\gamma_C^\mathrm{T}=\gamma_C$. It is
interesting to notice that the off-diagonal block matrix $\gamma_C$
has been left unchanged by the transformation.

The second step uses the singular-value-decomposition theorem. The
theorem states that, if $M$ is a two-dimensional real matrix, then it
can always be written as $M=U\Sigma V^\mathrm{T}$ where $U$ and $V$
are orthogonal (namely
$UU^\mathrm{T}=U^\mathrm{T}U=VV^\mathrm{T}=V^\mathrm{T}V=\setI_2$) and
$\Sigma $ is a diagonal matrix (in fact, the theorem is valid for
complex matrices of arbitrary dimension but we do not need this
general version here). The diagonal entries of $\Sigma$ are the
singular values of $M$ and are always positive. Since $U$ and $V$ are
orthogonal [they belong to the $\mathrm{O}(2)$ group], they have
determinant $+1$ or $-1$, hence they can be written in the form
  \begin{align}
  \label{eq:frak:M+:M-}
  \mathfrak{M}^+=
  \begin{pmatrix}
    \cos \psi & -\sin \psi \\
    \sin \psi & \cos \psi
  \end{pmatrix}
\quad\text{or} \quad
  \mathfrak{M}^-=
  \begin{pmatrix}
    -\cos \psi & \sin \psi \\
    \sin \psi & \cos \psi
  \end{pmatrix}
  \end{align}
depending on the value of their determinant. One can check that the
first matrix (with determinant $1$) is symplectic, \ie it satisfies
\Eq{eq:def:symplectic}, while the second matrix (with determinant
$-1$) is not, in agreement with the fact that symplectic matrices have
always determinant $+1$ (and in dimension $2$, being symplectic is
equivalent to having determinant $1$)~\cite{2002clme.book.....G}.

Our goal is to use the singular value decomposition theorem to define
a four-dimensional symplectic transformation, expressed in terms of
the two-dimensional matrices $U$ and $V$ that diagonalise $\gamma_C$,
without affecting the diagonal blocks of $\gamma'$.  We will show
that, quite intuitively, this is possible if $U$ and $V$ are
symplectic themselves, that is to say if they have determinant
$+1$. However, let us note that since \Eq{eq:def:gammaC} leads to
$\det\gamma_{C}=-[\gamma_{12}^2+(\gamma_{11}-\gamma_{22})^2/4]\sin^2(2\theta)<0$,
the equation $\gamma_C=U\Sigma V^\mathrm{T}$ implies that the
determinants of $U$ and $V$ are of opposite signs, hence they cannot
be both equal to $+1$. This issue can be dealt with by introducing the
matrix $\tilde{\gamma}_C=\sigma_3 \gamma_C$, where
$\sigma_3=\mathrm{diag}(1,-1)$ is the third Pauli matrix. Since $\det
\sigma_3=-1$, one has $\det \tilde{\gamma}_C>0$. One can then apply
the singular-value-decomposition theorem to $\tilde{\gamma}_C$, \ie
$\tilde{\gamma}_C=\tilde{U}\tilde{\Sigma} \tilde{V}^\mathrm{T}$, where
the determinants of $\tilde{U}$ and $\tilde{V}$ are now the same. If
they are both $-1$, one can simply multiply $\tilde{U}$ and
$\tilde{V}$ by $\sigma_3$ (which does not change the form of the
singular-value decomposition since $\sigma_3 \tilde{\Sigma} \sigma_3 =
\tilde{\Sigma}$ given that $\tilde{\Sigma}$ is diagonal) such that one
can assume that $\tilde{U}$ and $\tilde{V}$ have determinant $+1$
without loss of generality. This implies that they are symplectic and
that they satisfy \Eq{eq:def:symplectic}.  Let us then consider the
transformation generated by the matrix
\begin{align}
\label{eq:J:T:2}
  T=
  \begin{pmatrix}
    \tilde{U} & 0 \\
    0 & \tilde{V}^\mathrm{T}
  \end{pmatrix}\, .
  \end{align}
One can check that it is symplectic, given that $\tilde{U}$ and
$\tilde{V}$ are. Plugging \Eqs{eq:covgammap} and~\eqref{eq:J:T:2} into
\Eq{eq:transcov}, the covariance matrix becomes
\begin{align}
 \gamma''=
  \begin{pmatrix}
    \sigma(\theta) \setI_2 & \tilde{U}\gamma_C\tilde{V} \\
    \tilde{V}^\mathrm{T} \gamma_C \tilde{U}^\mathrm{T} &
    \sigma(\theta) \setI_2 
  \end{pmatrix}\, ,
  \end{align}
  where we have used that $\tilde{U}$ and $\tilde{V}$ are
  orthogonal. Since $\gamma_C=\sigma_3\tilde{\gamma}_C$ and
  $\tilde{\gamma}_C=\tilde{U}\tilde{\Sigma} \tilde{V}^\mathrm{T}$, the
  off-diagonal block reads $\tilde{U}\gamma_C\tilde{V} = \tilde{U}
  \sigma_3 \tilde{U} \tilde{\Sigma}$. One can then check explicitly
  that matrices of the form $\mathfrak{M}^+$ in \Eq{eq:frak:M+:M-}
  satisfy $\sigma_3 \mathfrak{M}^+ = (\mathfrak{M}^+)^\mathrm{T}
  \sigma_3$, so $\tilde{U} \sigma_3 \tilde{U}= \sigma_3$ and one
  obtains
\begin{align}
  \label{eq:finalcov}
  \gamma''=
  \begin{pmatrix}
    \sigma(\theta) \setI_2 & \sigma_3\tilde{\Sigma} \\
    \sigma_3\tilde{\Sigma} & \sigma(\theta) \setI_2 
  \end{pmatrix}.
 \end{align}
We have thus reached our goal, since $\gamma''$ is of the
form~\eqref{eq:covform} with
$\mathfrak{a}=\mathfrak{b}=\sigma(\theta)$ and
$\frak{C}=\sigma_3\tilde{\Sigma}$ is a diagonal matrix that we denote
$\mathrm{diag}(\mathfrak{c},\mathfrak{d})$.

Let us now explain how the numbers $\mathfrak{c}$ and $\mathfrak{d}$
can be obtained in practice. Since
$\tilde{\gamma}_C=\tilde{U}\tilde{\Sigma} \tilde{V}^\mathrm{T}$, the
eigenvalues of $\tilde{\gamma}_C \tilde{\gamma}_C^\mathrm{T}$ are the
same as those of $\tilde{\Sigma}^2$, \ie $\mathrm{Sp}(\tilde{\gamma}_C
\tilde{\gamma}_C^\mathrm{T})=\{
\mathfrak{c}^2,\mathfrak{d}^2\}$. Making use of \Eq{eq:def:gammaC},
the eigenvalues of $\tilde{\gamma}_C \tilde{\gamma}_C^\mathrm{T}$ can
be computed explicitly, and this leads to
%\footnote{Another way to reach the same result is the following. If $M$ is a two by two matrix, then its two singular values $s_\pm$ are given by \begin{align}s_\pm=& \Biggl\{\vert z_0\vert^2+\vert z_1\vert^2+\vert z_2\vert^2+\vert z_3\vert^2\pm \biggl[\left(\vert z_0\vert^2+\vert z_1\vert^2+\vert z_2\vert^2+\vert z_3\vert^2\right)^2\nonumber \\ &-\left \vert z_0^2-z_1^2-z_2^2-z_3^2\right \vert^2\biggr]^{1/2}\Biggr\}^{1/2}.\end{align}where $M=z_0\setI_2+z_1\sigma_1+z_2\sigma_2+z_3\sigma_3$, the$\sigma_i$'s being the Pauli matrices. One checks that thisexpression matches Eq.~(\ref{eq:mathfrakcd})}
%VV: Je commente cette footnote pour le moment car je la trouve un peu superflue, dans la mesure ou on balance une formule pour les singular values sans la justifier du tout. Et pour la justifier, il faut sans doute faire exactement ce qu'on a fait! 
\begin{align}
    \label{eq:mathfrakcd}
  \mathfrak{c}=-\mathfrak{d}=
  \frac12\sqrt{\left(\gamma_{11}-\gamma_{22}\right)^2+4\gamma_{12}^2}
  \left \vert \sin(2\theta)\right \vert.
  \end{align}
 Here, we have used the fact that $\mathfrak{c}$ and $\mathfrak{d}$
 are of opposite signs since, as mentioned already
 $\det\gamma_C<0$. In terms of the function $\sigma(\theta)$ given in
 \Eq{eq:sigma(theta)}, this can also be written as
 $\mathfrak{c}=-\mathfrak{d}=\sqrt{\sigma^2(\theta)-\sigma^2(0)}$.
  
The mutual information $\mathcal{J}$ for covariance matrices of the
form ~\eqref{eq:covform} is computed in \Refa{2010PhRvL.105c0501A},
where it is shown that the result depends on the sign of $(1+\det
\mathfrak{B})\det^2 \mathfrak{C}(\det\mathfrak{A} + \det\gamma) -
(\det\gamma - \det\mathfrak{A} \det\mathfrak{B})^2$. Using that
$\mathfrak{c}=-\mathfrak{d}$, this quantity is given by
$\mathfrak{c}^4
(\mathfrak{a}-\mathfrak{a}\mathfrak{b}^2+\mathfrak{b}\mathfrak{c}^2)^2$,
which is necessarily positive. In that case, $\mathcal{J}$ is given
by~\cite{2010PhRvL.105c0501A}
\begin{align}
\max_{\{\hat{\Pi}_i\}}\mathcal{J} =
\label{eq:J:result}
%f\left(\sqrt{\det\mathfrak{A}}\right) -f\left[\frac{\sqrt{2\det^2\mathfrak{C}+\left(\det\mathfrak{B}-1\right)\left(\det\gamma-\det\mathfrak{A}\right)+2\left\vert \det\mathfrak{C} \right\vert \sqrt{\det^2\mathfrak{C}+\left(\det\mathfrak{B}-1\right)\left(\det\gamma-\det\mathfrak{A}\right)}}}{\left\vert \det\mathfrak{B}-1\right\vert}\right] 
f\left[\sigma(\theta)\right]
  -f\left[\frac{\sigma^2(0)+\sigma(\theta)}{1+\sigma(\theta)}\right]\, ,
\end{align}
where we have used that $\sigma^2(0)>1$, see the discussion below
\Eq{eq:finalI}.

Plugging \Eqs{eq:finalI} and~\eqref{eq:J:result} into
\Eq{eq:discord:def}, one finally obtains the formula
\Eq{eq:finaldiscord} for the quantum discord.

\section{Covariance matrix for cosmological perturbations in
  the Caldeira-Leggett model}
\label{ap:sec:covmatcaldlegget}

In this appendix, we compute the covariance matrix of inflationary
perturbations, given by \Eqs{eq:exactgammadeco23} and
(\ref{eq:I:integrals:def}), (\ref{eq:J:integrals:def}),
(\ref{eq:K:integrals:def}) in the Caldeira-Leggett model described by
the ansatz~\eqref{eq:kG:def}.
\subsection{Exact calculation}
\label{subsec:exactapp}
Recalling that the mode function is given by
\Eq{eq:desittermodefunction}, the quantity $\Ima^2\left[v_{\bm
    k}(\eta)v_{\bm k}^*(\eta')\right]$ appearing in the integrand of
\Eq{eq:I:integrals:def} can be written as
\begin{align}
{\rm Im}^2\left[v_{\bm k}(\eta)
    v_{\bm k}^*(\eta')\right]
&=\frac{1}{k^4\eta^2\eta'^2}\biggl[k(\eta'-\eta)\cos(k\eta-k\eta')
%\nonumber \\ & 
  +(1+k^2\eta\eta')\sin(k\eta-k\eta')\biggr]^2.
\end{align}
Introduce the dimensionless variables $x=-k\eta$ and $x'=-k\eta'$ for
notational convenience, and recalling that $a=-1/(H\eta)$ where
$H=\mathcal{H}/a$ in the de Sitter space time,
\Eq{eq:exactgammadeco23} gives rise to 
  \begin{align}
   \label{eq:solgamma11}
    \gamma_{11}(\eta) =&v_{\bm k}(\eta)v_{\bm k}^*(\eta)
-2\left(\frac{k_\Gamma}{k}\right)^2
  \int_{1/(\ell_{E} H)}^{x}
  \left(\frac{x_*}{x'}\right)^{p-3}\frac{1}{x^2x'^2}
  \bigl[(x-x')\cos(x'-x)\nonumber \\
 & +(1+xx')\sin(x'-x) \bigr]^2 {\dd}x' \\
  =&v_{\bm k}(\eta)v_{\bm k}^*(\eta) -2
  \left(\frac{k_\Gamma}{k}\right)^2
  \left[I_{11}^{(1)}+I_{11}^{(2)}+I_{11}^{(3)}\right],     
\end{align}
with
 \begin{align}
 I_{11}^{(1)}&=
\frac{x_*^{p-3}(1+x^2)}{2x^2}\int _{1/(\ell_{E} H)}^{x}x'^{1-p}(1+x'^2){\dd}x'\, ,
\nonumber \\ 
I_{11}^{(2)}&=
\frac{x_*^{p-3}}{4x^2}e^{-2ix}\int _{1/(\ell_{E} H)}^{x}
e^{2ix'}x'^{1-p}\left[(x-x')^2-2i(x-x')(1+xx')-(1+xx')^2\right]{\dd}x'\, ,
\nonumber \\ 
I_{11}^{(3)}&=
\frac{x_*^{p-3}}{4x^2}e^{2ix}\int _{1/(\ell_{E} H)}^{x}
e^{-2ix'}x'^{1-p}\left[(x-x')^2+2i(x-x')(1+xx')-(1+xx')^2\right]{\dd}x'\, ,
 \end{align}
where the time $\eta_*$ and $k_*$ have been defined after
Eqs.~(\ref{eq:kG:def}) and~(\ref{eq:expansiong11}),
(\ref{eq:expansiong12}), (\ref{eq:expansiong22}). Our goal is now to
calculate the three above integrals.

The calculation of the first integral is straightforward and one
obtains the following expression
\begin{align}
I_{11}^{(1)}=\frac{x_*^{p-3}(1+x^2)}{2x^2} 
\left[\frac{x^{2-p}}{2-p}+\frac{x^{4-p}}{4-p}
  -\frac{(\ell_{E}H)^{p-2}}{2-p}
  -\frac{(\ell_{E}H)^{p-4}}{4-p}\right].
\end{align}
Of course, the result is not defined for the particular values $p=2$
or $p=4$. In these cases, instead of power law solutions, we just have
logarithms.

\begin{figure*}[t]
\begin{center}
\includegraphics[width=0.49\textwidth,clip=true]{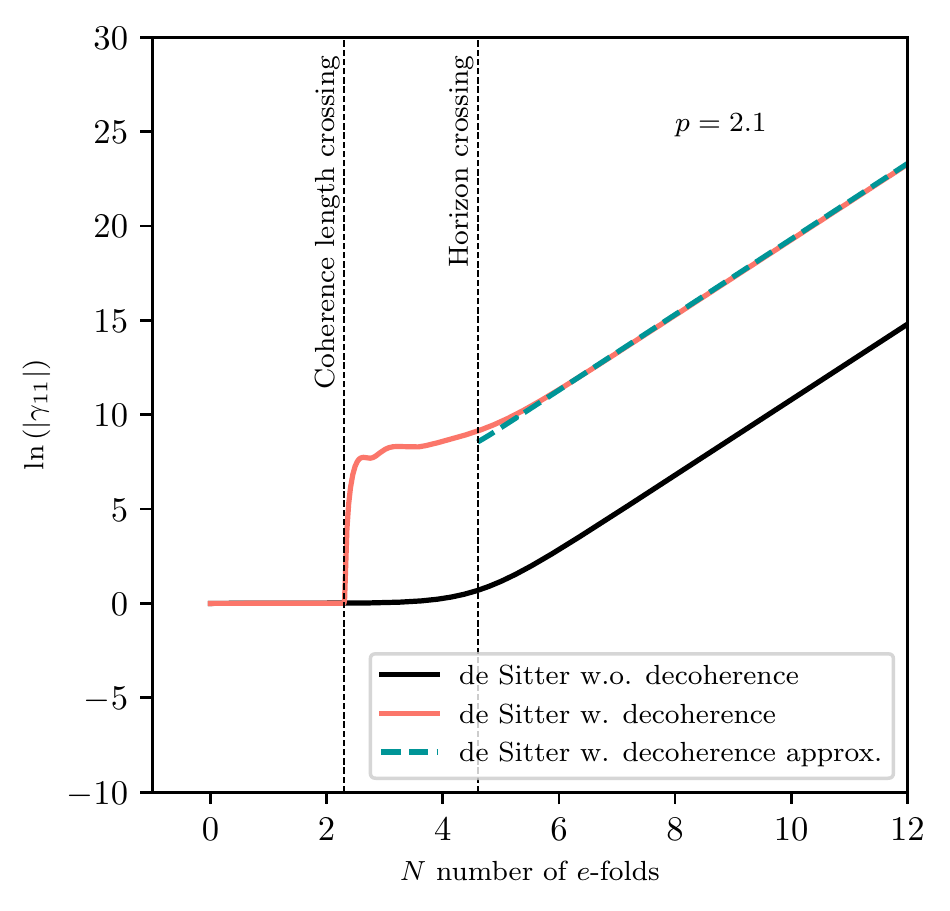}
\includegraphics[width=0.49\textwidth,clip=true]{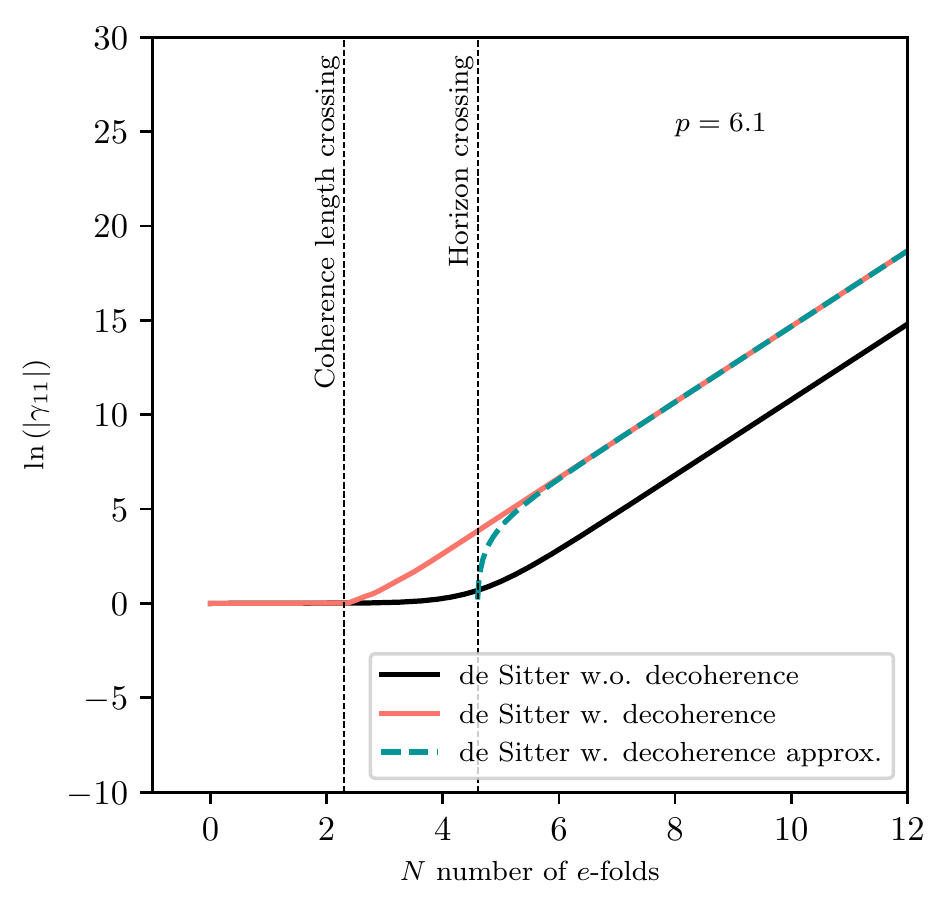}
\caption{$\ln \left( \vert \gamma_{11} \vert \right)$ with (pink) and
  without (black) decoherence for the de Sitter case. The approximated
  version (green dashed) is obtained using the first order
  approximations~(\ref{eq:expansiong11}). The first vertical dashed
  line shows the time when the mode $\bm{k}$ starts to decohere
  $\ell_{E} a / k = 1$, the second the time when the mode $\bm{k}$
  exits the Hubble radius. The parameters are $\ell_{E}H =0.1 $,
  $x_{\star} = 1$, $p = 2.1$ (left) or $p = 6.1$ (right), and
  $k_{\Gamma}/k = 10$.}
\label{fig:g11_decoh}
\end{center}
\end{figure*}

The calculation of the second term is more complicated but can still
be done in terms of special functions. After straightforward
manipulations, one arrives at
\begin{align}
  I_{11}^{(2)}=\frac{x_*^{p-3}}{4x^2}e^{-2ix}
  \left[(x^2-1-2ix)(A_{1-p}-A_{3-p})-(4x+2ix^2-2i)A_{2-p}\right],
\end{align}
with
\begin{align}
\label{eq:defAalpha}
  A_\alpha\equiv \int _{1/(\ell_{E} H)}^{x}e^{2ix'}x'^{\alpha}{\dd}x'&=-2^{-1-\alpha}
  (-i)^{-1-\alpha}\left[\Gamma\left(1+\alpha,-2ix\right)
    -\Gamma\left(1+\alpha,-\frac{2i}{\ell_{E}H}\right)\right],
\end{align}
where $\Gamma(a,z)=\int_z^{+\infty}t^{a-1}e^{-t}{\dd}{t}$ is the
incomplete Gamma
function~\cite{Gradshteyn:1965aa,abramowitz+stegun}. The third term,
$I_{11}^{(3)}$, is just given by $I_{11}^{(3)}=I_{11}^{(2)}{}^*$.  The
resulting time evolution of $\gamma_{11}(\eta)$, with the choices
$\ell_{E}H=0.1$, $x_*=0.1$, $p=2.1$ and $k_\Gamma/k=10$, is displayed
in \Fig{fig:g11_decoh}.

\begin{figure*}[t]
\begin{center}
\includegraphics[width=0.49\textwidth,clip=true]{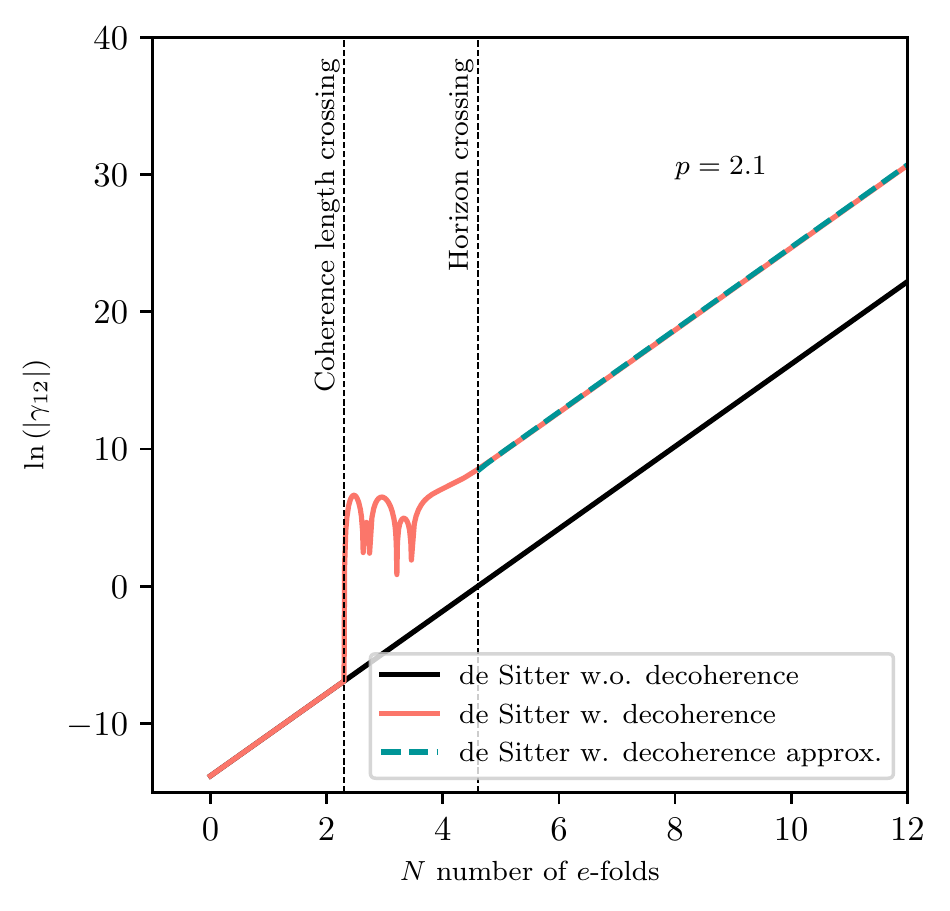}
\includegraphics[width=0.49\textwidth,clip=true]{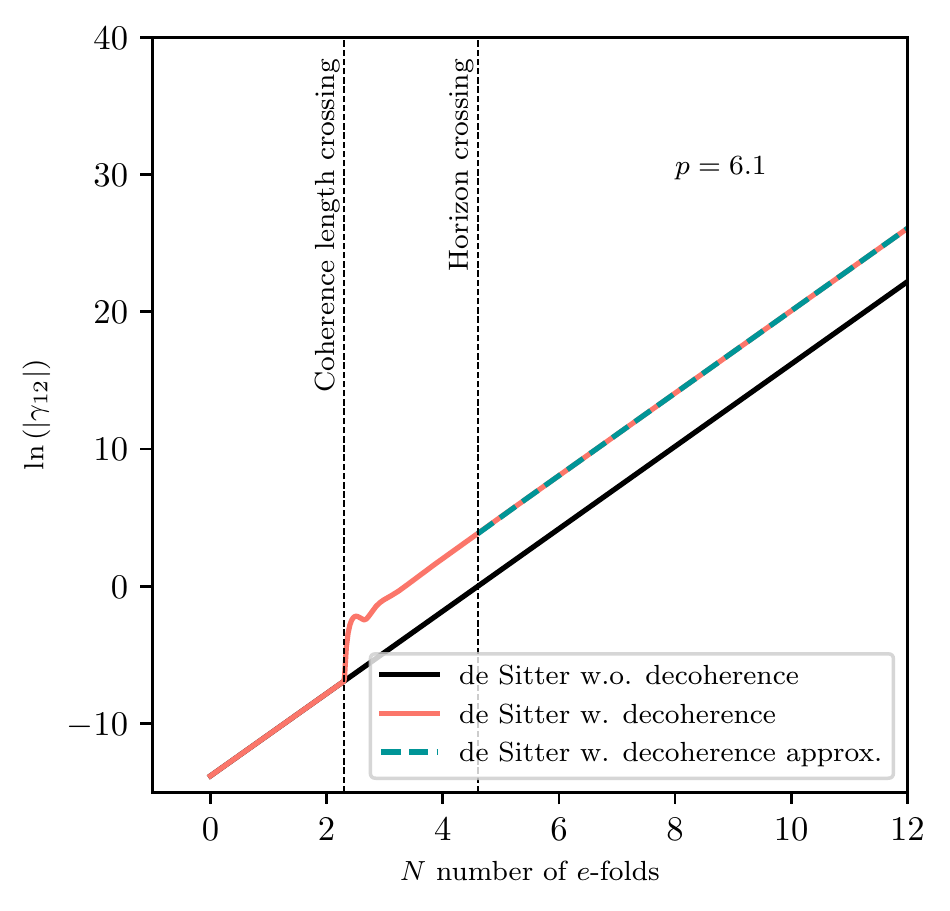}
\caption{$\ln \left( \vert \gamma_{12} \vert \right)$ with (pink) and
  without (black) decoherence for the de Sitter case. The approximated
  version (green dashed) is obtained using the first order
  approximations~(\ref{eq:expansiong12}). The first vertical dashed
  line shows the time when the mode $\bm{k}$ starts to decohere
  $\ell_{E} a / k = 1$, the second the time when the mode $\bm{k}$
  exits the Hubble radius. The parameters are $\ell_{E}H =0.1 $,
  $x_{\star} = 1$, $p = 2.1$ (left) or $p = 6.1$ (right), and
  $k_{\Gamma}/k = 10$.}
\label{fig:g12_decoh}
\end{center}
\end{figure*}

The next step consists in evaluating $\gamma_{12}$. Instead of using
\Eq{eq:exactgammadeco23} and performing a similar calculation as
above, one can use the transport equations~\eqref{eq:diffcov1deco},
(\ref{eq:diffcov2deco}) and (\ref{eq:diffcov3deco}). This leads to
\begin{align}
  \gamma_{12}=\frac{1}{2k}\frac{{\dd}}{{\dd}\eta}
  \left(v_{\bm k}v_{\bm k}^*\right)
+
\left(\frac{k_\Gamma}{k}\right)^2
  \int_{-\infty}^{\eta}
  \left(\frac{a}{a_*}\right)^{p-3}
  \mathrm{H} \left(1-\frac{k\ell_{E}}{a}\right)
\frac{\partial}{\partial \eta}{\rm Im}^2\left[v_{\bm k}(\eta)
  v_{\bm k}^*(\eta')\right]{\dd}\eta'.
\end{align}
Instead of ${\rm Im}^2\left[v_{\bm k}(\eta) v_{\bm k}^*(\eta')\right]$
in the integrand, as was the case for $\gamma_{11}$, we now have the
derivative of it. Explicitly, written in terms of the variables $x$
and $x'$, it can be expressed as
\begin{align}
\frac{\partial}{\partial \eta}{\rm Im}^2\left[v_{\bm k}(\eta)
  v_{\bm k}^*(\eta')\right]&=\frac{k}{x^3x'^2}(1+x'^2)
-\frac{i}{2x^3x'^2}e^{-2ix}e^{2ix'}k(-i+x)\left[-1+x(-i+x)\right]
\nonumber \\ & \times
(i+x')^2
%\nonumber \\ &
+\frac{i}{2x^3x'^2}e^{2ix}e^{-2ix'}k(i+x)\left[-1+x(i+x)\right]
(-i+x')^2.
\end{align}
As a consequence, $\gamma_{12}$ takes the following form 
\begin{align}
\label{eq:solgamma12}
\gamma_{12}=\frac{1}{2k}\frac{{\dd}}{{\dd}\eta}\left[
v_{\bm k}(\eta)v_{\bm k}^*(\eta)\right]
-2\left(\frac{k_\Gamma}{k}\right)^2
\left[I_{12}^{(1)}+I_{12}^{(2)}+I_{12}^{(3)}\right],
\end{align}
with
\begin{align}
    I_{12}^{(1)}&=\frac12 \frac{x_*^{p-3}}{x^3}
\int_{1/(\ell_{E}H)}^{x} x'^{1-p}(1+x'^2){\dd}x',
\\
I_{12}^{(2)}&=-\frac{ix_*^{p-3}}{4x^3}
e^{-2ix}(-i+x)\left[x(-i+x)-1\right]
\int_{1/(\ell_{E}H)}^{x} e^{2ix'}x'^{1-p}(i+x')^2{\dd}x',
\\
I_{12}^{(3)}&=\frac{ix_*^{p-3}}{4x^3}
e^{2ix}(i+x)\left[x(i+x)-1\right]
\int_{1/(\ell_{E}H)}^{x} e^{-2ix'}x'^{1-p}(-i+x')^2{\dd}x'.
\end{align}
These integrals are very similar to those appearing in the
expression~(\ref{eq:solgamma11}) of $\gamma_{11}$ and they can be
computed with the same techniques. We obtain
\begin{align}
\label{eq:I121explicit}
  I_{12}^{(1)}&=\frac12 \frac{x_*^{p-3}}{x^3}
\left[\frac{x^{2-p}}{2-p}+\frac{x^{4-p}}{4-p}
  -\frac{(\ell_{E}H)^{p-2}}{2-p}
  -\frac{(\ell_{E}H)^{p-4}}{4-p}\right],
\\
\label{eq:I122explicit}
I_{12}^{(2)}&=-\frac{ix_*^{p-3}}{4x^3}
e^{-2ix}(-i+x)\left[x(-i+x)-1\right]\left(-A_{1-p}+2iA_{2-p}+A_{3-p}\right),
\end{align}
and $I_{12}^{(3)}=I_{12}^{(2)}{}^*$. In \Fig{fig:g12_decoh}, we have
plotted $\gamma_{12}$ with the same parameter values as in
\Fig{fig:g11_decoh}, namely $\ell_{E}=0.1$, $x_*=0.1$, $p=2.1$ and
$k_\Gamma/k=10$.

\begin{figure*}[t]
\begin{center}
\includegraphics[width=0.49\textwidth,clip=true]{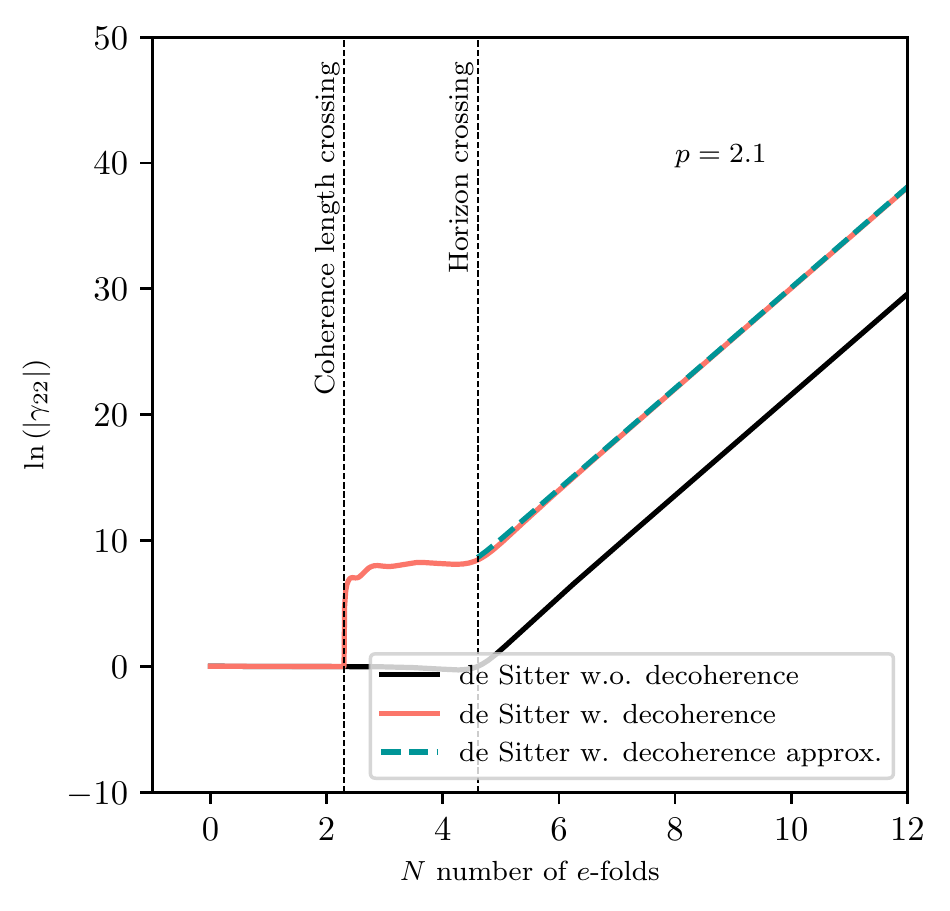}
\includegraphics[width=0.49\textwidth,clip=true]{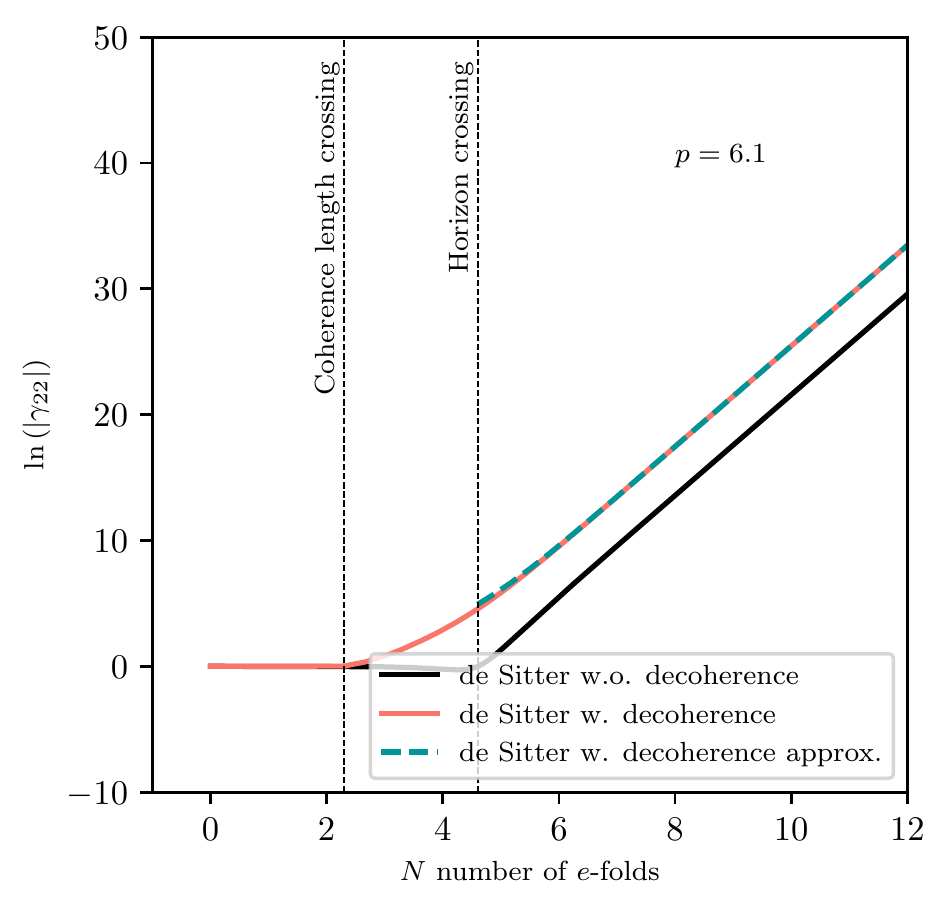}
\caption{$\ln \left( \vert \gamma_{22} \vert \right)$ with (pink) and
  without (black) decoherence for the de Sitter case. The approximated
  version (green dashed) is obtained using the first order
  approximations~(\ref{eq:expansiong22}). The first vertical dashed
  line shows the time when the mode $\bm{k}$ starts to decohere
  $\ell_{E} a / k = 1$, the second the time when the mode $\bm{k}$
  exits the Hubble radius. The parameters are $\ell_{E}H =0.1 $,
  $x_{\star} = 1$, $p = 2.1$ (left) or $p = 6.1$ (right), and
  $k_{\Gamma}/k = 10$.}
\label{fig:g22_decoh}
\end{center}
\end{figure*}

Finally, the component $\gamma_{22}$ remains to be evaluated. As done
above for the component $\gamma_{12}$, one can use the transport
equations to calculate $\gamma_{22}$ from the expression of
$\gamma_{11}$. This leads to the following expression 
\begin{align}
  \gamma_{22}=&\frac{1}{2k^2}\frac{{\dd}^2}{{\dd}\eta^2}
  \left(v_{\bm k}v_{\bm k}^*\right)
  +\frac{1}{k}
\left(\frac{k_\Gamma}{k}\right)^2
  \int_{-\infty}^{\eta}
  \left(\frac{a}{a_*}\right)^{p-3}
  \mathrm{H} \left(1-\frac{k\ell_{E}}{a}\right)
\frac{\partial^2}{\partial \eta^2}{\rm Im}^2\left[v_{\bm k}(\eta)
  v_{\bm k}^*(\eta')\right]{\dd}\eta'
  \nonumber \\ &
  +\frac{\omega^2}{k^2}\gamma_{11}.
\end{align}
As expected, the integrand in the above formula now contains the
second derivative of ${\rm Im}^2\left[v_{\bm k}(\eta) v_{\bm
    k}^*(\eta')\right]$. Concretely, this quantity can be written as
\begin{align}
\frac{\partial^2}{\partial \eta^2}{\rm Im}^2\left[v_{\bm k}(\eta)
  v_{\bm k}^*(\eta')\right]&=\frac{3k^2}{x^4x'^2}(1+x'^2)
\nonumber \\ &
+\frac{k^2}{2x^4x'^2}e^{-2ix}e^{2ix'}
\left\{3+2x\left[3i+x(-3-2ix+x^2)\right]\right\}(i+x')^2
\nonumber \\ &
+\frac{k^2}{2x^4x'^2}e^{2ix}e^{-2ix'}
\left\{3+2x\left[-3i+x(-3+2ix+x^2)\right]\right\}(-i+x')^2.
\end{align}
This leads to 
\begin{align}
\gamma_{22}=&\frac{1}{2k^2}\frac{{\dd}^2}{{\dd}\eta^2}\left[
v_{\bm k}(\eta)v_{\bm k}^*(\eta)\right]
-2\left(\frac{k_\Gamma}{k}\right)^2
\left[I_{22}^{(1)}+I_{22}^{(2)}+I_{22}^{(3)}\right]+\frac{\omega^2}{k^2}\gamma_{11}
\\
=&\frac{1}{2k^2}\frac{{\dd}^2}{{\dd}\eta^2}\left[
  v_{\bm k}(\eta)v_{\bm k}^*(\eta)\right]
+\frac{\omega^2}{k^2}v_{\bm k}(\eta)v_{\bm k}^*(\eta)
\nonumber \\ &
-2  \left(\frac{k_\Gamma}{k}\right)^2
\left\{I_{22}^{(1)}+I_{22}^{(2)}+I_{22}^{(3)}+
  +\frac{\omega^2}{k^2}
  \left[I_{11}^{(1)}+I_{11}^{(2)}+I_{11}^{(3)}\right]\right\},
\end{align}
with
\begin{align}
    I_{22}^{(1)}&=\frac{3}{2}
\frac{x_*^{p-3}}{x^4}
\int_{1/(\ell_{E}H)}^{x} x'^{1-p}(1+x'^2){\dd}x', 
\\
I_{22}^{(2)}&=\frac{x_*^{p-3}}{4x^4}
e^{-2ix}\left\{3+2x\left[3i+x(-3-2ix+x^2)\right]\right\}
\int_{1/(\ell_{E}H)}^{x} e^{2ix'}x'^{1-p}(i+x')^2{\dd}x',
\\
I_{22}^{(3)} &=\frac{x_*^{p-3}}{4x^4}
e^{2ix}\left\{3+2x\left[-3i+x(-3+2ix+x^2)\right]\right\}
\int_{1/(\ell_{E}H)}^{x} e^{-2ix'}x'^{1-p}(-i+x')^2{\dd}x'.
\end{align}
Again, the integrals $I_{22}^{(1)}$, $I_{22}^{(2)}$ and $I_{22}^{(3)}$
can be computed with the same tools used above. This leads to the
following expressions
\begin{align}
\label{eq:I221}
  I_{22}^{(1)}&=\frac32 \frac{x_*^{p-3}}{x^4}
\left[\frac{x^{2-p}}{2-p}+\frac{x^{4-p}}{4-p}
  -\frac{(\ell_{E}H)^{p-2}}{2-p}
  -\frac{(\ell_{E}H)^{p-4}}{4-p}\right]
\\
I_{22}^{(2)}&=\frac{x_*^{p-3}}{4x^4}
e^{-2ix}\left\{3+2x\left[3i+x(-3-2ix+x^2)\right]\right\}
\left(-A_{1-p}+2iA_{2-p}+A_{3-p}\right),
\end{align}
and $I_{22}^{(3)}=I_{22}^{(2)}{}^*$. The quantity $\gamma_{22}$ is
represented in Fig.~\ref{fig:g22_decoh} for the same values of the
parameters as above, that is to say $\ell_{E}H=0.1$, $x_*=1$, $p=2.1$,
and $k_\Gamma/k=10$.

\subsection{Approximations}
\label{ap:subsec:approxapp}

The above results allow one to fully study the time evolution of the
system, since they give exact and explicit expressions for the
elements of the covariance matrix. However, the corresponding formulas
are not particularly insightful and it is therefore useful to
approximate them and to extract the leading behaviours of
$\gamma_{11}$, $\gamma_{12}$ and $\gamma_{22}$ in the late time, \ie
super-Hubble, limit. This is the goal of this sub-section.

Let us start with the first component, $\gamma_{11}$. We write the
function $A_\alpha$ in terms of its real and imaginary parts,
$A_\alpha=A_\alpha^\mathrm{R}+iA_\alpha^\mathrm{I}$. Then, it follows
that
\begin{align}
  \label{eq:Irandi}
  I_{11}^{(2)}+I_{11}^{(3)}&=\frac{x_*^{p-3}}{2x^2}\bigl[
  (x^2-1)\left(A_{1-p}^\mathrm{R}
    -A_{3-p}^\mathrm{R}+2A_{2-p}^\mathrm{I}\right)
  \nonumber \\ &  
    +2x\left(A_{1-p}^\mathrm{I}
    -A_{3-p}^\mathrm{I}-2A_{2-p}^\mathrm{R}\right)\bigr]\cos(2x)
    \nonumber \\ &
+\frac{x_*^{p-3}}{2x^2}\biggl[(x^2-1)\left(A_{1-p}^\mathrm{I}
    -A_{3-p}^\mathrm{I}-2A_{2-p}^\mathrm{R}\right)
    \nonumber \\ &
    -2x\left(A_{1-p}^\mathrm{R}
    -A_{3-p}^\mathrm{R}+2A_{2-p}^\mathrm{I}\right)\biggr]\sin(2x).
\end{align}
Since $I_{11}^{(3)}=I_{11}^{(2)}{}^*$, the quantity
$I_{11}^{(2)}+I_{11}^{(3)}$ must be real and one can check that it is
indeed the case. Then, upon Taylor expanding the real and imaginary
parts of the function $A_\alpha$, defined in \Eq{eq:defAalpha}, around
$x=0$, one obtains
\begin{align}
  \label{eq:Ar}
  A_\alpha^\mathrm{R}&\simeq {\cal A}_\alpha^\mathrm{R}(\ell_{E}H)
  +x^{1+\alpha}\Biggl[\frac{1}{\alpha+1}-\frac{2x^2}{3+\alpha}
    +\frac{2x^4}{3(5+\alpha)}-\frac{4x^6}{45(7+\alpha)}
    \nonumber \\ &
+\frac{2x^8}{315(9+\alpha)}
-\frac{4x^{10}}{14175(11+\alpha)}
+\frac{4x^{12}}{467775(13+\alpha)}
-\frac{8x^{14}}{42567525(15+\alpha)}
\nonumber \\ &
+\frac{2x^{16}}{638512875(17+\alpha)}
-\frac{4x^{18}}{97692469875(19+\alpha)}
+\cdots \Biggr],
  \end{align}
where
\begin{align}
{\cal A}_\alpha^\mathrm{R}(\ell_{E}H)
&=2^{-1-\alpha}\Gamma(1+\alpha)\sin \left(\frac{\pi\alpha}{2}\right)
-i\, 2^{-2-\alpha}
\biggl[e^{-i\pi \alpha/2}\Gamma\left(1+\alpha,\frac{2i}{\ell_{E}H}
\right)
\nonumber \\ &
-e^{i\pi \alpha/2}\Gamma\left(1+\alpha,-\frac{2i}{\ell_{E}H}
\right)\biggr].
\end{align}
The same type of calculations lead to the following expression for the
imaginary part
\begin{align}
  \label{eq:Ai}
  A_\alpha^\mathrm{I}&\simeq {\cal A}_\alpha^\mathrm{I}(\ell_{E}H)
  +x^{2+\alpha}\Biggl[\frac{2}{2+\alpha}-\frac{4x^2}{3(4+\alpha)}
    +\frac{4 x^4}{15(6+\alpha)}
-\frac{8x^6}{315(8+\alpha)}
\nonumber \\ &
+\frac{4x^{8}}{2835(10+\alpha)}
-\frac{8x^{10}}{155925(12+\alpha)}
+\frac{8x^{12}}{6081075(14+\alpha)}
-\frac{16x^{14}}{638512875(16+\alpha)}
\nonumber \\ &
+\frac{4x^{16}}{10854718875(18+\alpha)}
-\frac{8x^{18}}{1856156927625(20+\alpha)}
+\cdots\Biggr],
\end{align}
where
\begin{align}
{\cal A}_\alpha^\mathrm{I}(\ell_{E}H)
&=-2^{-1-\alpha}\Gamma(1+\alpha)\cos \left(\frac{\pi\alpha}{2}\right)
+2^{-2-\alpha}
\biggl[e^{-i\pi \alpha/2}\Gamma\left(1+\alpha,\frac{2i}{\ell_{E}H}
\right)
\nonumber \\ &
+e^{i\pi \alpha/2}\Gamma\left(1+\alpha,-\frac{2i}{\ell_{E}H}
\right)\biggr].
\end{align}

The next step consists in using the above approximations for the real
and imaginary parts of $A_\alpha$ in Eq.~(\ref{eq:Irandi}) which,
together with the exact expressions of $I_{11}^{(1)}$ (which is
already given in terms of power-laws) leads to an approximation for
the term $I_{11}^{(1)}+I_{11}^{(2)}+I_{11}^{(3)}$. The corresponding
expression reads
\begin{align}
\label{eq:expansioncorr11}
  I_{11}^{(1)}+I_{11}^{(2)}+I_{11}^{(3)}
\simeq &
\, x^{-p}
\left[A_{11}x^6+{\cal O}\left(x^8\right)\right]
+\frac{B_{11}}{x^2}+ C_{11}+D_{11}x+E_{11}x^3+F_{11}x^4 \nonumber \\
& +G_{11}x^5+H_{11}x^6 +{\cal O}\left(x^7\right)\, ,
\end{align}
where
 \begin{align}
   \label{eq:A11}
A_{11} &= -\frac{2x_*^{p-3}}{(p-8)(p-5)(p-2)}
\\
    B_{11}&=\frac{x_*^{p-3}}{2}\left[\frac{(\ell_{E}H)^{p-4}}{p-4}
      +\frac{(\ell_{E}H)^{p-2}}{p-2}-{\cal A}_{1-p}^\mathrm{R}
      -2{\cal A}_{2-p}^\mathrm{I}+{\cal A}_{3-p}^\mathrm{R}\right],
    \\
    C_{11}&= B_{11},
    \\D_{11}&=\frac{x_*^{p-3}}{3}\left(-{\cal A}_{1-p}^\mathrm{I}
    +2{\cal A}_{2-p}^\mathrm{R}+{\cal A}_{3-p}^\mathrm{I}\right),
    \\
    E_{11}&=\frac25 D_{11},
    \\
    F_{11}&=\frac{x_*^{p-3}}{9}\left({\cal A}_{1-p}^\mathrm{R}
    +2{\cal A}_{2-p}^\mathrm{I}-{\cal A}_{3-p}^\mathrm{R}\right),
    \\
    G_{11}&= -\frac{6}{35}D_{11}, \quad
    H_{11}=-\frac{1}{5}F_{11}.
\end{align}
We combine the above with \Eq{eq:solgamma11} to obtain an
approximation of $\gamma_{11}$ which can be expressed as
\begin{equation}
\label{eq:g11approx}
\gamma_{11} = \frac{1}{x^2}\left[ 1 - 2
  \left(\frac{k_\Gamma}{k}\right)^2 B_{11} \right]
+ \mathcal{O} \left( x^0 \right)
- 2  \left(\frac{k_\Gamma}{k}\right)^2
A_{11}x^{6-p} + {\cal O}\left(x^{8-p}\right) \, .
\end{equation}
Which of the two terms in the expression dominates depends on the
value of $p$. This asymptotic expression of $\gamma_{11}$ is
represented by the green dashed line in Fig.~\ref{fig:g11_decoh}. We
see that it matches very well the exact result.

Let us now derive an approximation for the matrix element
$\gamma_{12}$. Compared to what has been done above for $\gamma_{11}$,
the calculation proceeds in a similar fashion. The expression of
$I_{12}^{(1)}$ is already explicit, see Eq.~(\ref{eq:I121explicit}).
The two remaining terms, $I_{12}^{(2)}$ and $I_{12}^{(3)}$, have
similar expressions in terms of the real and imaginary parts of
$A_{\alpha}$ as $I_{11}^{(2)}$ and $I_{11}^{(3)}$. This leads to
\begin{align}
  I_{12}^{(2)}+I_{12}^{(3)}&=\frac{x_*^{p-3}}{2x^3}
\biggl[(1-2x^2)\left(-A_{1-p}^\mathrm{R}-2A_{2-p}^\mathrm{I}
    +A_{3-p}^\mathrm{R}\right)
    \nonumber \\ &
    -x(2-x^2)\left(-A_{1-p}^\mathrm{I}
    +2A_{2-p}^\mathrm{R}+A_{3-p}^\mathrm{I}\right)\biggr]\cos(2x)
  \nonumber \\ &
  +\frac{x_*^{p-3}}{2x^3}
\biggl[(1-2x^2)\left(-A_{1-p}^\mathrm{I}+2A_{2-p}^\mathrm{R}
    +A_{3-p}^\mathrm{I}\right)
    \nonumber \\ &
    +x(2-x^2)\left(-A_{1-p}^\mathrm{R}
    -2A_{2-p}^\mathrm{I}+A_{3-p}^\mathrm{R}\right)\biggr]\sin(2x).
\end{align}
Using this result and expanding consistently the result, one obtains
the following expression
\begin{align}
\label{eq:expansioncorr12}
 I_{12}^{(1)}+I_{12}^{(2)}+I_{12}^{(3)}
&\simeq 
x^{-p}
\left[A_{12}x^5+{\cal O}\left(x^7\right)\right]
+\frac{B_{12}}{x^3}+C_{12}+D_{12}x^2+E_{12}x^3
\nonumber \\ &
+F_{12}x^4+G_{12}x^5+H_{12}x^6
+{\cal O}\left(x^7\right) ,
\end{align}
with 
\begin{align}
\label{eq:A12}
  A_{12}=-\frac{x_*^{p-3}(p-6)}{(p-8)(p-5)(p-2)},
\end{align}
and $B_{12}=B_{11}$, $C_{12}=-D_{11}/2$, $D_{12}=-3D_{11}/5$,
$E_{12}=-2F_{11}$, $F_{12}=3D_{11}/7$, $G_{12}=3F_{11}/5$ and
$H_{12}=-2D_{11}/27$. Using \Eq{eq:solgamma12} we get 
\begin{equation}
\label{eq:g12approx}
\gamma_{12} = \frac{1}{x^3} \left[ 1 - 2
  \left(\frac{k_\Gamma}{k}\right)^2 B_{12} \right]
+ {\cal O}\left(x^0\right) -
2\left(\frac{k_\Gamma}{k}\right)^2
\left[A_{12}x^{5-p} +{\cal O}\left(x^{7-p} \right) \right] \, ,
\end{equation}
Again, which of the two terms dominates in the above equation depends
on the value of $p$. The approximation~(\ref{eq:g12approx}) is
represented in Fig.~\ref{fig:g12_decoh} and we notice that, in its domain
of validity (namely, on large scales), it is very accurate.

Let us finally consider the matrix element $\gamma_{22}$. In order to
establish its large scale expansion, the considerations presented
before can be repeated once more. The term $I_{22}^{(1)}$ has already
the adequate form, see \Eq{eq:I221}. As a consequence, the only
calculation that is needed is to express $I_{22}^{(2)}+I_{22}^{(3)}$
in terms of the real and imaginary parts of $A_{\alpha}$. One obtains
\begin{align}
  I_{22}^{(2)}+I_{22}^{(3)}&=\frac{x_*^{p-3}}{2x^4}
\biggl[(3-6x^2+2x^4)\left(-A_{1-p}^\mathrm{R}-2A_{2-p}^\mathrm{I}
    +A_{3-p}^\mathrm{R}\right)
    \nonumber \\ &
    -(6x-4x^3)\left(-A_{1-p}^\mathrm{I}
    +2A_{2-p}^\mathrm{R}+A_{3-p}^\mathrm{I}\right)\biggr]\cos(2x)
  \nonumber \\ &
  +\frac{x_*^{p-3}}{2x^4}
\biggl[(3-6x^2+2x^4)\left(-A_{1-p}^\mathrm{I}+2A_{2-p}^\mathrm{R}
    +A_{3-p}^\mathrm{I}\right)
    \nonumber \\ &
    +(6x-4x^3)\left(-A_{1-p}^\mathrm{R}
    -2A_{2-p}^\mathrm{I}+A_{3-p}^\mathrm{R}\right)\biggr]\sin(2x).
\end{align}
The next step consists in inserting the expressions~(\ref{eq:Ar})
and~(\ref{eq:Ai}) of the real and imaginary parts of $A_\alpha$ in the
above formula. This leads to the following equation for the correction
\begin{align}
\label{eq:expansioncorr22}
I_{22}^{(1)}&+I_{22}^{(2)}+I_{22}^{(3)}+
  \frac{\omega^2}{k^2}
  \left[I_{11}^{(1)}+I_{11}^{(2)}+I_{11}^{(3)}\right]
  =  x^{-p} \left[A_{22}x^4 +{\cal O}\left(x^6\right) \right]
  +\frac{B_{22}}{x^4}
  +\frac{C_{22}}{x^2}
\nonumber \\ &
  +\frac{D_{22}}{x}+E_{22}+F_{22}x
  +G_{22}x^2+H_{22}x^3+I_{22}x^4+J_{22}x^5+K_{22}x^6 +{\cal O}\left(x^7\right)
\, ,
\end{align}
with 
\begin{align}
  \label{eq:A22}
    A_{22}&=-\frac{[26+p(p-11)]x_*^{p-3}}{(p-8)(p-5)(p-2)},
\end{align}
and $B_{22}=B_{11}$, $C_{22}=-B_{11}$, $D_{22}=-2D_{11}$,
$E_{22}=B_{11}$, $F_{22}=7D_{11}/5$, $G_{22}=4F_{11}$,
$H_{22}=-34D_{11}/35$, $I_{22}=-8F_{11}/5$, $J_{22}=218 D_{11}/945$
and $K_{22}=43F_{11}/175$. Finally, we obtain the following
approximation for $\gamma_{22}$ 
\begin{align}
\label{eq:g22approx}
\gamma_{22}(\eta)&= \frac{1}{x^4} \left[ 1
  - 2  \left(\frac{k_\Gamma}{k}\right)^2
  B_{22}\right]
+{\cal O}\left( \frac{1}{x^2} \right)
  - 2  \left(\frac{k_\Gamma}{k}\right)^2
  A_{22}x^{4-p}
  +{\cal O}\left(x^{6-p}\right) \, .
\end{align}
This approximation~(\ref{eq:g22approx}) is plotted in
Fig.~\ref{fig:g22_decoh} and we notice that it fits very well the
exact result. Summarising, we have obtained, for each component of the
covariant matrix, precise and simple approximations valid on large
scales.

An interesting feature of the above calculations is the
  relationships that exist between the coefficients of the expansions
  of $\gamma_{11}$, $\gamma_{12}$ and $\gamma_{22}$. This can be
  understood as follows. Combining \Eqs{eq:diffcov1deco} and
  (\ref{eq:diffcov3deco}), one has
\begin{align}
    -\frac{\dd \gamma_{22}}{\dd x}=\frac{\omega^2}{k^2}
    \frac{\dd \gamma_{11}}{\dd x}+2 \left(\frac{k_\Gamma}{k}
    \right)^2x_*^{p-3}x^{3-p}.
    \end{align}
Then, one can insert \Eqs{eq:g11approx} and (\ref{eq:g22approx}) in
the above formula and this leads to
\begin{align}
  \frac{4}{x^5}-&2\left(\frac{k_\Gamma}{k}\right)^2\left[\frac{4B_{22}}{x^5}
  -(4-p)A_{22}x^{3-p}\right]=-\frac{2}{x^3}+\frac{4}{x^5}
  -2\left(\frac{k_\Gamma}{k}\right)^2\biggl[-\frac{2B_{11}}{x^3}
\nonumber \\ &
    +(6-p)A_{11}x^{5-p}
  +\frac{4B_{11}}{x^5}-2(6-p)A_{11}x^{3-p}\biggr]
+2\left(\frac{k_\Gamma}{k}\right)^2x_*^{p-3}x^{3-p} .
    \end{align}
At this stage, one has to remember that the expressions used above are
valid in the long time limit only. Therefore, the term $-2/x^3$ (first
term on the right hand side) can be neglected compared to $4/x^5$
(second term in the right hand side) and, indeed, the equation is
satisfied in the limit $k_\Gamma \rightarrow 0$. Applying the same
reasoning for the terms proportional to $k_\Gamma ^2$, one deduces
that $B_{22}=B_{11}$, a relation already established before but whose
origin is now understood, and
\begin{align}
  (4-p)A_{22}=2(6-p)A_{11}+x_*^{p-3}.
\end{align}
One checks that this equation is satisfied by $A_{11}$ and $A_{22}$
given in \Eqs{eq:A11} and~(\ref{eq:A22}). Of course, the above
considerations are just an example illustrating the origin of the
relationships between the coefficients. A systematic generalisation of
these calculations, with more terms in the expansions, would allow us
to derive all the relationships among the coefficients.

\bibliographystyle{JHEP}
\bibliography{biblio}

\end{document}

%% file: newcommands.tex
\newcommand{\ie}{{i.e.~}}

\let\oldsqrt\sqrt
\def\sqrt{\mathpalette\DHLhksqrt}
\def\DHLhksqrt#1#2{%
\setbox0=\hbox{$#1\oldsqrt{#2\,}$}\dimen0=\ht0
\advance\dimen0-0.2\ht0
\setbox2=\hbox{\vrule height\ht0 depth -\dimen0}%
{\box0\lower0.4pt\box2}}

\newcommand{\dd}{\mathrm{d}}
\newcommand{\ee}{e}

\newcommand{\boldmathsymbol}[1]{{\ensuremath{\boldsymbol{#1}}}}

\newcommand{\uin}{\mathrm{in}}

\newcommand{\uc}{\mathrm{c}}

\newcommand{\bmk}{\boldmathsymbol{k}}

\newcommand{\calH}{\mathcal{H}}
\newcommand{\calP}{\mathcal{P}}

\newcommand{\setR}{\mathbb{R}}

\newcommand{\Rea}{\Re \mathrm{e}\,}
\newcommand{\Ima}{\Im \mathrm{m}\,}

\newcommand{\efolds}{$e$-folds}

\newcommand{\beq}{\begin{equation}}
\newcommand{\eeq}{\end{equation}}
\newcommand{\bea}{\begin{equation}\begin{aligned}}
\newcommand{\eea}{\end{aligned}\end{equation}}

\newlength{\wsingfig}
\newlength{\wdblefig}
\newlength{\wquadfig}
\newlength{\wtriplefig}
\newcommand{\Eq}[1]{Eq.~(\ref{#1})}
\newcommand{\Eqs}[1]{Eqs.~(\ref{#1})}
\newcommand{\Fig}[1]{Fig.~{\ref{#1}}}

\newcommand{\Refa}[1]{Ref.~{\cite{#1}}}
\newcommand{\Refs}[1]{Refs.~{\cite{#1}}}
\newcommand{\Sec}[1]{Sec.~\ref{#1}}
\newcommand{\Secs}[1]{Secs.~\ref{#1}}
\newcommand{\App}[1]{Appendix~\ref{#1}}